\documentstyle[psfig]{l-aa}

\psfigurepath{/gagax4/evol/TEX/ARTICLES/AGB96.2/}

\newcommand {\chem}[2] {$\rm{}^{#2}\kern-0.8pt#1$}
\newcommand {\mass}[1] {$\rm #1\,M_{\sun}$}
\newcommand {\lum}[1] {$\rm #1\,L_{\sun}$}
\newcommand {\teff}[0] {$\rm T_{\rm e \mkern -1.8mu f \mkern -1.8mu f}\,$}
\newcommand {\reac}[6] {$\rm\,{}^{#2}\kern-0.8pt{#1}\,({#3}\,,{#4})
  \,{}^{#6}\kern-0.8pt{#5}\,$}

\def\dot#1{\hbox{$#1$ \kern -1.8ex \raisebox{1.7ex}.}\,}

\makeatletter
\def\cal{\fam\tw@}
\makeatother

\begin{document}

\title{Nucleosynthesis of light elements inside thermally pulsing AGB stars}

\subtitle{I: the case of intermediate-mass stars}

\author{M.~Forestini \inst{1} \and
  C.~Charbonnel \inst{2}}

\offprints{Manuel~Forestini}

\institute{Laboratoire d'Astrophysique, Observatoire de Grenoble,
  Universit\'e Joseph Fourier, BP 53, F--38041 Grenoble Cedex 9,
  France \and
  Laboratoire d'Astrophysique de Toulouse, CNRS UMR 5572, 
  Toulouse, France}

\date{Received ; Accepted for publication in Astronomy \& Astrophysics
Supplement Series}

\thesaurus{08.01.1;08.03.2;08.05.3;08.16.4}

\maketitle

\markboth{Nucleosynthesis in intermediate-mass AGB stars} {Forestini,
  Charbonnel}

\begin{abstract}

  The structural and nucleosynthetic evolution of 3, 4, 5, 6 and \mass{7}
  stars with two metallicities ($Z = 0.005$ and 0.02) has been computed in
  detail, from the early pre-main sequence phase up to the thermally pulsing
  (TP) AGB phase or the onset of off-center carbon burning. Typically 10 to
  20 thermal pulses have been followed for each TP-AGB object. This
  homogeneous and quite large set of models allows us to present an overview
  of the thermal pulse properties as well as of the nucleosynthesis
  accompanying the TP-AGB phase of intermediate-mass stars.

  More specifically, after a brief description of the previous evolutionary
  stages, predictions are given for the isotopic ratios involving C, N, O,
  Ne, Mg, Al and Si. Also the surface abundances of \chem{Li}{7},
  \chem{F}{19} and \chem{Na}{23} are reported. As the asymptotic phase of
  the thermal pulses has been reached for each star, we also indicate how
  these abundances will probably evolve until the stars completely loose
  their envelope, by including the evolution of the nucleosynthesis itself.

  This article, in its paper form, has been shortened at a level of roughly
  60 \% as required by directives coming from the A\&A editors. The complete
  article (50 pages containing 37 figures) is only available in electronic
  form.


\end{abstract}

\section {Introduction}

At the end of their life, stars of low- and intermediate-mass evolve along
the Asymptotic Giant Branch, where they experience recurrent thermal
instabilities and substantial mass loss. During this phase, the stars
undergo a very rich and unique nucleosynthesis. Moreover, recurrent
dredge-up events enrich the stellar surface with the freshly synthesized
nuclides which are then ejected into the interstellar medium through the
strong winds. The thermally pulsing AGB stars thus play a crucial role in
the chemical evolution of galaxies.

On the other hand, AGB stars are important contributors to the integrated
luminosity of stellar systems in a wide range of ages. A coverage of this
phase is needed in particular for the treatment of the luminosity functions
and color magnitude diagrams of star clusters. Last but not least, the
corresponding stellar models are basic tools to study carbon stars, OH/IR
stars, planetary nebulae and their central objects (white dwarfs).

In the last 15 years, extensive observational data has become available that
changed and highly constrained the theoretical view of the AGB evolution.
Most of the progresses came from Magellanic Cloud studies, which revealed
for example the absence of very luminous carbon stars (which were previously
predicted to originate from stars with initial masses between 5 and
\mass{8}), and the unexpected existence of relatively faint and lower mass
carbon stars (Blanco et al. 1980, Mould \& Aaronson 1982, 1986; Wood et al.
1983; Reid \& Mould 1984; Aaronson \& Mould 1985; Smith \& Lambert 1989).
This so-called ``carbon star mystery'' (Iben 1981) has led to the concept of
hot-bottom convective envelope burning (Iben 1975; Sackmann et al. 1974;
Scalo et al. 1975; Sackmann \& Boothroyd 1992). This mechanism, which
reconverts C-rich to O-rich envelopes and produces \chem{Li}{7}, was
confirmed by the existence of lithium enriched AGB stars (Smith \& Lambert
1989, 1990b; Plez et al. 1993; Smith et al. 1995). On the other hand, the
discovery, inside meteorites, of interstellar grains which have been formed
in the wind of cool carbon stars (see e.g. Zinner et al. 1991a),
complemented the stellar abundance analysis, and brought crucial
informations on AGB star nucleosynthesis. In spite of important
improvements, many important issues of AGB evolution still remain to be
addressed, observationally as well as theoretically.

This paper is the first of a series aimed at examining all the physical
processes that may induce abundance anomalies in the surface of evolved AGB
stars. In the present work, we specifically investigate the case of
intermediate-mass stars. Stars with initial masses between 3 and \mass{7}
and with metallicities $Z$ of 0.02 and 0.005 have been evolved, starting
from the pre-main sequence until they had gone through a number of
helium-shell flashes along the TP-AGB. The \mass{7} (6 and \mass{7}) stars
with $Z = 0.02$ (0.005) ignite off-center carbon burning and will not be
discussed in detail. For each other stellar mass, full evolutionary models
were computed up to the fourth ``full amplitude'' thermal pulse in the
so-called ``asymptotic regime''. Synthetic evolution was then used to follow
global surface properties and nucleosynthesis up to the planetary nebula
ejection. This self-consistent and extended set of models is particularly
relevant for population synthesis and chemical evolution models. Very
complete figures and tables containing informations about the center and
surface evolution of the structure and chemical composition for all our
models are available.

In the present computations, which include the latest input physics, the
evolution of the chemical composition is followed by only taking into
account nuclear reactions and mixing inside convective zones (we assume
instantaneous mixing, except when hot-bottom burning occurs, which requires
a time-dependent convective diffusion algorithm). We neglect all other
mixing mechanism of non-convective origin which may contribute to transport
chemical elements in radiative regions, like rotation-induced mixing or
diffusion. In the stellar mass range we consider, this approximation is
supported by the observed surface abundances at different evolutionary
stages. This is however certainly not the case for lower mass stars, for
which some ``non-standard'' mixing processes have to be invoked to explain
various chemical anomalies.

The physical ingredients of our models and numerical aspects of the
computations are presented in Sects. 2 and 3, respectively. The general
properties of the stars and their chemical evolution during the phases prior
to the TP-AGB one are summarized in Sects. 4 and 5, respectively.

In Sect. 6, we briefly present the structural features of the thermal pulses
and discuss the occurrence of the third dredge-up in our models. We derive a
core-mass luminosity relation for our less massive stars that do not undergo
hot-bottom burning. We then present the extrapolation procedure that we use
to follow both the evolution and the nucleosynthesis in the ``asymptotic
regime" up to the TP-AGB tip. We also give the initial mass--final mass
relation resulting from our models, that appears to be in good agreement
with the observed one.

In Sect. 7, we review the various nucleosynthesis sites inside TP-AGB stars,
namely the burning shells (He, \chem{C}{13} and H), the convective tongue
that develops during a thermal runaway and the base of hot convective
envelopes. We investigate in detail the nucleosynthetic processes that
involve the Li, CNO, F, NeNa and MgAl elements and neutrons. We then discuss
which specific nuclear region, inside a TP-AGB star, mainly contributes to
the surface abundance change of each nuclide when a third dredge-up occurs.

In Sect. 8, we finally present our predictions concerning {\it (i)} the
evolution of all the surface isotopic ratios by including our evolutionary
models and synthetic extrapolations up to the convective envelope removal
and {\it (ii)} the chemical yields ejected at different ages for all the
species up to \chem{Si}{28}. As in the previous sections, comparisons
between our predictions and observations are attentively performed.

In this paper version, Sects. 2, 3, 5 and 8 have been somewhat reduced.

\section {Physics of the models}

\subsection {Structure}

\subsubsection {Equation of state}

Our equation of state is analytic. It includes

\begin{itemize}

\item
  the ionization of H, He, C, N and O. Complete ionization is assumed when
  OIX/OVIII $> 50$;

\item
  the electrostatic corrections through the statistical model of Debye-H\"uckel
  for partially ionized regions and through the interpolation formalism of
  Koester (1976), between the Debye-H\"uckel and the Thomas-Fermi models,
  for the completely ionized matter;

\item
  the electronic degeneracy as soon as its contribution to the total
  pressure exceeds $0.1 \%$;

\item
  the production of neutrino pairs through four different processes, namely
  the photo-neutrinos, plasma neutrinos, $(e^+ \, e^-)$ pair annihilations
  and bremsstrahlung neutrinos. Energy loss contributions comes from
  Munakata et al. (1985), Itoh et al. (1992) and Kohyama et al. (1993).

\end{itemize}

\subsubsection {Opacities}

At low temperatures, i.e. below 8000 K, we use the Alexander \& Ferguson
(1994) opacity tables, very well suited for cool red giant envelopes and
atmospheres.

At temperatures above 8000 K, we use the OPAL opacity tables computed by
Rogers \& Iglesias (1992). The abundance distribution of the heavy elements
being scaled from the solar composition given by Anders \& Grevesse (1989)
and updated by Grevesse (1991), as for low-temperature opacities.

For both sets of opacity tables, we first interpolate bi-linearly in $T$
and $R = \log (\rho/T_6^3)$ for each table the composition of which is
close to the model composition and then linearly in $X$, $Z$ (and $X($O$)$)
to obtain the right opacity coefficient corresponding to each shell.

Finally, we used the Hubbard \& Lampe (1969) program to generate conductive
opacity tables corresponding to the same chemical compositions and $(R,T)$
grid points as for the OPAL tables.

\subsubsection {Atmosphere treatment}

The stellar structure is integrated from the center to a very low
optical depth ($\tau = 0.001$) in the atmosphere. We constraint the
atmospheric temperature profile to correspond to those coming from
realistic atmosphere models obtained by integrating the radiative
transfer equation. More specifically, we use atmosphere models {\it (1)}
from Plez (1992) for \teff $< 3900$ K, {\it (2)} from Eriksson (1994,
private communication, using a physics similar to Bell et al. 1976) up to
5500 K and {\it (3)} computed with the Kurucz atmosphere program above
5500 K. At each time step, we linearly interpolate a $T(\tau)$ profile
corresponding to our model \teff, $\log g$ and $Z$ from these grids of
atmosphere models and then correspondingly modify the radiative pressure
and gradient.

\subsubsection {Convection}

The structure of the convective regions is computed using the classical
Mixing-Length Theory (MLT), as prescribed by Kippenhahn et al. (1968).  Our
models are standard in the sense that {\it (1)} the Schwarzschild criterion
is considered to delimit the convective zones and {\it (2)} neither
overshooting nor semi-convection have been considered in the present
computations. The ratio $\alpha$ of the mixing-length free parameter over
the pressure scale height has been put to a value of 1.5, i.e. rather close
to $\alpha_{\sun} = 1.64$ with which we fit the solar structure.

\subsubsection {Mass Loss rates}

>From the Main Sequence up to the thermally pulsing AGB phase, we used the
empirical Reimers (1975) formula

\begin{equation}
  \dot M = -3.98 \, 10^{-13} \eta {L R \over M} \,\rm{M_{\sun} \,
    yr^{-1}} \,,
\end{equation}

\noindent with $\eta = 0.5$ up to the central He exhaustion.

>From the beginning of the AGB phase, we adopted variable $\eta$ values,
depending on the total mass of the star. More specifically, we took $\eta
= 2.5$, 3, 3.5 or 4 for our 3, 4, 5 or \mass{6} models respectively,
whatever $Z$. This $\eta$ increase with the total mass was found by Bryan
et al. (1990) to give rather good results along the AGB. However, later
during the thermally pulsing phase, Bl\"ocker (1995) recently showed that
the Bryan et al. (1990) formula underestimates the observed mass loss
rate. This leads us to further increase $\eta$ (see Sect. 6.3).
However, as mentioned by Vassiliadis \& Wood (1993), whatever empirical
formula, the mass loss rate increase used in the models along the
thermally pulsing AGB phase still remain uncertain by more than a factor
of two. As demonstrated in Sect. 8, this important uncertainty on the mass
loss rate formula conditions the AGB phase duration and consequently, the
surface composition evolution through the successive third dredge-up events.

\subsubsection {Rate of gravitational energy change}

We note that along the thermally pulsing AGB phase, the thermal pulse
features slightly depend on the way one estimates the rate
$\varepsilon_{grav}$ of gravitational energy release or gain (see e.g.
Kippenhahn \& Weigert 1991). So, let us specify that we wrote

\begin{equation}
  \varepsilon_{grav} = - {D u \over D t} + {P \over \rho^2} {D \rho \over
    D t} \,,
\end{equation}

\noindent where $u$ is the specific (i.e. by unit mass) internal energy.

\subsection {Nucleosynthesis}

\subsubsection {Selected nuclides and network}

We follow the abundance evolution, through the whole stars, of 45 nuclides,
namely the neutrons, all the 31 stable nuclides up to \chem{S}{33} as well
as \chem{Be}{7}, \chem{B}{8}, \chem{C}{11}, \chem{C}{14}, \chem{N}{13},
\chem{O}{15}, \chem{F}{18}, \chem{Na}{22}, \chem{Al^g}{26}, \chem{Al^m}{26},
\chem{Si}{27}, \chem{P}{30} and \chem{S}{31}.

These nuclides interact through a network containing 172 nuclear (neutron,
proton and \chem{He}{4} captures) and decay reactions. Many nuclear reaction
rates are taken from Caughlan \& Fowler (1988). Let us just mention the
nuclear reactions for which we selected another rate prescription.

\begin{itemize}

\item
  \reac{C}{12}{n}{\gamma}{}{} from Nagai et al. (1991),

\item
  \reac{C}{12}{\alpha}{\gamma}{}{} from Caughlan et al. (1985),

\item
  \reac{C}{13}{n}{\gamma}{}{} from Raman et al. (1990),

\item
  \reac{C}{14}{n}{\gamma}{}{} from Beer et al. (1991),

\item
  \reac{C}{14}{p}{\gamma}{}{} from Wiesher et al. (1990),

\item
  \reac{C}{14}{\alpha}{\gamma}{}{} from Funck \& Langanke (1989),

\item
  \reac{C}{14}{\alpha}{n}{}{}, \reac{N}{14}{n}{\gamma}{}{},
  \reac{N}{15}{n}{\gamma}{}{} and \reac{O}{17}{n}{\alpha}{}{} from Fowler
  et al. (1967),

\item
  \reac{N}{14}{n}{p}{}{} from Brehm et al. (1988),

\item
  \reac{N}{15}{\alpha}{\gamma}{}{} from De Oliveira (1995),

\item
  \reac{O}{16}{n}{\gamma}{}{} from Allen et al. (1971),

\item
  \reac{O}{17}{n}{\gamma}{}{} from Wagoner (1969),

\item
  \reac{O}{17}{p}{\gamma}{}{} and \reac{O}{17}{p}{\alpha}{}{} from Landr\'e
  et al. (1990),

\item
  \reac{O}{18}{n}{\gamma}{}{} from Thielemann (1991),

\item
  \reac{F}{19}{n}{\gamma}{}{} and \reac{Ne}{21}{n}{\gamma}{}{} from Bao \&
  K\"appeler (1987),

\item
  \reac{F}{19}{p}{\alpha}{}{} from Kious (1990),

\item
  \reac{Ne}{20}{n}{\gamma}{}{}, \reac{Ne}{22}{n}{\gamma}{}{},
  \reac{Na}{23}{n}{\gamma}{}{}, \reac{Mg}{24}{n}{\gamma}{}{},
  \reac{Mg}{25}{n}{\gamma}{}{} and \reac{Mg}{26}{n}{\gamma}{}{} from Beer
  et al. (1991),

\item
  \reac{Ne}{22}{\alpha}{\gamma}{}{} and \reac{Ne}{22}{\alpha}{n}{}{} from
  Hammer (1991),

\item
  \reac{Na}{22}{n}{\gamma}{}{}, \reac{Na}{23}{\alpha}{p}{}{},
  \reac{Al^g}{26}{n}{\gamma}{}{} and \reac{Al^m}{26}{n}{\gamma}{}{} from
  Woosley et al. (1978),

\item
  \reac{Mg}{24}{\alpha}{p}{}{} and \reac{Al}{27}{p}{\alpha}{}{} from
  Champagne et al. (1988),

\item
  \reac{Mg}{25}{p}{\gamma}{}{} and \reac{Mg}{26}{p}{\gamma}{}{} from Illiadis
  (1990),

\item
  and \reac{Al^g}{26}{p}{\gamma}{}{} and $\rm{{}^{26}Al^g (\gamma)
    {}^{26}Al^m}$ from Vogelaar (1989).

\end{itemize}

On the other hand, the nuclear reaction rates for \reac{F}{18}{n}{p}{}{},
\reac{F}{18}{n}{\alpha}{}{}, \reac{Mg}{25}{\alpha}{\gamma}{}{},
\reac{Mg}{25}{\alpha}{p}{}{}, \reac{Mg}{26}{\alpha}{\gamma}{}{},
\reac{Mg}{26}{\alpha}{p}{}{}, \reac{Mg}{26}{\alpha}{n}{}{},
\reac{Al}{27}{\alpha}{n}{}{} as well as the 45 nuclear reactions involving
Si, P and S are fitted with a procedure developed by Rayet (1993, private
communication). Let us finally mention recent improvements not yet
incorporated in our network: various $\rm{(n , \gamma)}$ reactions rates on
light nuclei (Nagai et al. 1995) and $\alpha$ capture rates on \chem{O}{18}
by K\"appeler et al. (1994).

Finally, the nuclear screening factors are parameterized by using the
Graboske et al. (1973) formalism, including weak, intermediate and strong
screening cases.

\subsubsection {Neutron abundance}

Neutrons can be produced during the thermally pulsing AGB phase. They can,
in particular, be captured by nuclides heavier than \chem{S}{33}, mainly
through $\rm{(n , \gamma)}$ reactions. As we do not follow the abundance of
such elements, we created an additional nuclide, called
\chem{S^{\Sigma}}{34}, the mass fraction of which is set to

\begin{equation}
  X(\rm{{}^{34}S^{\Sigma}}) = \sum_{a = 34}^{209} \, X_a \,.
\end{equation}

\noindent This sum counts in fact 253 stable nuclides up to \chem{Bi}{209}.
In order to have a good estimation of the free neutron abundance, we follow
a method described by Jorissen \& Arnould (1989).

\subsubsection {Mixing inside convective zones}

If nuclear reactions occur inside a convective zone, most of them proceed in
general with a longer time scale than the turn-over time scale associated to
the motion of convective cells. In such a case, one usually consider the
convective mixing as instantaneous. This allows to treat nucleosynthesis in
one shot, by taking mass-weighted averages of the number abundances and
nuclear reaction rates over the whole convective region.

However, if some key nuclides nuclearly evolve more rapidly than they are
mixed, we have to consider the convection transport together with the
nuclear reactions, so that nucleosynthesis equations become diffusion
equations of the form

\begin{equation}
  {\partial Y_i \over \partial t} = {1 \over \rho r^2} {\partial \over
    \partial r} \left[ \rho r^2 D_{CONV} {\partial Y_i \over \partial r}
  \right] \, + \, \makebox{nuclear terms} \,,
\end{equation}

\noindent where $Y_i = X_i / A_i$, the ratio of the mass fraction over
the atomic mass of each nuclide $i$, and $D_{CONV} = \bar{v}^2 \tau_{CONV}$,
the convective diffusion coefficient, with $\bar{v}$ and $\tau_{CONV}$ being
the mean velocity and turn-over time of the convective cells, respectively.
It is very CPU time consuming to treat the nucleosynthesis and convective
mixing together through such diffusion equations for each nuclide in each
convective shell for all the time steps. Consequently, as usual,
instantaneous mixing has been assumed by default for our stellar evolution
computations, except in two cases that potentially require a time-dependent
treatment of the convective mixing: {\it (1)} the convective tongue
associated with thermal pulses and {\it (2)} the base of the convective
envelope when it is hot enough, both along the AGB phase.

We verified that inside thermal pulses, the elements that are the most
affected by a diffusion treatment of the convective mixing are the very
unstable nuclides (like \chem{Na}{22}), neutrons and protons. However, this
do not change very significantly the nucleosynthesis global results (see
however Sect. 7.5).

On the contrary, the mean turn-over time scale of the convective motions is
much longer inside the convective envelope than inside a thermal pulse.  If
nuclear burning occurs at the base of the convective envelope, some
important nuclear reactions can, depending on the bottom temperature, occur
faster than the convective mixing so that a diffusion treatment is required.
This is in particular the case for some of the reactions involved in the
\chem{Li}{7} synthesis (see Sect. 7.2).

\subsection {Initial models}

We have built initial models of 3, 4, 5, 6 and \mass{7} allowing us to start
our computations at the beginning of the pre-main sequence phase (along the
Hayashi track), when central temperatures are below $5 \, 10^5$ K.

For our $Z = 0.005$ (0.02) models, the hydrogen mass fraction has been set
to $X = 0.745$ (0.687) and that of helium to $Y = 0.250$ (0.293). The
abundances of the heavier elements have been scaled on the Anders \&
Grevesse (1989) abundance distribution.

\section {Numerical aspects}

We solve the five one-dimensional hydrodynamical equations for the evolution
of the stellar structure with the mass coordinate as independent variable.

\subsection {Structure}

We first convert the five non-linear stellar structure differential
equations into difference equations following the Henyey's scheme by
defining discrete shells in mass. We then solve the resulting linearized
system of equations by the Newton-Raphson relaxation procedure (see e.g.
Press et al. 1986). Typically, 600 shells are required for a model on the
main sequence phase, while along the AGB, 1000 to 1300 shells are needed
(200 to 500 more shells being added during a thermal pulse).

On the other hand, typically 200 (700) models are necessary to model the
central hydrogen (helium) burning phase. During the AGB phase, 2500 to 4500
models are required to model the time separating two successive thermal
pulses, while the thermal pulse itself requires 100 to 350 models.

Let us stress that these rather small numbers of mass shells and time steps
have been chosen in order to save computer time and present a very large set
of nucleosynthesis predictions. Straniero et al. (1996) very recently showed
that this is probably in the disadvantage of the occurrence of the third
dredge-up along the TP-AGB phase.

Finally, the mass $\Delta M = \dot M \Delta t$ lost during each time step
$\Delta t$ is suppressed proportionally to the shell masses of the whole
envelope above the burning regions.

\subsection {Nucleosynthesis}

The nucleosynthesis equations are solved using the Wagoner (1969) numerical
technique, well suited for our purposes. However, if a diffusion treatment
is required to simultaneously compute the time-dependent convective mixing
and the nuclear burning, the diffusion Eqs. (4) are then solved with the
same Newton-Raphson algorithm.

\section {Summary of the global evolution previous to TP-AGB phase}

\subsection {Some results}

\begin{figure*}
  \psfig{file=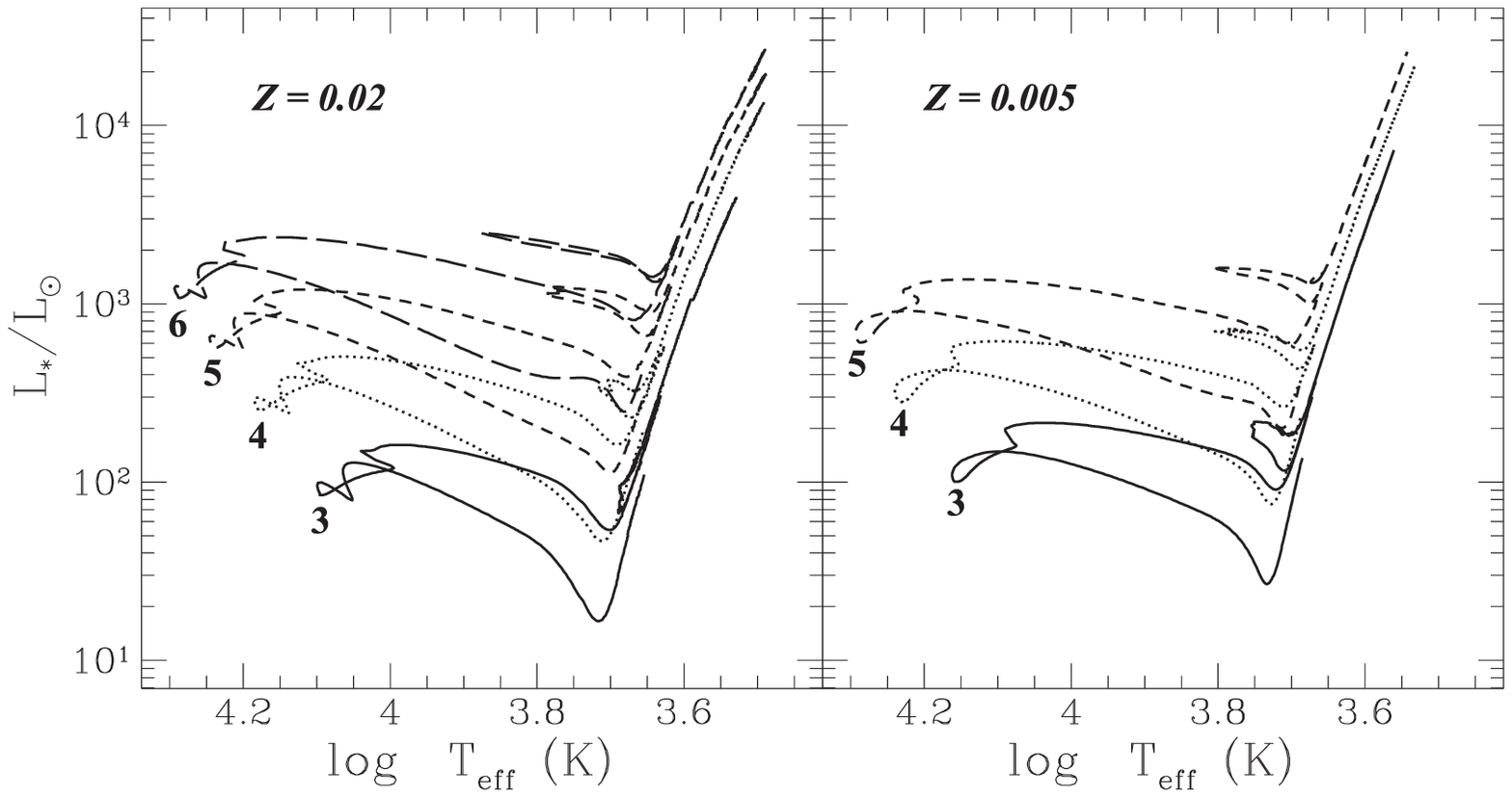,width=18.0cm}
  \caption[ ]{\label{hrd}
    Theoretical HRD, from the beginning of the PMS phase up to the beginning
    of the first thermal pulse along the AGB phase, for $Z = 0.02$ (left) and
    $Z = 0.005$ (right)
    }
\end{figure*}

Theoretical Hertzsprung-Russell diagrams (HRD) for the ensemble of our
models are shown in Fig. \ref{hrd}, from the early PMS phase up to the
beginning of the first thermal pulse (or He-shell flash) along the AGB
phase.

The lifetimes are given in Table \ref{times} for the PMS, the H-burning
phase (from the zero-age-main sequence to central H-exhaustion), the RGB
phase (from central H-exhaustion to central He-ignition), the He-burning
phase (up to central He-exhaustion), the early (E) AGB phase (up to the
first thermal pulse) and the thermally pulsing (TP) AGB phase. This last
duration includes our extrapolation procedure up to the convective envelope
exhaustion (see Sect. 6.3).

\begin{table*}
  \caption[ ]{\label{times}
    Lifetimes in all the evolutionary phases for the modeled stars which
    undergo thermal pulses
    }
  \begin{flushleft}
    \begin{tabular}{lccccccc}
      \noalign{\smallskip}
      \hline
      \noalign{\smallskip}
      &\multicolumn{4}{c}{$Z = 0.02$}&\multicolumn{3}{c}{$Z = 0.005$} \\
      \noalign{\vspace{-2.0mm}}
      &\multicolumn{4}{c}{\makebox[62mm]{\downbracefill}}&\multicolumn{3}{c}{\makebox[44mm]{\downbracefill}}\\
      &\mass{3}&\mass{4}&\mass{5}&\mass{6}&\mass{3}&\mass{4}&\mass{5} \\
      \noalign{\smallskip}
      \hline \hline
      \noalign{\smallskip}
      $\Delta t$ [PMS] $\, (\rm{yr})$&1.20$\, 10^{ 7}$&5.88$\, 10^{ 6}$&3.70$\, 10^{ 6}$&2.19$\, 10^{ 6}$&1.10$\, 10^{ 7}$&5.82$\, 10^{ 6}$&3.50$\, 10^{ 6}$ \\
      $\Delta t$ [MS] $\, (\rm{yr})$&2.80$\, 10^{ 8}$&1.34$\, 10^{ 8}$&7.69$\, 10^{ 7}$&5.09$\, 10^{ 7}$&2.59$\, 10^{ 8}$&1.30$\, 10^{ 8}$&7.88$\, 10^{ 7}$ \\
      $\Delta t$ [RG] $\, (\rm{yr})$&1.78$\, 10^{ 7}$&6.39$\, 10^{ 6}$&2.86$\, 10^{ 6}$&1.22$\, 10^{ 6}$&1.45$\, 10^{ 7}$&5.17$\, 10^{ 6}$&2.38$\, 10^{ 6}$ \\
      $\Delta t$ [HB] $\, (\rm{yr})$&1.55$\, 10^{ 8}$&5.26$\, 10^{ 7}$&2.64$\, 10^{ 7}$&1.08$\, 10^{ 7}$&9.37$\, 10^{ 7}$&4.30$\, 10^{ 7}$&2.59$\, 10^{ 7}$ \\
      $\Delta t$ [E-AGB] $\, (\rm{yr})$&6.16$\, 10^{ 6}$&2.88$\, 10^{ 6}$&1.25$\, 10^{ 6}$&5.98$\, 10^{ 5}$&4.29$\, 10^{ 6}$&1.69$\, 10^{ 6}$&6.41$\, 10^{ 5}$ \\
      $\Delta t$ [TP-AGB] $\, (\rm{yr})$&1.22$\, 10^{ 6}$&2.26$\, 10^{ 5}$&1.33$\, 10^{ 5}$&9.12$\, 10^{ 4}$&9.03$\, 10^{ 5}$&1.59$\, 10^{ 5}$&7.97$\, 10^{ 4}$ \\
      \noalign{\smallskip}
      \hline
      \noalign{\smallskip}
    \end{tabular}
  \end{flushleft}
\end{table*}

The occurrence of the first thermal pulse marks the end of the E-AGB phase
and the beginning of the TP-AGB phase. This precise time, marked by the
maximum surface luminosity just preceding the first major thermal
instability, will be called $t_0$ in the following. Table \ref{t0} presents
some characteristics of our models at $t_0$.

\begin{table*}
  \caption[ ]{\label{t0}
    Main features concerning the internal and surface structure of our
    models at the end of the E-AGB phase.  $[He]$ ($[H]$) specifies values
    taken at the maximum energy production rate in the helium (hydrogen)
    burning shell.  $[H-He]$ refers to the inter-shell region (i.e. between
    the HeBS top and the HBS base) and $[env]$ to the convective envelope
    bottom. $\Delta m$ is the mass thickness of a region }
  \begin{flushleft}
    \begin{tabular}{lccccccc}
      \noalign{\smallskip}
      \hline
      \noalign{\smallskip}
      &\multicolumn{4}{c}{$Z = 0.02$}&\multicolumn{3}{c}{$Z = 0.005$} \\
      \noalign{\vspace{-2.0mm}}
      &\multicolumn{4}{c}{\makebox[67mm]{\downbracefill}}&\multicolumn{3}{c}{\makebox[48mm]{\downbracefill}}\\
      &\mass{3}&\mass{4}&\mass{5}&\mass{6}&\mass{3}&\mass{4}&\mass{5} \\
      \noalign{\smallskip}
      \hline \hline
      \noalign{\smallskip}
      $t_{0} \, (\rm{yr})$&4.739$\, 10^{ 8}$&2.034$\, 10^{ 8}$&1.123$\, 10^{ 8}$&6.624$\, 10^{ 7}$&3.851$\, 10^{ 8}$&1.880$\, 10^{ 8}$&1.117$\, 10^{ 8}$ \\
      \noalign{\smallskip}
      $L \, (\rm{L_{\odot}})$& 3921&13274&19367&25743& 5420&14761&21479 \\
      $R \, (\rm{R_{\odot}})$&182&398&486&554&182&339&387 \\
      $\Delta R[atmos.] \, (\rm{R_{\odot}})$& 3&15&17&18& 3&10&10 \\
      $M \, (\rm{M_{\odot}})$&2.92&3.77&4.70&5.72&2.81&3.51&4.31 \\
      $\dot M \, (\rm{M_{\odot} \, yr^{-1}})$&2.0$\, 10^{-7}$&1.5$\, 10^{-6}$&2.5$\, 10^{-6}$&3.6$\, 10^{-6}$&3.6$\, 10^{-7}$&1.8$\, 10^{-6}$&2.8$\, 10^{-6}$ \\
      $\rm{T_{eff} \, (K)}$&3381&3102&3088&3104&3667&3450&3550 \\
      $\rm{\log g_{eff} \, (g \, cm^{-3})}$&  0.38& -0.19& -0.26& -0.29&  0.36& -0.08& -0.10 \\
      \noalign{\smallskip}
      $L_{nucl}/L$& 0.96& 0.87& 0.80& 0.71& 0.94& 0.79& 0.66 \\
      $L_{He}/L_{H}$& 0.09& 0.08& 0.07& 0.06& 0.06& 0.13& 0.10 \\
      $\eta_{c}$&20&30&37&39&20&32&39 \\
      $T_{max} \, (K)$&2.11$\, 10^{ 8}$&3.48$\, 10^{ 8}$&3.97$\, 10^{ 8}$&4.60$\, 10^{ 8}$&2.44$\, 10^{ 8}$&3.91$\, 10^{ 8}$&4.54$\, 10^{ 8}$ \\
      $m_{Tmax} \, (\rm{M_{\odot}})$& 0.320& 0.521& 0.643& 0.745& 0.373& 0.608& 0.736 \\
      \noalign{\smallskip}
      $r[He] \, (\rm{R_{\odot}})$& 0.00008& 0.00003& 0.00003& 0.00002& 0.00008& 0.00004& 0.00003 \\
      $m[He] \, (\rm{M_{\odot}})$& 0.520& 0.732& 0.815& 0.894& 0.594& 0.814& 0.903 \\
      $\Delta m[He] \, (\rm{M_{\odot}})$& 0.08& 0.03& 0.02& 0.24& 0.04& 0.01& 0.23 \\
      $T[He] \, \rm{(K)}$&1.43$\, 10^{ 8}$&1.61$\, 10^{ 8}$&1.64$\, 10^{ 8}$&1.68$\, 10^{ 8}$&1.37$\, 10^{ 8}$&1.64$\, 10^{ 8}$&1.68$\, 10^{ 8}$ \\
      $\rho[He] \, \rm{(g \, cm^{-3})}$&13.8& 7.7& 5.9& 5.9&11.2& 5.3& 4.6 \\
      \noalign{\smallskip}
      $\Delta m[H-He] \, (\rm{M_{\odot}})$&4.0$\, 10^{ -3}$&4.7$\, 10^{ -4}$&1.5$\, 10^{ -4}$&6.0$\, 10^{ -5}$&3.3$\, 10^{ -3}$&3.8$\, 10^{ -4}$&1.2$\, 10^{ -4}$ \\
      \noalign{\smallskip}
      $r[H] \, (\rm{R_{\odot}})$& 0.00016& 0.00006& 0.00004& 0.00003& 0.00015& 0.00006& 0.00005 \\
      $m[H] \, (\rm{M_{\odot}})$& 0.542& 0.738& 0.818& 0.895& 0.609& 0.818& 0.905 \\
      $\Delta m[H] \, (\rm{M_{\odot}})$&1.2$\, 10^{-3}$&3.2$\, 10^{-4}$&1.8$\, 10^{-4}$&5.5$\, 10^{-3}$&9.7$\, 10^{-4}$&3.4$\, 10^{-4}$&2.2$\, 10^{-4}$ \\
      $T[H] \, \rm{(K)}$&5.07$\, 10^{ 7}$&6.43$\, 10^{ 7}$&7.10$\, 10^{ 7}$&7.77$\, 10^{ 7}$&5.75$\, 10^{ 7}$&7.00$\, 10^{ 7}$&7.74$\, 10^{ 7}$ \\
      $\rho[H] \, \rm{(g \, cm^{-3})}$&2.0$\, 10^{-2}$&1.4$\, 10^{-2}$&1.3$\, 10^{-2}$&1.1$\, 10^{-2}$&2.4$\, 10^{-2}$&1.7$\, 10^{-2}$&1.6$\, 10^{-2}$ \\
      \noalign{\smallskip}
      $r[env] \, (\rm{R_{\odot}})$& 0.0042& 0.0008& 0.0002& 0.0001& 0.0044& 0.0008& 0.0002 \\
      $m[env] \, (\rm{M_{\odot}})$& 0.544& 0.738& 0.818& 0.896& 0.612& 0.819& 0.905 \\
      $T[env] \, \rm{(K)}$&2.68$\, 10^{ 6}$&5.72$\, 10^{ 6}$&1.58$\, 10^{ 7}$&4.27$\, 10^{ 7}$&2.71$\, 10^{ 6}$&6.53$\, 10^{ 6}$&2.25$\, 10^{ 7}$ \\
      $\rho[env] \, \rm{(g \, cm^{-3})}$&1.1$\, 10^{-3}$&4.1$\, 10^{-3}$&5.7$\, 10^{-2}$& 0.9&1.0$\, 10^{-3}$&6.0$\, 10^{-3}$& 0.2 \\
      \noalign{\smallskip}
      \hline
      \noalign{\smallskip}
    \end{tabular}
  \end{flushleft}
\end{table*}

\subsection {Comparisons with other works}

For the main sequence duration, our results for $Z = 0.02$ are in better
agreement with Bressan et al. (1993) predictions than with the slightly
higher values given by Schaller et al. (1992). The central helium burning
lifetime crucially depends on the amount of overshooting. As our stellar
models are computed without overshooting, we can only compare our
predictions at this phase with those of the Padova group calculated in the
same conditions; we obtain very similar results.

The other important quantity for the AGB phase is the core mass (i.e.
the mass contained up to the HBS) at the end of the central He-burning
phase. Indeed, its value determines the surface luminosity (through the
core mass--luminosity relation; see e.g. Boothroyd \& Sackmann 1988a), the
mass loss rate, and consequently the AGB phase duration. Last but not
least, along the TP-AGB phase, the thermal pulses are stronger and the
third dredge-up deeper when the core mass is higher. Our \mass{3} star
with $Z = 0.02$ enter the AGB phase with a core mass of \mass{0.542} (see
Table \ref{t0}), a value very close to that of the comparable model of
Boothroyd \& Sackmann (1988a) or Lattanzio (1986), also computed without
overshooting. Always for a reference \mass{3} star, the TP-AGB phase
of the Straniero et al. (1995) models begins with a core mass of
\mass{0.53} (2 \% lower). At a comparable evolution stage however,
the \mass{5} TP-AGB star of Vassiliadis \& Wood (1993), also computed
without any kind of extra-mixing, has a core mass $\sim 4$ \% higher than
ours. This could be due to the somewhat lower metallicity of their models
($Z = 0.016$). In conclusion, the core mass at the beginning of the
TP-AGB phase, although very important for this phase, still depends on
various uncertainties related to the treatment of convection during the
previous evolutionary phases.

\section {Surface abundances prior to TP-AGB phase}

\subsection {First and second dredge-up events}

During the first and second dredges-up, convective mixing and induced
dilution modify the surface abundances which were unaltered until this phase
(Iben 1964). Table \ref{dup12} presents the resulting changes for the
principal elements altered by H-burning. Let us emphasize the main points.

A second dredge-up occurs in our $\geq$ \mass{4} models, also leading to
surface abundance changes. For our $\geq$ \mass{5} models, the convective
envelope penetrates so deep that it pushes down the H--He discontinuity.

\begin{table*}
  \caption[ ]{\label{dup12}
    Surface abundances and isotopic ratios consecutive to the first and
    second dredge-up episodes (i.e. respectively at the top RG and time
    $t_0$ defined in the text) }
  \begin{flushleft}
    \begin{tabular}{llccccccc}
      \noalign{\smallskip}
      \hline
      \noalign{\smallskip}
      & &\multicolumn{4}{c}{$Z = 0.02$}&\multicolumn{3}{c}{$Z = 0.005$} \\
      \noalign{\vspace{-2.0mm}}
      & &\multicolumn{4}{c}{\makebox[73mm]{\downbracefill}}&\multicolumn{3}{c}{\makebox[52mm]{\downbracefill}}\\
      & &\mass{3}&\mass{4}&\mass{5}&\mass{6}&\mass{3}&\mass{4}&\mass{5} \\
      \noalign{\smallskip}
      \hline \hline
      \noalign{\smallskip}
      \chem{He}{3}&{\it initial}&3.22$\, 10^{-5}$&3.19$\, 10^{-5}$&3.13$\, 10^{-5}$&3.12$\, 10^{-5}$&2.71$\, 10^{-5}$&2.71$\, 10^{-5}$&2.66$\, 10^{-5}$ \\
      &{\it top RG}&1.55$\, 10^{-4}$&1.05$\, 10^{-4}$&8.52$\, 10^{-5}$&7.32$\, 10^{-5}$&1.71$\, 10^{-4}$&1.28$\, 10^{-4}$&1.08$\, 10^{-4}$ \\
      &{\it at $t_{0}$}&1.54$\, 10^{-4}$&1.03$\, 10^{-4}$&7.79$\, 10^{-5}$&6.65$\, 10^{-5}$&1.66$\, 10^{-4}$&1.13$\, 10^{-4}$&8.56$\, 10^{-5}$ \\
      \noalign{\smallskip}
      \chem{He}{4}&{\it initial}& 0.2932& 0.2932& 0.2932& 0.2932& 0.2495& 0.2495& 0.2495 \\
      &{\it top RG}& 0.3047& 0.3039& 0.2995& 0.3017& 0.2585& 0.2505& 0.2498 \\
      &{\it at $t_{0}$}& 0.3054& 0.3052& 0.3367& 0.3421& 0.2624& 0.2629& 0.3014 \\
      \noalign{\smallskip}
      \chem{Li}{7}&{\it initial}&9.90$\, 10^{ -9}$&9.90$\, 10^{ -9}$&9.90$\, 10^{ -9}$&9.90$\, 10^{ -9}$&2.47$\, 10^{ -9}$&2.47$\, 10^{ -9}$&2.47$\, 10^{ -9}$ \\
      &{\it top RG}&1.44$\, 10^{-10}$&1.44$\, 10^{-10}$&1.43$\, 10^{-10}$&1.44$\, 10^{-10}$&3.14$\, 10^{-11}$&3.51$\, 10^{-11}$&3.86$\, 10^{-11}$ \\
      &{\it at $t_{0}$}&1.20$\, 10^{-10}$&9.85$\, 10^{-11}$&7.39$\, 10^{-11}$&7.53$\, 10^{-17}$&1.49$\, 10^{-11}$&1.68$\, 10^{-11}$&3.59$\, 10^{-16}$ \\
      \noalign{\smallskip}
      \chem{C}{12}/\chem{C}{13}&{\it initial}&83.1&83.1&83.1&83.1&83.1&83.1&83.1 \\
      &{\it top RG}&19.3&19.0&18.8&18.5&18.7&18.4&18.3 \\
      &{\it at $t_{0}$}&19.1&18.6&17.8&17.5&18.1&17.5&17.4 \\
      \noalign{\smallskip}
      \chem{C}{12}/\chem{N}{14}&{\it initial}&2.745&2.745&2.745&2.745&2.745&2.745&2.745 \\
      &{\it top RG}& 0.725& 0.714& 0.769& 0.735& 0.636& 0.888&1.100 \\
      &{\it at $t_{0}$}& 0.712& 0.684& 0.565& 0.538& 0.581& 0.610& 0.496 \\
      \noalign{\smallskip}
      \chem{N}{14}/\chem{N}{15}&{\it initial}&  253&  253&  253&  253&  253&  253&  253 \\
      &{\it top RG}& 1243& 1281& 1210& 1279& 1397& 1033&  856 \\
      &{\it at $t_{0}$}& 1270& 1346& 1654& 1774& 1546& 1510& 1841 \\
      \noalign{\smallskip}
      \chem{C}{12}/\chem{O}{16}&{\it initial}& 0.316& 0.316& 0.316& 0.316& 0.316& 0.316& 0.316 \\
      &{\it top RG}& 0.205& 0.207& 0.210& 0.211& 0.194& 0.208& 0.229 \\
      &{\it at $t_{0}$}& 0.204& 0.205& 0.206& 0.207& 0.191& 0.192& 0.197 \\
      \noalign{\smallskip}
      \chem{O}{16}/\chem{O}{17}&{\it initial}&2467&2467&2467&2467&2467&2467&2467 \\
      &{\it top RG}& 314& 409& 562& 617& 258& 869&1814 \\
      &{\it at $t_{0}$}& 311& 391& 508& 579& 245& 323& 358 \\
      \noalign{\smallskip}
      \chem{O}{16}/\chem{O}{18}&{\it initial}& 442& 442& 442& 442& 442& 442& 442 \\
      &{\it top RG}& 589& 582& 577& 577& 616& 573& 522 \\
      &{\it at $t_{0}$}& 591& 588& 588& 588& 623& 619& 611 \\
      \noalign{\smallskip}
      \hline
      \noalign{\smallskip}
    \end{tabular}
  \end{flushleft}
\end{table*}

\subsubsection {\chem{He}{3}}

\chem{He}{3} is produced at the beginning of the T Tauri phase from the
initial \chem{H}{2} burning. On the main sequence, a peak of \chem{He}{3}
appears, due to the competition of the reactions \reac{p}{}{p}{e^+ \,
  \nu_e}{H}{2}\reac{}{}{p}{\gamma}{He}{3} and
\reac{He}{3}{{}^{3}He}{2p}{He}{4}.  During the first dredge-up, the surface
abundance of \chem{He}{3} thus increases. However, the higher the stellar
mass, the lower the \chem{He}{3} peak, and the lower the surface enhancement
of \chem{He}{3} after the first dredge-up.

\subsubsection {CNO elements}

The first (and second) dredge-up phase(s) lead to a decrease of the
\chem{C}{12}/\chem{C}{13}, \chem{C}{12}/\chem{N}{14},
\chem{N}{15}/\chem{N}{14} and \chem{O}{16}/\chem{O}{17} isotopic ratios, the
extent of which depends on both stellar mass and metallicity.

\begin{itemize}

\item
  From an initial value equal to 83, the \chem{C}{12}/\chem{C}{13} ratio
  decreases  to $\simeq 18-19$. The higher the stellar mass, the lower the
  final carbon isotopic ratio.

  This ratio slightly further decreases after the second dredge-up.

\item
  For $Z = 0.02$, the maximum depth of the convective envelope reaches
  the deepest step of the \chem{N}{14} profile, for $\geq$ \mass{3} stars.
  This implies that \chem{C}{12}/\chem{N}{14}, as well as
  \chem{N}{15}/\chem{N}{14} have almost the same value in these models after
  the first dredge-up. On the contrary, for the $Z = 0.005$ objects, 
  the lower the stellar mass, the less efficient the dredge-up, and
  consequently, the less \chem{N}{14} is up-heaved to the surface. This
  explains the trend for the final $^{12}$C/$^{14}$N ratio versus stellar
  mass (see Charbonnel 1994 for more details).

  After the second dredge-up, whatever $Z$, the convective envelope of 
  $\geq$ \mass{4} stars penetrates deeper for increasing stellar mass as
  attested by both \chem{C}{12}/\chem{N}{14} and \chem{N}{15}/\chem{N}{14}
  ratios. 

\item
  The theoretical post dredge-up value of \chem{O}{16}/\chem{O}{17}
  increases with the stellar mass. Moreover, since for $Z = 0.005$ the
  deepening convective envelope of our 4 and \mass{5} stars does not reach
  the regions where the ON cycle operated, the post dredge-up values of
  \chem{O}{16}/\chem{O}{17} are higher than for $Z = 0.02$.

  After the second dredge-up however, \chem{O}{17} enriched matter is
  up-heaved in the convective envelope leading to very low
  \chem{O}{16}/\chem{O}{17}. 

\item
  The predicted post dredge-up value of \chem{O}{16}/\chem{O}{18} slightly
  decreases with increasing stellar mass. 

\end{itemize}

\subsection {Comparisons with other works}

Our \chem{He}{3} surface abundance increases are in rather good agreement
with the predictions by Bressan et al. (1993).

Concerning the CNO elements, our results are in general good agreement with
the predictions by Schaller et al. (1992), Bressan et al. (1993) and El Eid
(1994). The only important discrepancy between theoretical predictions of
the different groups for intermediate-mass stars concerns the post dredge-up
\chem{O}{16}/\chem{O}{17} ratio, which value highly depends on the adopted
\reac{O}{17}{p}{\alpha}{N}{14} and \reac{O}{17}{p}{\gamma}{F}{18} reaction
rates. We use the rates given by Landr\'e et al. (1990), as do Schaller et
al. (1992), and we obtain \chem{O}{16}/\chem{O}{17} ratios very similar to
those of the Geneva group. On the other hand, Bressan et al. (1993) use the
lower \chem{O}{17} proton capture rates given by Caughlan \& Fowler (1988),
and obtain much lower \chem{O}{16}/\chem{O}{17} ratios [see also El Eid
(1994) for a discussion of the influence of the adopted \chem{O}{17}
destruction rate on the resulting \chem{O}{16}/\chem{O}{17} ratio].

\subsection {Comparisons with observations}

In our intermediate-mass stars, the theoretical post dredge-up values of the
carbon isotopic ratio are slightly lower than the observations by Gilroy
(1989) in galactic cluster giants.

Red giants in this evolutionary phase present \chem{O}{16}/\chem{O}{17}
ratios between 300 to 1000 and \chem{O}{16}/\chem{O}{18} ratios in the range
400 to 600 (Harris et al. 1988; Smith \& Lambert 1990a).  This is in very
good agreement with our predictions. In particular, our prediction is in
perfect agreement with the observed value in $\alpha$ UMa (Harris et al.
1988), which estimated mass is roughly \mass{3}. This point is in favor of
high \chem{O}{17} destruction rates.

This agreement between standard models and observations confirms that no
extra-mixing (diffusion, rotation-induced mixing, \dots) is expected to
occur in intermediate-mass stars, as already discussed in Charbonnel (1994,
1995).

\subsection {Internet Tables}

Very complete tables containing informations about the center and surface
evolution of the structure and chemical composition for our 3, 4, 5, 6 and
\mass{7} models with $Z = 0.005$ and 0.02, from the beginning of the PMS
phase up to the end of the E-AGB phase, can be found through the internet
network at the following address and path:

\begin{center}
  {\sf http://gag.observ-gr.fr/liens/starevol/evol.html}
\end{center}

More specifically, we present, for each star, the evolution of

\begin{itemize}

\item
  the age, mass ,mass loss rate, surface luminosity, bolometric magnitude,
  effective temperature, total radius, surface density and gravity;

\item
  the central temperature, density, pressure, degeneracy parameter and the
  mass fraction of \chem{H}{1}, \chem{He}{4}, \chem{C}{12} and \chem{O}{16};

\item
  the mass and radius positions of the central convective core boundary and
  of the bottom of the convective envelope as well as its temperature and
  density;

\item
  the nuclear luminosities of the H- and He-burning shells;

\item
  the mass and radius positions of the H- and He-burning shells;

\item
  the surface mass fractions of \chem{H}{1}, \chem{He}{3}, \chem{He}{4},
  \chem{Li}{7}, \chem{Be}{9}, \chem{C}{12}, \chem{C}{13}, \chem{N}{14},
  \chem{N}{15}, \chem{O}{16}, \chem{O}{17}, \chem{O}{18}, \chem{F}{19},
  \chem{Ne}{20}, \chem{Ne}{21}, \chem{Ne}{22}, \chem{Na}{23},
  \chem{Mg}{24}, \chem{Mg}{25}, \chem{Mg}{26}, \chem{Al^g}{26},
  \chem{Al}{27}

\end{itemize}

\section {Structural properties of the thermal pulses}

\subsection {Evolutionary properties of the thermal pulses as a function
  of $M$ and $Z$}

As the stars are ascending the AGB, the features of the thermal pulses
evolve, and some of the main characterizing quantities of these events reach
``asymptotic values". The most important feature is the thermal pulse
intensity that can be defined in terms of the HeBS luminosity $L^{max}_{He}$
reached at the maximum of the He-shell flash.  $L^{max}_{He}$ is quite low
during the first thermal pulses, but it rapidly increases with core mass (or
flash number). The increasing rate is greater for lower metallicities or
higher stellar masses. After typically 10 (5) thermal pulses in our $Z =
0.02$ (0.005) models, the thermal pulse intensity is very high and only
slightly increases from pulse to pulse, monotonously with the core mass.
This second part of the TP-AGB phase is usually called the ``asymptotic
regime'' or ``full amplitude'' of the thermal pulses.

For each of the modeled stars, we have computed at least four full amplitude
thermal pulses. Let us briefly discuss the main evolutionary properties of
these thermal runaways.

The thermal runaways are stronger inside stars of lower $Z$, which
facilitates dredge-up events (see Sect. 6.2). The intensities of our
He-shell flashes are comparable to other published values for stars of about
the same core masses (see e.g. Boothroyd \& Sackmann 1988b and Table 4).

While the stars evolve along the TP-AGB, the HeBS and HBS advance in mass,
and the mean inter-shell mass slightly decreases with time. From one pulse
to another, the top of the convective tongue at its maximum extension gets
closer from the HBS, even if contact between both regions has never been
found due to the fact that the HBS is still active at that time. Moreover,
for all the thermal runaways we have computed, the overlap in mass between
two successive thermal pulses is ranged between 0.6 and 0.4 for AGB stars of
increasing total mass. This overlap is slightly larger for our lower $Z$
models.

The maximum (minimum) temperature $T_b^{max}$ ($T_t^{min}$) and density
$\rho_b^{max}$ ($\rho_t^{min}$) reached at the base (top) of a thermal
pulse both increase (decrease) from pulse to pulse. For a given He-shell
flash, $T_b^{max}$ increases with stellar mass and metallicity, while
$\rho_b^{max}$ decreases with stellar mass but increases with metallicity. 
In the asymptotic regime, the growth rate of $T_b^{max}$ also tends to
an asymptotic value. By comparison with other works for the \mass{3} (Z =
0.02) star, we found that our $T_b^{max}$ are typically 4 \% higher than
those of Boothroyd \& Sackmann (1988b) and Straniero et al. (1996). This
could be mainly due to {\it (i)} different prescriptions to compute
$\varepsilon_{grav}$, {\it (ii)} different envelope masses due to
different mass loss rates and {\it (iii)} different MLT treatments (see
Sect. 2.1.4). The temperature and density differences between the base
and top of each convective tongue are rather large. This point is important
for nucleosynthesis purposes (see Sect. 7).

The duration of the convective tongue associated with each thermal pulse
($\Delta t_p$) and, to a lower extent, that of the inter-pulse phase
($\Delta t_{ip}$), are crucial quantities for the nucleosynthesis too.  The
stronger a He-flash, the shorter it is and the quickest it deactivates.
Both quantities are thus smaller for higher core masses and lower
metallicities. Moreover, $\Delta t_p$ tends to decrease as the stars evolve
along the TP-AGB. $\Delta t_{ip}$ also decreases after the first He-shell
flashes, but it then reaches an almost constant value during the asymptotic
regime, only very slightly decreasing with the increasing core mass.

\subsection {Third dredge-up}

Along the TP-AGB phase, the bottom of the convective envelope can penetrate
deep inside regions that have been nuclearly processed just after a thermal
pulse, if it is strong enough (in order to give rise to a large enough
entropy increase in the inter-shell region). Contrarily to the first and
second dredges-up, this so-called third dredge-up, when it occurs,
successively mixes up to the surface {\it (i)} material that has experienced
H-burning in the thin HBS but also {\it (ii)} part of the region where the
thermal pulse nucleosynthesis operated just before. Both regions give rise
to very different -- and sometimes opposite -- chemical pollutions of the
convective envelope (see Sects. 7.1 and 7.6).

In our $Z = 0.02$ models, the HBS region is already dredged-up from the
14th, 5th and 3rd thermal pulse for the 3, 4 and \mass{5} stars,
respectively, while for the $Z = 0.005$ models, such a mixing already occurs
from the 2nd and 1st thermal pulse for the 3 and \mass{4} stars,
respectively. The \mass{6} (\mass{5}) star with $Z = 0.02$ (0.005)
continuously mixes H-burning products up to the surface as it experiences
H-burning at the bottom of its convective envelope (the so-called
``Hot-Bottom Burning'' or HBB; see Sect. 7.1). This kind of dredge-up
already modifies the surface composition for some species (like
\chem{Al^g}{26}; see Sects. 7.4 and 7.6).

The complete third dredge-up (one usually simply calls the third
dredge-up, 3DUP hereafter), i.e. reaching the upper part of the previous
thermal pulse region, is more difficult to obtain numerically. Following
models (Wood 1981; Iben 1983; Lattanzio 1987; Boothroyd \& Sackmann 1988c),
the dredge-up of thermal pulse material is easier for stars of lower
metallicity $Z$. This is supported by observations (see e.g. Blanco et al.
1978). In our models, we just found a 3DUP from the 9th thermal pulse of
the \mass{4} with $Z = 0.005$ star. However, based on the evolution of
the post-thermal pulse entropy profile from pulse to pulse, we can
predict that such a dredge-up will also occur, within the next two
thermal pulses, for our 5 and \mass{6} stars with $Z = 0.02$ and for the
\mass{5} star with $Z = 0.005$ (see Table \ref{endagb} for the number of
computed thermal pulses for each object). Some additional thermal pulses
are still needed for lower mass objects. Several years ago, the 3DUP was
found to occur too late along the TP-AGB phase for AGB stars of decreasing
mass and/or increasing metallicity (see e.g. Iben 1976, Fujimoto \&
Sugimoto 1979). This was a common problem (see Lattanzio 1989 for a
review). Recent works showed that with the new OPAL opacities, the
3DUP can now be obtained even for low-mass Pop I objects (see e.g. Frost
\& Lattanzio 1996 or Straniero et al. 1995).

One of the clearest signature of the 3DUP is the significant \chem{C}{12}
pollution of the convective envelope, brought into the inter-shell region by
the convective tongue of a thermal pulse (see Sect. 8.1). As the thermal
pulse intensity (and consequently the probability of 3DUP occurrence)
increases with time, as do the core mass and the total luminosity, the
models tend predict the formation of carbon (C) stars (for which the surface
\chem{C}{12}/\chem{O}{16} becomes greater than unity thanks to the
repetitive 3DUP events) at too high luminosities, i.e. to late during the
TP-AGB phase, especially for solar metallicity stars. Solutions to increase
the depth of the convective envelope during a dredge-up, e.g.  by treating
in some way semi-convection (see e.g. Lattanzio 1986) or by invoking
undershooting (see Alongi et al. 1991) have been suggested. They indeed help
the 3DUP to occur but do not yet solve completely the problem.
Fundamentally, this could be due to our bad knowledge of convection inside
stars. Very recently, Straniero et al. (1996) suggested that by considerably
increasing the number of mass shells and especially the number of time steps
to model thermal pulses, the 3DUP occurs earlier during the
TP-AGB phase. This consists in a real numerical progress in that field. The
fact we find 3DUP latter than them along the AGB phase can be due to our
lower number of time steps (typically by a factor of 3).

\subsection {Towards the end of TP-AGB phase}

\subsubsection {Core Mass--Luminosity relation}

It was first discovered by Paczy\'nski (1975) that during the asymptotic
TP-AGB phase, there is a linear relation between the surface luminosity $L$
and the core mass $M_C \equiv m\rm{[H]}$ (see Table \ref{t0}) which is only
slightly dependent on the total mass of the AGB star. Bl\"ocker \&
Sch\"onberner (1991) however found that this kind of relation does not exist
for the most massive AGB stars for which the bottom of the convective
envelope goes down to the HBS so that part of the H-burning occurs inside
the convective envelope (HBB).

>From our models for the last four thermal pulses of the 3 and \mass{4} with
$Z = 0.02$ and the \mass{3} with $Z = 0.005$, we have derived a core
mass--luminosity relation in the form

\begin{eqnarray}
  {L \over L_{\sun}} & = & \alpha \left( {M \over M_{\sun}} \right)^{\beta}
  \left( {Z \over 0.02} \right)^{\gamma} \left[ {M_C \over M_{\sun}} -
  \right. \nonumber \\
  & & \delta \left . \left( {M \over M_{\sun}} \right)^{\varepsilon} \left(
      {Z \over 0.02} \right)^{\zeta} \right] \,,
\end{eqnarray}

\noindent where $\alpha = 1233 L_{\sun}$, $\beta = 3.566$, $\gamma =
0.519$, $\delta = 0.1057$, $\varepsilon = 1.312$ and $\zeta = 0.203$. It
is in good agreement with other $M_C - L$ relations, for $0.50 < M_C /
M_{\sun} < 1.10$ and $0.005 \le Z \le 0.02$ (see e.g. Lattanzio 1986 or
Boothroyd \& Sackmann 1988a). The more massive stars experience HBB and do
not fit, consequently, to the same relation. In Table \ref{endagb}, we
however give a linear relation of the form $L / L_{\sun} = \alpha (M_C /
M_{\sun} - \beta)$ for the \mass{5} (\mass{4}) AGB star with $Z = 0.02$
(0.005) that do not experience a very strong HBB. Let us stress that
during the TP-AGB phase, $L$ suffers very important variations during and
after each thermal pulse. Frequently, the core mass--luminosity relation
is derived by taking solely account of the maximum surface luminosity
preceding each thermal pulse. However, the probability to observe an AGB
star just at that time is very weak ($\sim \Delta t_p / \Delta t_{ip} < 1
\%$; see Table \ref{endagb}). We thus preferred to derive our relations with
a more realistic surface luminosity that is averaged over each
pulse-inter-pulse cycle.

On the other hand, it is well known that the rate of mass advance of a
burning shell (here $M_C$ for the HBS) is related to the rate of nuclear
energy production by

\begin{equation}
  {d M_C \over d t} \simeq {L_H \over \overline{X} E_H} \,,
\end{equation}

\noindent with $L_H$ being the luminosity produced by the HBS,
$\overline{X}$ the mean hydrogen mass fraction over this burning shell
and $E_H = 6.35 \, 10^{18} \, {\rm erg \, g^{-1}}$ the amount of
energy produced by the conversion of 1 g of \chem{H}{1} to \chem{He}{4}
through the CNO bi-cycle. Along the TP-AGB, most of the surface
luminosity $L$ is produced by the HBS. As indicated in Table \ref{t0},
the HeBS contributes much less than the HBS to the nuclear luminosity
(except during a thermal runaway). Gravitation also constitutes a
relatively small energy source, especially for the less massive AGB
stars. In conclusion, we can write $L_H \simeq \xi L$ with $0.6 < \xi <
0.9$ depending on $M$. By eliminating $L_H$ in Eq. (6) with the core
mass--luminosity relation, one finds that the core mass $M_C$ increases
exponentially with time during the asymptotic TP-AGB phase. The time
scale for the core mass growth is $\sim 10^6$ ($\sim 10^5$) yr for a
\mass{3} (\mass{6}) AGB star, that is, interestingly, comparable to the
total duration of the TP-AGB phase. Table \ref{mcl} demonstrates that
Eq. (6) predicts rates of HBS advance in mass that are very similar to
those coming from our numerical models, except for the \mass{6}
(\mass{5}) object with $Z = 0.02$ (0.005), probably due to the strong HBB.

\begin{table*}
  \caption[ ]{\label{endagb}
    Some features of our TP-AGB models at the end of the complete stellar
    evolution computations. We indicate the total number $n_p$ of computed
    thermal pulses. The core mass ($M_C$) -- luminosity ($L$) relations
    $L = \alpha (M_C - \beta)$ indicated for the \mass{5} (\mass{4}) star with
    $Z = 0.02$ (0.005) is marginally valid as these stars already experience
    significant hot-bottom burning. No such a relation is given for the
    \mass{6} (\mass{5}) object with $Z = 0.02$ (0.005). We refer to the text
    for the fitting core mass--luminosity relation found for the other
    stars. $M_{env}$ is the envelope mass. We also indicate the intensity of
    the last computed thermal pulse (in term of the maximum luminosity
    $L_{He}^{max}$ produced by the HeBS during this thermal runaway), the
    duration of the last inter-pulse ($\Delta t_{ip}$) and pulse
    ($\Delta t_p$), the time $\Delta t_{pm}$ elapsed between the maximum
    extent of the convective tongue of the last thermal pulse and its
    disappearance, the maximum mass extension of the convective tongue
    ($\Delta m_p$) and the mass overlap between the last two thermal pulses
    ($\Delta m_{pr}$). We finally mention the occurrence of the third
    dredge-up
    }
  \begin{flushleft}
    \begin{tabular}{lccccccc}
      \noalign{\smallskip}
      \hline
      \noalign{\smallskip}
      &\multicolumn{4}{c}{$Z = 0.02$}&\multicolumn{3}{c}{$Z = 0.005$} \\
      \noalign{\vspace{-2.0mm}}
      &\multicolumn{4}{c}{\makebox[62mm]{\downbracefill}}&\multicolumn{3}{c}{\makebox[45mm]{\downbracefill}}\\
      &\mass{3}&\mass{4}&\mass{5}&\mass{6}&\mass{3}&\mass{4}&\mass{5} \\
      \noalign{\smallskip}
      \hline \hline
      \noalign{\smallskip}
      $t_{end} - t_{0} \, (\rm{yr})$&9.94$\, 10^{ 5}$&1.13$\, 10^{ 5}$&4.56$\, 10^{ 4}$&2.32$\, 10^{ 4}$&4.39$\, 10^{ 5}$&5.53$\, 10^{ 4}$&2.13$\, 10^{ 4}$ \\
      $n_{p}$&19&13&13&14&8&9&9 \\
      \noalign{\smallskip}
      $L \, (\rm{L_{\odot}})$& 9750&18200&24150&33650& 9250&19700&27600 \\
      $M_{env} \, (\rm{M_{\odot}})$&1.662&2.718&3.642&4.575&1.869&2.547&3.124 \\
      $M_{C} \, (\rm{M_{\odot}})$&0.6052&0.7563&0.8290&0.9022&0.6411&0.8277&0.9101 \\
      $\alpha$&see rel.&see rel.&264550&--&see rel.&383550&-- \\
      $\beta$&see rel.&see rel.&0.73729&--&see rel.&0.77564&-- \\
      \noalign{\smallskip}
      $L_{He}^{max} \, (\rm{L_{\odot}})$&2.1$\, 10^{ 7}$&1.2$\, 10^{ 7}$&5.1$\, 10^{ 6}$&1.9$\, 10^{ 6}$&2.4$\, 10^{ 7}$&6.2$\, 10^{ 6} $&1.1$\, 10^{ 6}$ \\
      \noalign{\smallskip}
      $\Delta t_{ip} \, (\rm{yr})$& 38000& 9200& 3700& 1600& 45000& 7500& 2500 \\
      $\Delta t_{p} \, (\rm{yr})$& 95& 24& 14& 9& 104& 20& 13 \\
      $\Delta t_{pm} \, (\rm{yr})$& 19& 10& 7& 6& 20& 9& 8 \\
      \noalign{\smallskip}
      $\Delta m_{p} \, (\rm{M_{\odot}})$& 0.0115& 0.0038& 0.0021& 0.0011& 0.0119& 0.0031& 0.0015 \\
      $\Delta m_{pr} \, (\rm{M_{\odot}})$& 0.0066& 0.0019& 0.0010& 0.0005& 0.0069& 0.0017& 0.0008 \\
      \noalign{\smallskip}
      3DUP ?& no \dots& no \dots& no& no& no \dots& yes& no \\
      \noalign{\smallskip}
      \hline
      \noalign{\smallskip}
    \end{tabular}
  \end{flushleft}
\end{table*}

\begin{table*}
  \caption[ ]{\label{mcl}
    Comparison between predictions for the rate of HBS advance in mass from
    Eq. (6) and those coming from our computed models. The mean hydrogen
    mass fraction $\overline{X}$ and HBS luminosity $L_H$ are taken during
    the inter-pulse phase of the last computed thermal pulse for each star.
    $d M_C / d t$ is correspondingly calculated over the same inter-pulse
    phase }
  \begin{flushleft}
    \begin{tabular}{lccccccc}
      \noalign{\smallskip}
      \hline
      \noalign{\smallskip}
      &\multicolumn{4}{c}{$Z = 0.02$}&\multicolumn{3}{c}{$Z = 0.005$} \\
      \noalign{\vspace{-2.0mm}}
      &\multicolumn{4}{c}{\makebox[63mm]{\downbracefill}}&\multicolumn{3}{c}{\makebox[45mm]{\downbracefill}}\\
      &\mass{3}&\mass{4}&\mass{5}&\mass{6}&\mass{3}&\mass{4}&\mass{5} \\
      \noalign{\smallskip}
      \hline \hline
      \noalign{\smallskip}
      $\overline{X}$&%
      0.49&0.54&0.52&0.64&0.54&0.58&0.69 \\
      \noalign{\smallskip}
      $L_H \, \rm{(L_{\odot})}$&%
      3450&10690&14480&17250&4600&10320&12890 \\
      \noalign{\smallskip}
      Eq. (6) prediction&%
      6.7$\, 10^{-8}$&1.9$\, 10^{-7}$&2.7$\, 10^{-7}$&2.6$\, 10^{-7}$&8.1$\, 10^{-8}$&1.7$\, 10^{-7}$&1.8$\, 10^{-7}$ \\
      \noalign{\smallskip}
      $d M_C / d t \, \rm{(M_{\odot} \, yr^{-1})}$&%
      6.4$\, 10^{-8}$&1.7$\, 10^{-7}$&2.5$\, 10^{-7}$&3.1$\, 10^{-7}$&7.8$\, 10^{-8}$&1.8$\, 10^{-7}$&2.4$\, 10^{-7}$ \\
      \noalign{\smallskip}
      \hline
      \noalign{\smallskip}
    \end{tabular}
  \end{flushleft}
\end{table*}

\subsubsection {Extrapolation procedure}

One of our ambitions is to extrapolate our predictions coming from the
complete evolution models not only for the global surface properties but
also for the chemical composition. Eqs. (5) and (6) already allow to
extrapolate rather accurately the core mass and surface luminosity from our
last computed models towards the end of the TP-AGB phase. For the \mass{6}
(\mass{5}) star with $Z = 0.02$ (0.005), independent fits of $L$ and $M_C$
as a function of time have been calculated. We also need fits for the
surface radius [in order to follow the mass loss rate increase through Eq.
(1)] as well as the temperature and density at the base of the convective
envelope (to follow the HBB nucleosynthesis). We found, on grounds of the
last four computed thermal pulses for each star, very good fitting relations
for these quantities, which are presented in Table \ref{fits}. Note that
they are linear with time during the last part of the TP-AGB phase,
contrarily to $M_C$ and $L$ that increase exponentially during this phase.

The main interest of such an extrapolation procedure is that very detailed
stellar evolution computations (including a very complete description of the
nucleosynthesis) would take prohibitive computation time. However, we had to
make two important approximations (compared to full evolution models) that
must be mentioned.

\begin{itemize}

\item
  The successive 3DUP events progressively pollute the convective envelope
  in \chem{He}{4}, \chem{C}{12} and heavier elements to the detriment of
  \chem{H}{1}. This increases the mean molecular weight (or $Z$) and the
  mean opacity in the sub-photospheric region, probably leading to an
  overestimation of the surface luminosity.

\item
  Very few computations are existing for the very end of the TP-AGB phase
  (Vassiliadis \& Wood 1993; Bl\"ocker 1995), that cannot maintain to give
  accurate predictions concerning the evolution of the surface isotopic
  composition. One of their interests is to indicate that the last few thermal
  pulses seem to have different properties than the full amplitude ones (e.g.
  longer inter-pulse durations). On the other hand, it has been noted by
  Boothroyd \& Sackmann (1988a) that the increasing mass loss rate and $Z$
  possibly prevent the 3DUP to occur after these last thermal pulses. As a
  consequence, assuming the full amplitude regime up to the AGB tip probably
  makes our following predictions quite indefinite concerning the very end
  of the TP-AGB phase.

\end{itemize}

\begin{table*}
  \caption[ ]{\label{fits}
    Fitting parameters from the last four computed asymptotic thermal pulses
    towards the end of the TP-AGB phase, for the mean total radius and
    bottom temperature and density of the convective envelope. For each
    quantity $A$, the relation reads $A(t) = \alpha (t-t_0) + \beta$, with
    time in yr }
  \begin{flushleft}
    \begin{tabular}{llccccccc}
      \noalign{\smallskip}
      \hline
      \noalign{\smallskip}
      & &\multicolumn{4}{c}{$Z = 0.02$}&\multicolumn{3}{c}{$Z = 0.005$} \\
      \noalign{\vspace{-2.0mm}}
      & &\multicolumn{4}{c}{\makebox[69mm]{\downbracefill}}&\multicolumn{3}{c}{\makebox[49mm]{\downbracefill}}\\
      & &\mass{3}&\mass{4}&\mass{5}&\mass{6}&\mass{3}&\mass{4}&\mass{5} \\
      \noalign{\smallskip}
      \hline \hline
      \noalign{\smallskip}
      $\overline{R}\, (t-t_{0}) \, (\rm{R_{\odot}})$&$\alpha$&%
      3.17$\, 10^{-4}$&1.16$\, 10^{-3}$&1.76$\, 10^{-3}$&1.95$\, 10^{-3}$&9.03$\, 10^{-5}$&1.33$\, 10^{-3}$&3.48$\, 10^{-3}$ \\
      &$\beta$&%
      8.53$\, 10^{1}$&4.11$\, 10^{2}$&5.11$\, 10^{2}$&6.35$\, 10^{2}$&2.37$\, 10^{2}$&3.57$\, 10^{2}$&4.06$\, 10^{2}$ \\
      \noalign{\smallskip}
      $\overline{T_{cb}}\, (t-t_{0}) \, (\rm{K})$&$\alpha$&%
      1.81$\, 10^{-1}$&2.88$\, 10^{1}$&5.13$\, 10^{2}$&4.21$\, 10^{2}$&3.53&2.81$\, 10^{2}$&1.07$\, 10^{3}$ \\
      &$\beta$&%
      4.05$\, 10^{6}$&1.14$\, 10^{7}$&2.05$\, 10^{7}$&6.34$\, 10^{7}$&3.16$\, 10^{6}$&8.82$\, 10^{6}$&4.50$\, 10^{7}$ \\
      \noalign{\smallskip}
      $\overline{\rho_{cb}}\, (t-t_{0}) \, (\rm{cgs})$&$\alpha$&%
      1.16$\, 10^{-10}$&2.51$\, 10^{-7}$&1.56$\, 10^{-5}$&3.32$\, 10^{-5}$&4.71$\, 10^{-9}$&3.86$\, 10^{-6}$&1.24$\, 10^{-4}$ \\
      &$\beta$&%
      1.72$\, 10^{-3}$&2.28$\, 10^{-2}$&5.40$\, 10^{-2}$&2.47&9.82$\, 10^{-4}$&1.04$\, 10^{-3}$&3.53$\, 10^{-1}$ \\
      \noalign{\smallskip}
      \hline
      \noalign{\smallskip}
    \end{tabular}
  \end{flushleft}
\end{table*}

Such extrapolation procedures have already been realized by Groenewegen \&
de Jong (1993 and following papers). These synthetic evolution models are
very powerful to compare observations with computed models concerning the
evolution of the global surface properties of AGB stars. However, they
contain shortcomings that do not allow to predict in detail the evolution of
the surface isotopic composition.  Mainly, {\it (i)} only the most abundant
elements are followed (\chem{He}{4}, \chem{C}{12}, \chem{C}{13},
\chem{N}{14} and \chem{O}{16}) and {\it (ii)} the HBB nucleosynthesis is not
followed in detail whereas it becomes a very influent process concerning the
convective envelope composition towards the TP-AGB tip. Recent progresses
have been made by Busso et al. (1995) and Lambert et al. (1995) to follow in
detail the nucleosynthesis as late as possible along the AGB phase.

In Sect. 8, we both present the nucleosynthesis encountered in our full
evolution models and the extension of the predictions concerning the
convective envelope composition towards the end of the AGB phase
corresponding to the seven computed stars. For this extrapolation, we
proceed as follows.

\begin{enumerate}

\item
  From the last computed inter-pulse phase for each star, we integrate, during
  the successive inter-pulse phases, the evolution of the nucleosynthesis
  occuring inside the convective envelope by solving Eqs. (4), suited for the
  HBB, for each nuclide while imposing a temperature and density increases
  at the bottom of the convective envelope given by the fits presented in
  Table \ref{fits}. We took the inter-pulse durations of Table \ref{endagb};
  indeed, they are only very slightly decreasing with the core mass
  increase along the asymptotic TP-AGB phase.

\item
  Since the 3DUP is almost ready to occur in our models (see Sect. 6.2), we
  mix the envelope with the HBS and inter-shell regions recurrently, every
  $\Delta t_{ip}$. To quantify the adopted depth of each 3DUP, let us
  remember the definition of the usual parameter

  \begin{equation}
    \lambda = {\Delta M_{DUP} \over \Delta M_C} \,,
  \end{equation}

  \noindent where $\Delta M_{DUP}$ represents the amount of mass that is
  dredged-up (difference between the normal depth of the convective
  envelope and its maximum downwards penetration during a dredge-up) and
  $\Delta M_C$ the increase in mass of the HBS between two successive
  thermal pulses. We adopted a 3DUP depth corresponding to $\lambda = 0.6$
  for all the stars. This value has been suggested by Groenewegen \& de Jong
  (1993) because it reproduces rather well the carbon star luminosity
  function of the LMC (that cannot be fitted on grounds of self-consistent
  evolution models only; see Sect. 6.2).

  During such a dredge-up, we assumed that the material coming from the
  convective tongue of a thermal pulse has abundance profiles corresponding
  to end of the last computed thermal pulse. Of course, we cannot do better
  without further computing thermal pulses with detailed nucleosynthesis up
  to the end of the TP-AGB phase that is, again, not yet realistic. However,
  this appears to be rather justified for many important chemical elements as
  they reach almost constant abundances after the few last computed thermal
  pulses (see discussion in Sect. 7).

\end{enumerate}

Taking a constant value for $\lambda$ is probably not strictly right. As far
as the surface composition is concerned, increasing $\lambda$ more
efficiently pollutes the convective envelope from pulse to pulse.

There is another quantity that greatly influences the evolution of the
envelope composition: the mass loss rate. Indeed, decreasing the mass loss
rate allows to increase the total number of thermal pulses (and consequently
the total number of 3DUP episodes) before the end of the TP-AGB phase. We
had to increase our $\eta$ values in Eq. (1) for each star, in order to
reach the highest mass loss rates (called super-winds) observed among very
evolved AGB stars (see the discussion in Sect. 2.1.5; see also Zuckerman et
al. 1986; Wannier \& Sahai 1986; Knapp et al. 1989): we adopted $\eta = 5$,
10, 15 and 20 for the 3, 4, 5 and \mass{6} AGB models, respectively,
whatever $Z$. As the exact way following which the mass loss rate increases
with time along the asymptotic TP-AGB phase is still questionable by more
than a factor of two (see Sect. 2.1.5), the resulting uncertainty on the
evolution of the surface composition up to the AGB tip dominates the effect
of $\lambda$ (see our discussion in Sect. 8.1, where our mass loss rates are
varied by a factor of four).

In Table \ref{mlos}, we present the global mass loss increases we obtained,
for the seven TP -AGB stars, from the beginning of the TP-AGB phase up to
its end (i.e. including our extrapolation computations).

\begin{table*}
  \caption[ ]{\label{mlos}
    Increase of the mass loss rates during the whole TP-AGB phase
    }
  \begin{flushleft}
    \begin{tabular}{llccccccc}
      \noalign{\smallskip}
      \hline
      \noalign{\smallskip}
      & &\multicolumn{4}{c}{$Z = 0.02$}&\multicolumn{3}{c}{$Z = 0.005$} \\
      \noalign{\vspace{-2.0mm}}
      & &\multicolumn{4}{c}{\makebox[63mm]{\downbracefill}}&\multicolumn{3}{c}{\makebox[45mm]{\downbracefill}}\\
      & &\mass{3}&\mass{4}&\mass{5}&\mass{6}&\mass{3}&\mass{4}&\mass{5} \\
      \noalign{\smallskip}
      \hline \hline
      \noalign{\smallskip}
      $\dot M \, (\rm{M_{\odot} \, yr^{-1}})$&{\it at $t_{0}$}&%
      2.0$\, 10^{-7}$&1.5$\, 10^{-6}$&2.5$\, 10^{-6}$&3.6$\, 10^{-6}$&3.6$\, 10^{-7}$&1.8$\, 10^{-6}$&2.8$\, 10^{-6}$ \\
      \noalign{\smallskip}
      &{\it AGB tip}&%
      7.6$\, 10^{-6}$&4.3$\, 10^{-5}$&9.4$\, 10^{-5}$&1.5$\, 10^{-4}$&5.6$\, 10^{-6}$&4.0$\, 10^{-5}$&1.0$\, 10^{-4}$ \\
      \noalign{\smallskip}
      \hline
      \noalign{\smallskip}
    \end{tabular}
  \end{flushleft}
\end{table*}

\subsubsection {Initial mass--Final mass relation}

\begin{figure}
  \psfig{file=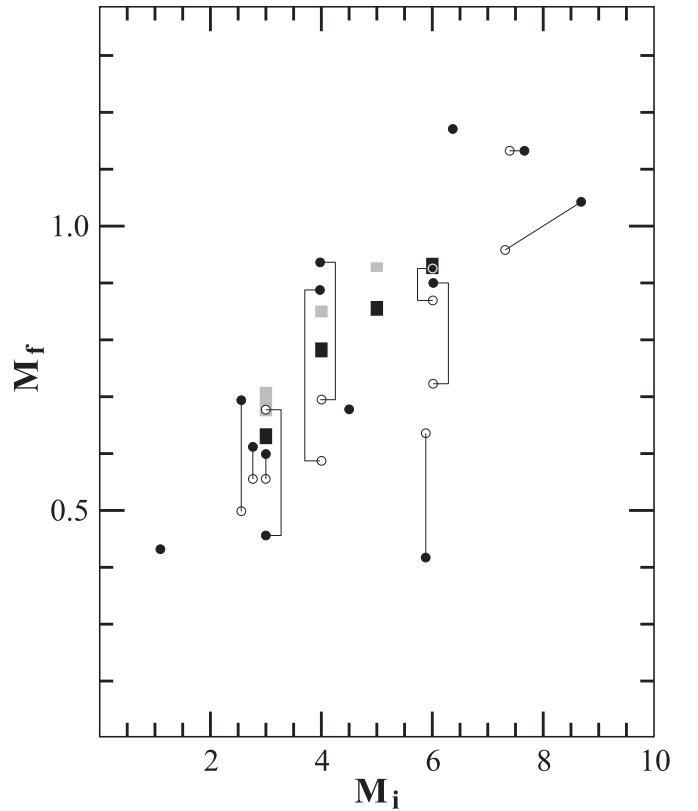,width=8.8cm}
  \caption[ ]{\label{mimf}
    Relation between the initial mass of stars and the final white dwarf
    (WD) mass at the end of the AGB phase. Observational final masses are
    derived either from the surface gravity of the WD (filled circle) or
    from their radius (open circles); lines are connecting both
    determinations for identical objects. Our predictions are the black and
    gray rectangles, for the $Z = 0.02$ and $Z = 0.005$ modeled stars,
    respectively. Their extension is the result of varying the mass loss
    rate by a factor of four during the asymptotic TP-AGB phase }
\end{figure}

One of the most important constraint for evolutionary models towards the AGB
tip is the reproduction of the observed correlation between the initial mass
of stars ($M_i$) and their final one ($M_f$), i.e. their core mass (become a
hot white dwarf) when the convective envelope has been completely removed
after the planetary nebula (PN) ejection. We present in Fig. \ref{mimf} the
comparison between observations analyzed by Weidemann \& Koester (1983) and
Weidemann (1987) and the final core masses we obtained after our
evolutionary plus extrapolations computations. We note that agreement with
observations is slightly better than for Vassiliadis \& Wood (1993),
especially around $M_i =$ \mass{3} where they only reproduce the upper
limits. On the contrary, the final core mass obtained by Boothroyd \&
Sackmann (1988b) is rather low compared to observations.  Note we have not
taken into account of the core mass reduction following each 3DUP event.
However, as explained below, such a reduction, even if it increases the
agreement with the observed relation, is rather negligible compared to other
uncertainties (core mass at the first thermal pulse, thermal pulse number
depending on the adopted mass loss rate, \dots ).

We observe that changing the mass loss rate by a factor of four during the
asymptotic TP-AGB phase (see also Sect. 8.1 concerning surface abundance
predictions) does not significantly change the agreement with observations.
We claim that this is due to the fact that while almost 50\% of the C--O
core mass is built during the central He-burning phase, roughly 35 to 45\%
(in mass) are added during the E-AGB phase. More precisely, during the whole
TP-AGB phase, we found that our core masses increased by 14 (12), 7 (4), 5
(3) and 4\% for the 3, 4, 5, and \mass{6} stars with $Z = 0.02$ (0.005),
respectively. The final core mass is consequently almost decided before the
super-wind occurs and this observed relation is not very compulsive for the
empirical mass loss rate to adopt during the TP-AGB phase.

\section {Nucleosynthesis during TP-AGB phase}

>From the pioneer work of Iben (1975), one realized that inside TP-AGB stars,
a very rich nucleosynthesis occurs. During the recurrent third dredge-up
events, the synthesized nuclides are mixed inside the convective envelope
and then ejected into the interstellar medium (ISM) through the strong mass
loss. In conclusion, TP-AGB stars play a crucial role for the chemical
evolution of galaxies, especially for some elements like \chem{He}{3},
\chem{Li}{7}, \chem{F}{19}, \chem{Al}{26}, nuclides heavier than
\chem{Fe}{56}, \dots It is the reason why we will give, in Sect. 8.3, the
yields resulting from our intermediate-mass star computations.

\subsection {Overview of the various nucleosynthesis sites}

Let us analyze the various nucleosynthesis processes occuring inside
intermediate-mass TP-AGB stars. In fact, three very different sites have to
be distinguished.

\subsubsection {Burning shells}

As indicated in Table \ref{t0}, most of the nuclear energy release on the
AGB is due to the HBS, mainly through the CNO bi-cycle. More precisely, the
six most important nuclear reactions that energetically supply the star are
\reac{C}{13}{p}{\gamma}{}{} ($\sim 26 \%$), \reac{N}{14}{p}{\gamma}{}{}
($\sim 24 \%$), \reac{N}{15}{p}{\alpha}{}{} ($\sim 17 \%$), \chem{O}{15}
$\, (\beta^+)$ ($\sim 9 \%$), \chem{N}{13} $\, (\beta^+)$ ($\sim 7 \%$)
and \reac{C}{12}{p}{\gamma}{}{} ($\sim 6 \%$), all occuring inside the
HBS. These numbers are quite similar (within a few \%) for all our
modeled stars. Whatever the AGB star total mass, the HBS slightly depletes
\chem{C}{12} and \chem{N}{15} and significantly depletes \chem{O}{17},
\chem{O}{18} and \chem{F}{19}, while \chem{N}{14} is produced. The case
of \chem{C}{13} is more difficult: while its abundance is very slightly
increased in the HBS of \mass{3} stars (whatever $Z$), it is somewhat
decreased in the HBS of \mass{6} stars.

The rather high temperature inside the HBS also allows the activation of the
NeNa and MgAl chains, even if they are much less influent for the global
energetics. The most interesting consequences of the NeNa chain are the
production of \chem{Ne}{22} and \chem{Na}{23} in $\le$ \mass{5} stars and
their slight destruction in more massive AGB stars. The MgAl chain is very
sensitive to the AGB total mass. In $<$ \mass{5} objects, \chem{Mg}{25} is
destroyed to the benefit of \chem{Al}{26} (roughly half in \chem{Al^g}{26}
and half in \chem{Al^m}{26} that quite instantaneously decays in
\chem{Mg}{26}); a small quantity of \chem{Mg}{26} is also destroyed in
\chem{Al}{27}. In more massive AGB stars, the HBS temperature is high enough
to substantially produce \chem{Mg}{25} through the
\reac{Mg}{24}{p}{\gamma}{}{} reaction. This in turn leads to a much
important \chem{Al^g}{26} production. However, as the HBS mean temperature
always increases with time, \chem{Al^g}{26} begins to be partially destroyed
by $\rm{(p , \gamma)}$ reactions late along the TP-AGB. We refer to
Forestini et al. (1991) for the first detailed study of \chem{Al^g}{26}
production inside AGB stars, by radiative proton burning (i.e. in the HBS).
In Guelin et al. (1995), more details can be found, including the
possibility to synthesize \chem{Al^g}{26} by proton burning at the bottom of
the convective envelope (i.e. by HBB). It was however a parametric study.
First predictions concerning the enrichment of the AGB star surfaces, on
grounds of full evolutionary models are presented in Sects. 7.6 and 8.1.

The comparatively small amount of nuclear energy supplied by the HeBS comes
from the $3 \alpha$ ($\sim 70 \%$) and \reac{C}{12}{\alpha}{\gamma}{}{}
($\sim 29 \%$) reactions. Consequently, the HeBS mainly produces
\chem{C}{12} and, to a lower extent, \chem{O}{16}, while during the previous
central He-burning stage, the \chem{O}{16} production was preponderant. At
the beginning of a thermal pulse, it is the $3 \alpha$ reaction that becomes
the dominant energy source (see also Sect. 7.1.3 for the thermal pulse
energetics).

\begin{figure}
  \psfig{file=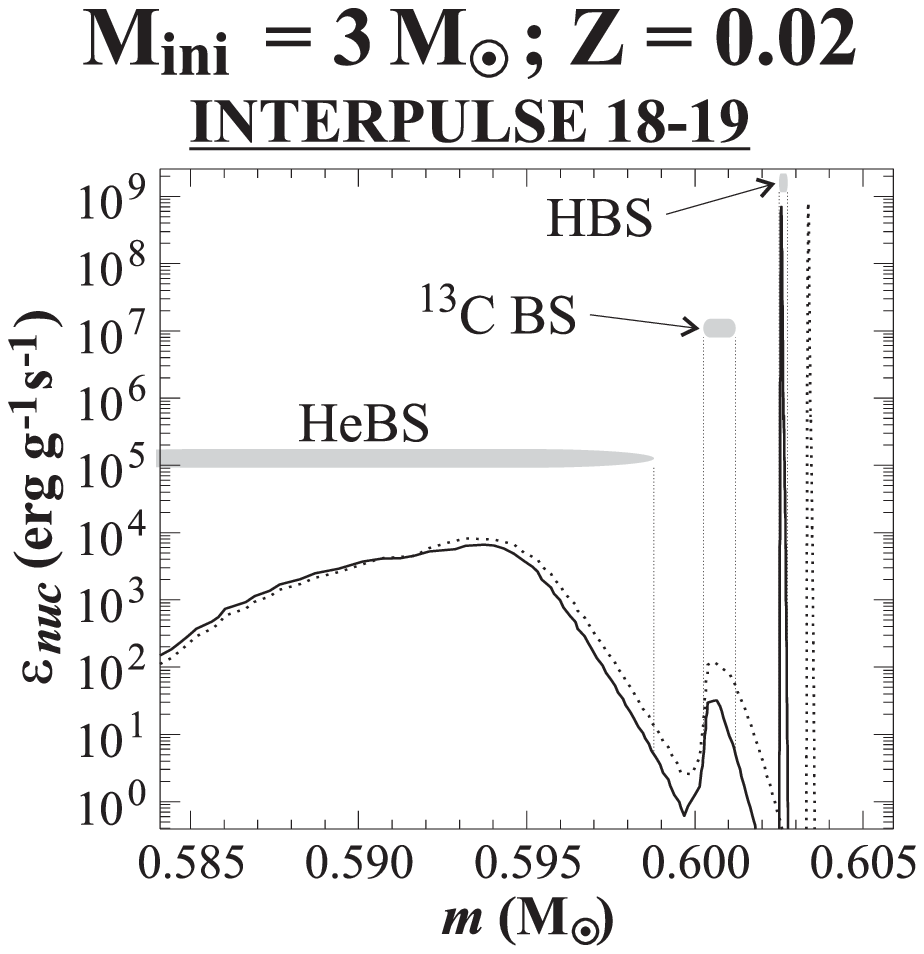,width=8.8cm}
  \caption[ ]{\label{13csh}
    Profile of the nuclear energy production rate per unit mass,
    $\varepsilon_{nuc}$, in the region of the burning shells, in the case of
    the \mass{3} ($Z = 0.02$) star, during the inter-pulse phase between the
    18th and the 19th thermal runaways. The dotted curve corresponds to a
    time just before the 19th thermal pulse, while the solid one is plotted
    roughly at the middle of the inter-pulse phase (see the text for a
    discussion) }
\end{figure}

During the long inter-pulse phase that follows a 3DUP episode, the
\chem{C}{13} coming out of the HBS can be burnt in the inter-shell region,
close to the HeBS top, mainly through the \reac{C}{13}{\alpha}{n}{O}{16}
reaction. This reaction is highly energetic. The first published work
reporting the existence of this inter-shell \chem{C}{13} burning is
Straniero et al. (1995). The produced neutrons are then captured by light
species up to \chem{Fe}{56} mainly, but part on them are also used to
synthesize $s$ elements. By decreasing order of amplitude, the nuclear
energy liberated in the inter-shell region during the inter-pulse phase
comes from these $\rm{(n , \gamma)}$ reactions up to \chem{Fe}{56} ($\sim 29
\%$), \reac{C}{13}{\alpha}{n}{}{} ($\sim 21 \%$), the so-called $s-$ process
reactions ($\sim 12\%$), \reac{Al^g}{26}{n}{p}{Mg}{26} ($\sim 7 \%$) and
\reac{N}{14}{n}{p}{C}{14} ($\sim 2.5 \%$). Note the important nuclear energy
released by the neutron captures. This energy production creates a third
burning shell between the HeBS and HBS, even if its nuclear activity is
considerably lower than that of both the HeBS and HBS. It continues to exist
until \chem{C}{13} has been destroyed there. This happens, in general,
almost when the following thermal pulse begins. As a consequence, when the
convective tongue of a thermal pulse engulfs the inter-shell matter, the
amount of \chem{C}{13} that is mixed is reduced compared to what was
previously thought, i.e. $\Delta m [H - He] X$(\chem{C}{13}) $[H - He]$,
where $[H - He]$ designates, as before, the inter-shell region. We show, in
Fig. \ref{13csh}, the nuclear energy production rate ($\varepsilon_{nuc}$)
profile in the case of our \mass{3} AGB star. This clearly demonstrates the
three-burning shell structure.  Such a structure is also found in the more
massive AGB stars. Let us emphasize that this third burning shell could at
least partially explain the production of the $s$ nuclides (see Sect. 7.5)
in a radiative zone.  This has been suggested by Straniero et al. (1995),
even if their amount of \chem{C}{13} in the inter-shell was artificially
increased in order to study the $s-$process nucleosynthesis.

\subsubsection {Hot-Bottom Burning}

If the temperature $T_{cb}$ at the base of the deep convective envelope is
high enough (i.e. $> 20 \, 10^6$ K), i.e. if the base of the convective
envelope reaches the HBS, proton burning occurs inside it. This HBB occurs
in models with initial masses $\ge$ \mass{5} for $Z = 0.02$ and $\ge$
\mass{4} for $Z = 0.005$. These lower mass limits and this $Z$ dependence
are common features of all the AGB models (see e.g. Boothroyd et al.  1993).
As mixing up to the cool surface is very efficient (the mean turn-over time
$\tau_{CONV}$ of the convective envelope is roughly 0.5 yr), the effective
nuclear reaction rates are lower than in radiative H-Burning, leading to
specific nucleosynthesis signatures.

For $T_{cb} > 40 \, 10^6$ K already, \chem{Li}{7} can be produced from
\chem{He}{3} following a scenario that cannot operate in radiative H-burning
(see Sect. 7.2 for more details). This Li synthesis has been qualitatively
suggested by Cameron \& Fowler (1971) and quantitatively first demonstrated
on grounds of evolutionary models by Sackmann \& Boothroyd (1992).

For $T_{cb} > 60 \, 10^6$ K, the CN cycle begins to operate partially, i.e.
\chem{C}{12} is converted to \chem{C}{13} that accumulates. As first
investigated by Boothroyd et al. (1993), this conversion can efficiently
reduce the \chem{C}{12}/\chem{O}{16} ratio and eventually, if HBB is strong
enough, this can prevent the formation of C stars (see also Sects. 8.1 and
8.2).

If $T_{cb}$ $>$ $80 \, 10^6$ K typically, the CN cycle almost operates
at equilibrium through the whole envelope, leading to a
\chem{C}{12}/\chem{C}{13} isotopic ratio lower than 10 (depending on
$T_{cb}$). In our models, this happens for the \mass{6} (\mass{5}) AGB
star with $Z = 0.02$ (0.005) and also, later along the TP-AGB, for our
\mass{5} star with $Z = 0.02$ (see Sect. 8.1). At very high $T_{cb}$,
other reactions begin to occur; in particular, in extreme cases (i.e. the
end of the TP-AGB phase of our \mass{6} models with $Z = 0.02$),
\chem{Al^g}{26} can eventually be directly produced inside the convective
envelope through the \reac{Mg}{25}{p}{\gamma}{}{} reaction on
\chem{Mg}{25} supplied by \reac{Mg}{24}{p}{\gamma}{}{} (see Sect. 8.1).

\subsubsection {Convective tongue of thermal pulses}

The thermal runaway, inside the HeBS, leading to form a convective tongue
is principally due to the $3 \alpha$ reaction excitement. When this
convective tongue finally penetrates the inter-shell region, the HBS ashes
are ingested, i.e. not only \chem{He}{4} but also \chem{C}{13}, \chem{N}{14}
and \chem{Al^g}{26} mainly. As these elements are partially burnt in the
bottom of the inter-shell region during the inter-pulse phase (globally,
we find that half of the \chem{C}{13} is so destroyed), they mainly
survive in the upper part of the inter-shell. They are consequently
engulfed by the convective tongue when it reaches its maximum extension.
At that time, the $3 \alpha$ reaction is already dominant for the
energetics. We consequently find that the energetics associated with the
\chem{C}{13} burning through the \reac{C}{13}{\alpha}{n}{}{} reaction as
well as the neutron captures that follow is rather negligible (less than
1 \%). This is also found by Straniero et al. (1995). This conclusion
could however be modified if more \chem{C}{13} was synthesized in the
inter-shell region, as required for the $s-$process (see Sect. 7.5). The 
important energetic effects reported by Bazan \& Lattanzio (1993)
concerning the \chem{C}{13} burning inside thermal pulses was probably
due to {\it (i)} the fact they did not give account of partial inter-shell
\chem{C}{13} burning and {\it (ii)} the very high \chem{C}{13} amount
artificially ingested into their thermal pulses in order to study the
consequent $s-$process.

Part of produced neutrons participate to $\rm{(n , p)}$ reactions, mainly on
\chem{Al^g}{26} and, to a lower extent, on \chem{N}{14}. The resulting
protons, at these high temperatures characterizing He-burning during a
thermal pulse ($> 200 \, 10^6$ K at the base of the convective tongue), are
rapidly captured by various elements (see Sects.  7.3 to 7.5 for more
details). Let us stress that the simultaneous presence, inside a mixed
He-burning region (the convective tongue), of neutrons and protons leads to
a very specific (and unique) nucleosynthesis.

We will analyze in more details the nucleosynthesis processes associated
with the thermal pulses in Sects. 7.3 to 7.5.

\subsection {Lithium production}

\begin{figure}
  \psfig{file=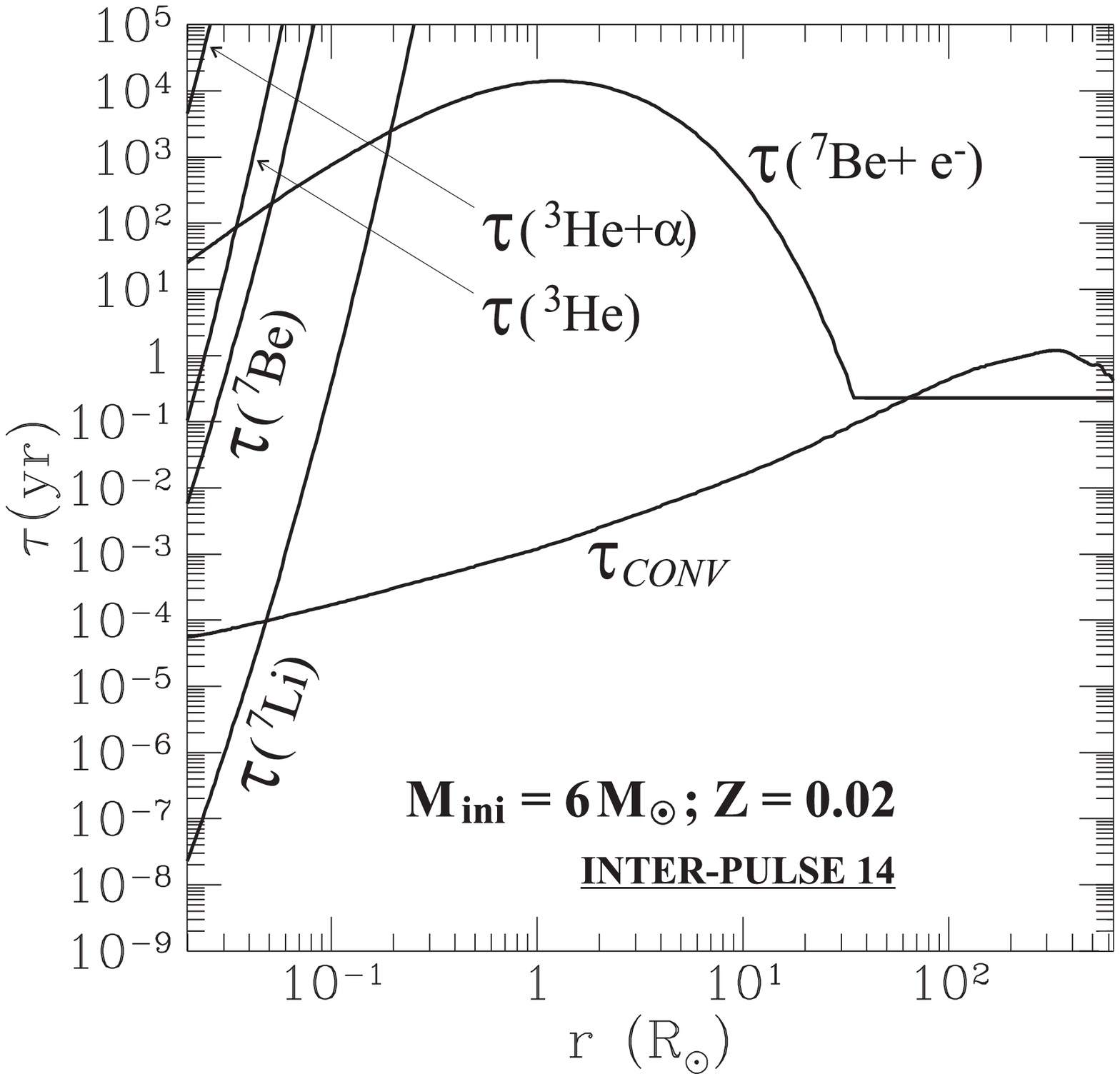,width=8.8cm}
  \caption[ ]{\label{reacthp}
    Nuclear reaction time scales for the reactions involved in the
    Cameron-Fowler scenario, inside the convective envelope of the \mass{6}
    ($Z = 0.02$) star during the 14th inter-pulse.  At that time, its base
    temperature is $\simeq 75 \, 10^6$ K. $\tau$(\chem{Be}{7}) and
    $\tau$(\chem{Li}{7}) correspond to the proton capture on \chem{Be}{7}
    and \chem{Li}{7}, respectively. $\tau$(\chem{He}{3}) and
    $\tau$(\chem{He}{3}$+\alpha$) relate to the \chem{He}{3} destruction
    by \chem{He}{3} and \chem{He}{4} capture, respectively. Finally,
    $\tau$(\chem{Be}{7}$+e^-$) is the electron capture rate on \chem{Be}{7}.
    The mean turn-over time of the convective motions is also plotted
    ($\tau_{CONV}$)
    }
\end{figure}

Sackmann \& Boothroyd (1992) were the first ones to quantitatively
demonstrate how the Cameron \& Fowler (1971) scenario leads to \chem{Li}{7}
synthesis in intermediate-mass AGB stars. This major result was obtained by
using a time-dependent convective diffusion algorithm suited for hot-bottom
convective envelopes. Indeed, at the base of such envelopes, some nuclear
reactions involved in \chem{Li}{7} production occur faster than mean
turn-over time associated with the convective motions (see Fig.
\ref{reacthp}).

\begin{figure}
  \psfig{file=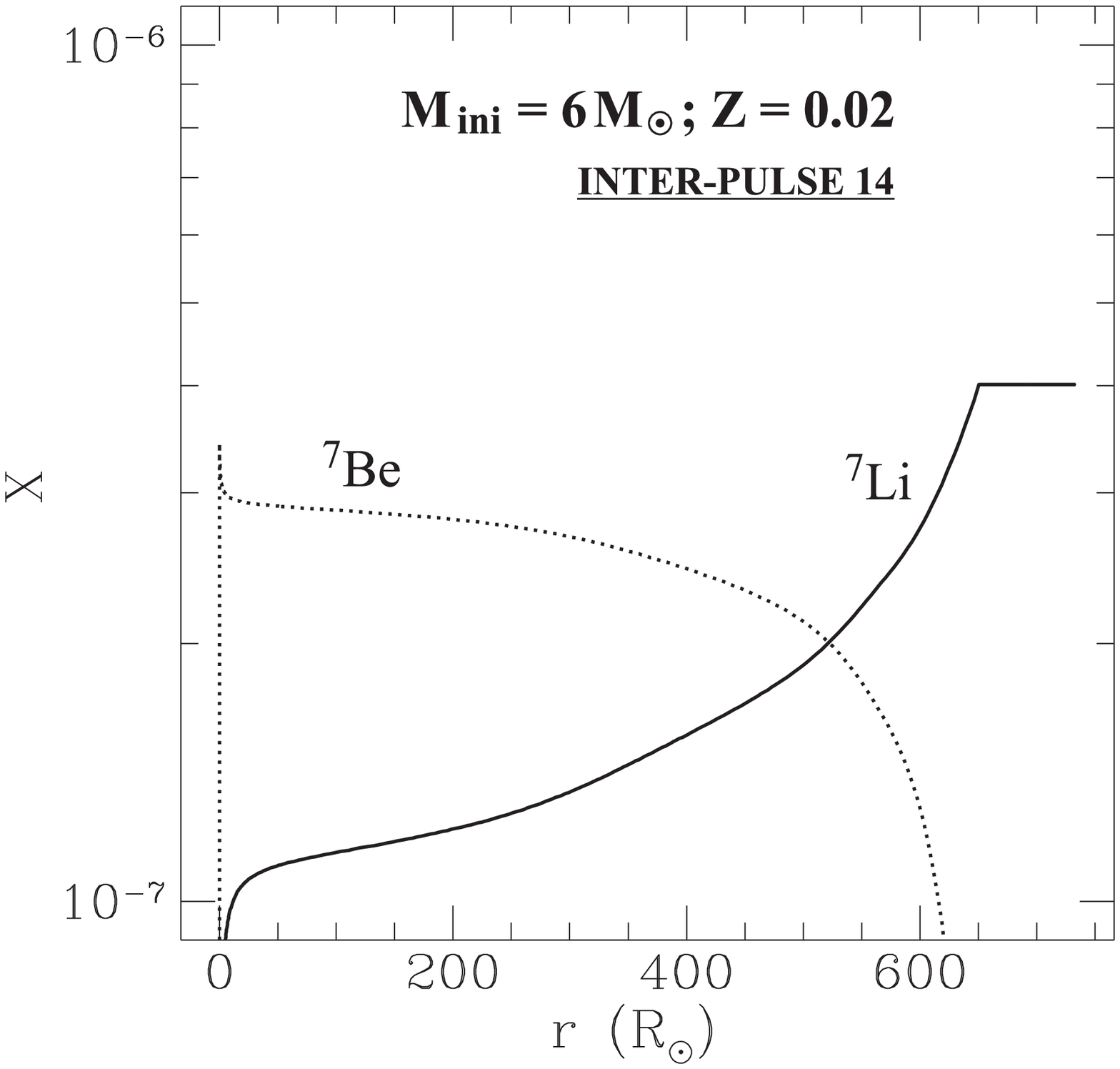,width=8.8cm}
  \caption[ ]{\label{libe}
    \chem{Be}{7} and \chem{Li}{7} profiles inside the same convective
    envelope as in Fig. \ref{reacthp}
    }
\end{figure}

We also use a time-dependent treatment to study the \chem{Li}{7} synthesis
[see Eqs. (4) in Sect. 2.2.3]. We show in Fig. \ref{libe} the typical
resulting \chem{Be}{7} and \chem{Li}{7} profiles inside the convective
envelope of our \mass{6} AGB models with $Z = 0.02$. When HBB occurs (see
Sect. 7.1.2), \chem{He}{3} first begins to burn through the
\reac{He}{3}{\rm{{}^3He}}{2p}{}{} and \reac{He}{3}{\alpha}{\gamma}{Be}{7}
reactions, the second one being significantly slower than the first one.  As
in deep stellar interiors, the \chem{Be}{7} destruction by electron capture
and proton capture are much longer than the time scale associated with the
convective mixing, the produced \chem{Be}{7} is efficiently up-heaved
towards cooler regions where it progressively decays in \chem{Li}{7}. Let us
emphasize that without fast convective mixing, \chem{Li}{7} could not
accumulate at the surface. Indeed, at the base of the convective envelope,
{\it (i)} \chem{Be}{7} would be partially destroyed by electron and proton
captures and {\it (ii)} the \chem{Li}{7} resulting from \chem{Be}{7} decay
would be quite instantaneously destroyed by proton capture too.

\begin{figure}
  \psfig{file=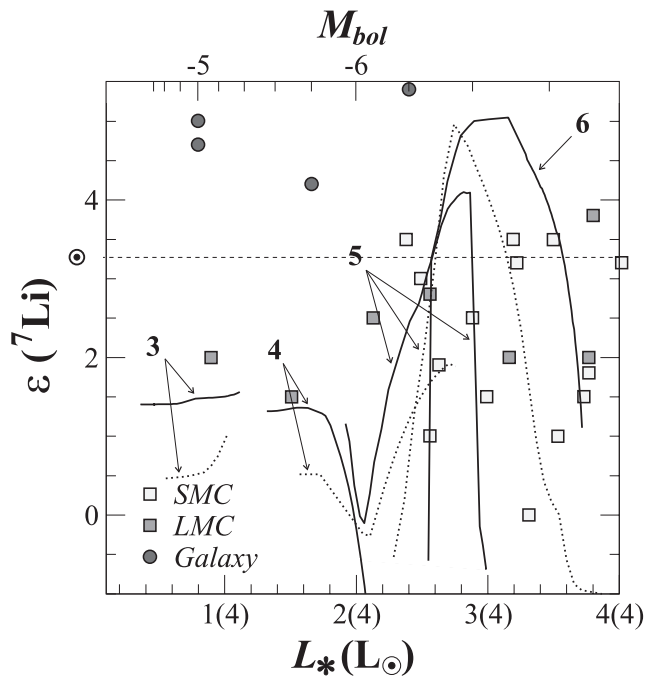,width=8.8cm}
  \caption[ ]{\label{lli}
    Resulting surface \chem{Li}{7} abundances, in term of
    $\varepsilon$(\chem{Li}{7}), for the seven computed TP-AGB phases, as a
    function of the stellar surface luminosity. The solid (dotted) lines
    refer to the $Z = 0.02$ (0.005) models. Observations are shown for
    bright AGB stars of the SMC and LMC (Smith et al. 1995) and of the
    Galaxy (Abia et al. 1991) }
\end{figure}

Let us now compare our predictions with Sackmann \& Boothroyd (1992) results
and with observations. A direct comparison with their Figs. 2a and 2b
reveals nearly identical features. In Fig. \ref{lli}, we present the
evolution of the surface \chem{Li}{7} abundance during the TP-AGB phase of
all our modeled stars. In our \mass{3} models, the temperature at the base
of the convective envelope is never high enough for \chem{Be}{7} to be
produced efficiently, and the surface \chem{Li}{7} abundance remains almost
constant at a very low value resulting from the first dredge-up. We also
see that our \mass{4} star with $Z = 0.005$ begins to produce some
\chem{Li}{7} when it enters the asymptotic regime, but the surface abundance
of this light element always stays at least one order of magnitude below
the interstellar medium (or cosmic) value $\varepsilon$(\chem{Li}{7})
$= \log {\rm{Li} \over \rm{H}} + 12 \simeq 3.3$. Note the \chem{Li}{7}
destruction inside the \mass{4} star with $Z = 0.02$ attesting that its
convective bottom temperature is high enough to burn \chem{Li}{7} through
\reac{Li}{7}{p}{\alpha}{}{} but not enough to produce \chem{Be}{7} by
\chem{He}{3} burning. Finally, our 5 and \mass{6} models present real high
\chem{Li}{7} surface enhancements, compared to the cosmic value. The
maximum abundances we obtain range between $\varepsilon$(\chem{Li}{7})
$\simeq 4.2$ to 5.1, depending on the stellar mass and metallicity. Our
super-lithium-rich stars are predicted to appear in a luminosity range
between $2.6 \, 10^4$ and \lum{3.6 \, 10^4}, i.e. for $M_{bol}$ between
-6.25 and -6.65. These values are in good agreement with those found by
Sackmann \& Boothroyd (1992).

As the evolution proceeds, the surface \chem{Li}{7} abundances finally
decrease again very rapidly. This indicates that \chem{He}{3} has been
almost completely burned in the envelope [remember that the
\reac{He}{3}{\rm{{}^3He}}{2p}{}{} is dominant], so that the \chem{Li}{7}
production can no more be supported and it is then destroyed by proton
capture. This fact clearly puts an upper limit in luminosity to observe
super-lithium-rich AGB stars. This last feature is also present in Sackmann
\& Boothroyd (1992), even if it appears somewhat later during the TP-AGB
phase of their most massive objects. This could be due to different mass
loss rate prescriptions.

How do these predictions compare with observational data (see Fig.
\ref{lli})? In the Magellanic Clouds, the large majority of lithium-rich
giants are observed within a narrow luminosity range, $M_{bol} \simeq -6$
to $\simeq -6.9$ (Smith \& Lambert 1989, 1990b; Plez et al. 1993; Smith
et al. 1995), in very good agreement with the predictions. The observed
abundance range is also well reproduced, even if no star has been observed
with $\varepsilon$(\chem{Li}{7}) higher than 4.5 in the Clouds. Moreover,
the mass estimate from pulsation theory for some lithium-rich giants of
both Clouds is consistent with the predicted mass range for surface
\chem{Li}{7} enhancement (Smith et al. 1995). Comparisons with \chem{Li}{7}
in galactic giants are more delicate because of the larger observational
uncertainties on luminosity and mass. Galactic super-lithium-rich AGB stars
can exhibit $\varepsilon$(\chem{Li}{7}) values as high as 5.4 (Abia et al.
1991), but these carbon stars have $M_{bol}$ laying between -5 and -6.2,
i.e. rather lower than the predicted luminosities. This could be due to
a bad luminosity determination, as it seems difficult to synthesize
\chem{Li}{7} inside the convective envelope of AGB stars with such
relatively low luminosities. However, most of the galactic
super-lithium-rich AGB stars have lower masses than the ones we model here,
and they will be addressed in a forthcoming paper.

\subsection {CNO nucleosynthesis and Fluorine production}

We now review all the species (from C) by indicating in which nuclear
reactions they are implicated inside the convective tongue of the successive
thermal pulses, as a function of the initial mass $M$ and metallicity $Z$ of
the computed TP-AGB models. For the discussion, we refer to Figs. 7 to 9 of
the electronic version of the paper. The mean mass fractions
$\overline{X_a}$ are calculated by taking the average of the mass fraction
profile of each nuclide $a$ over the maximum region covered by the
convective tongue of each thermal pulse, just after its disappearance, i.e.

\begin{equation}
  \overline{X_a} = {1 \over \Delta M_{conv. \atop tongue}} \, \int_{conv.
    \atop tongue} \, X_a \, dm \,.
\end{equation}

\begin{figure}
  \psfig{file=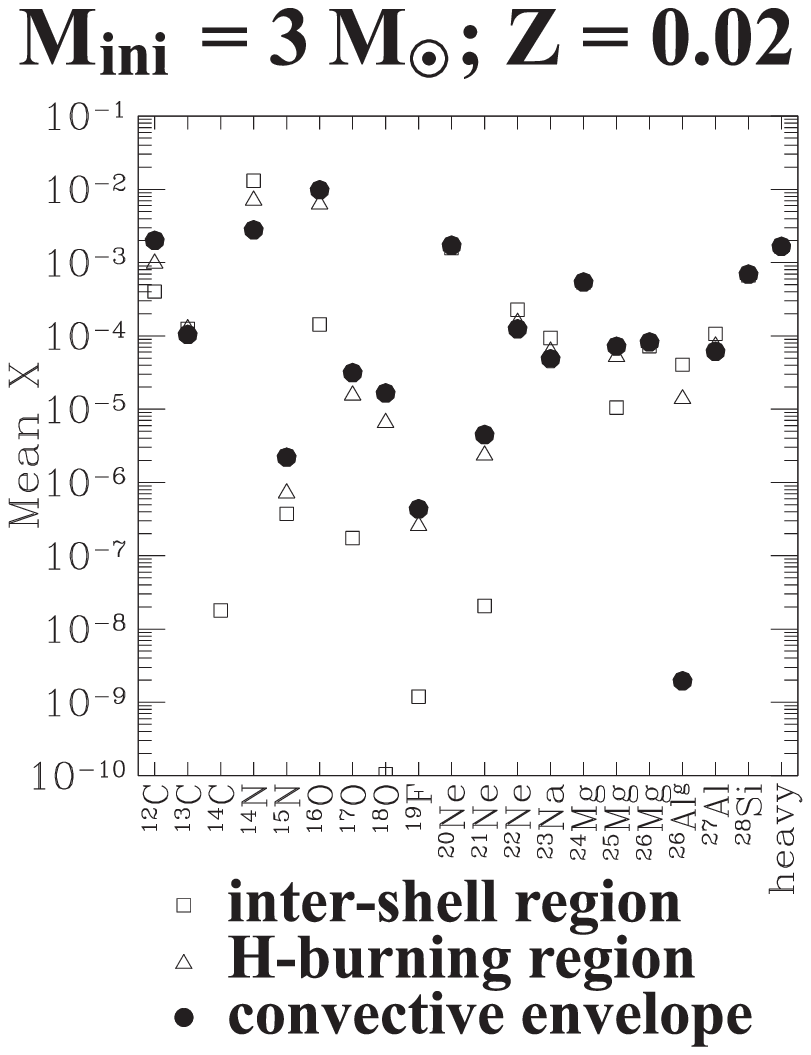,width=8.8cm}
  \caption[ ]{\label{xenv302}
    Mean mass fractions of the 20 most relevant nuclides inside the
    convective envelope, the HBS and the inter-shell region, for the
    \mass{3} ($Z = 0.02$) star }
\end{figure}

\begin{figure*}
  \psfig{file=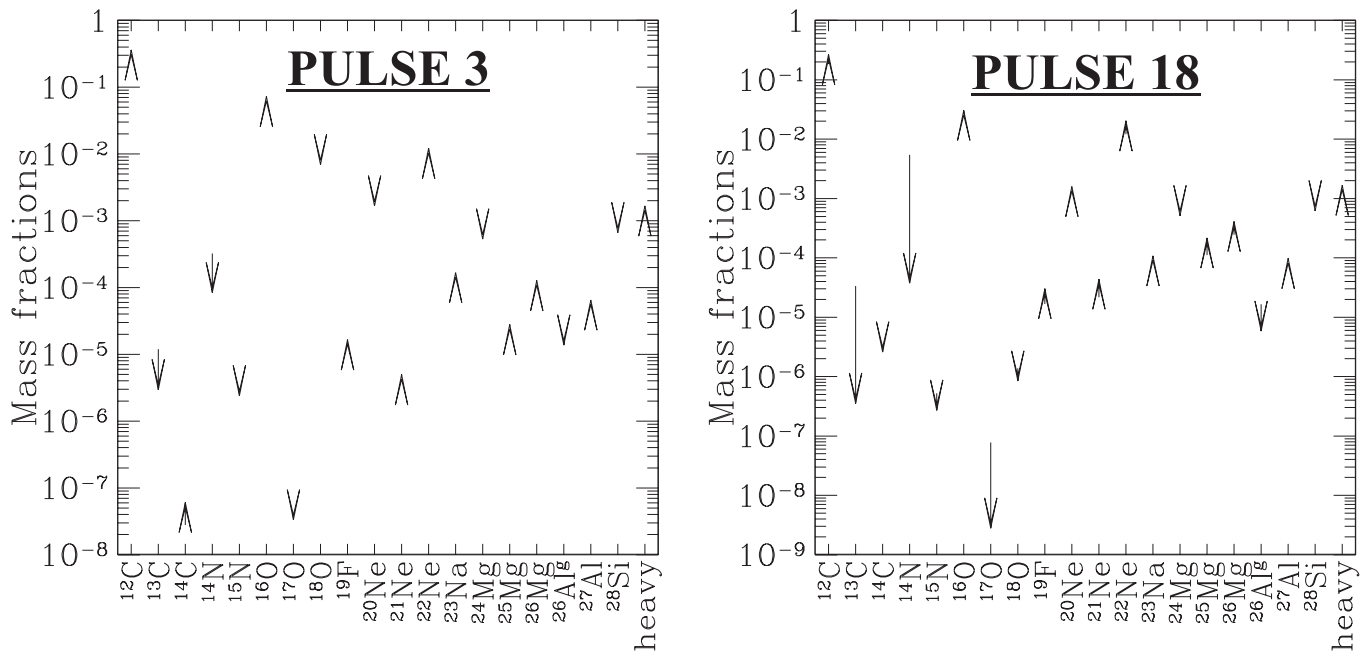,width=18.0cm}
  \caption[ ]{\label{xthp302}
    Variation of the mean mass fractions of the same 20 nuclides as in Fig.
    \ref{xenv302}, due to the thermal pulse nucleosynthesis, for the
    \mass{3} ($Z = 0.02$) star, inside an early thermal pulse (the 3d) and
    an asymptotic one (the 18th). Arrows indicate the mean abundance
    evolution from the beginning of a thermal pulse to its end }
\end{figure*}

We also refer to Fig. \ref{xenv302} that presents the mean mass fractions of
key nuclides (calculated as for the convective tongue) in each burning
region (i.e. the HBS and inter-shell regions) compared to their convective
envelope content, for the \mass{3} star with $Z = 0.02$. Similar figures for
the other six AGB stars are only available in the electronic version.
Similarly, Fig. \ref{xthp302} presents the abundance variations of the same
nuclides inside typical convective thermal pulses. Again, similar figures
for the other six AGB stars can solely be found in the electronic form. All
these figures are supporting the following nucleosynthesis descriptions.




During the growth of the convective tongue, the mean \chem{C}{12} mass
fraction first decreases due to the \chem{He}{4} mixing from the inter-shell
region. Then, at the thermal pulse maximum, \chem{C}{12} is produced by the
$3 \alpha$ reaction and also enhanced due to the downwards penetration of
the convective tongue into the He-burnt shell. Although this penetration is
commonly found, its amplitude seems somewhat more important in our case,
bringing more \chem{C}{12} into the thermal pulses. Note this increased
amount of \chem{C}{12} makes the 3DUP episodes more efficient to convert an
AGB star into a C star. During the first thermal pulses, the amount of
\chem{C}{12} that is produced increases due to the higher temperatures
reached at the base of the convective tongue. This bottom temperature
increase from pulse to pulse is a common feature of all the stars and it
traduces that the thermal pulse intensity increases. The last computed
thermal pulses of the \mass{3} object with $Z = 0.02$ are even hot enough to
partially destroy \chem{C}{12} by the \reac{C}{12}{\alpha}{\gamma}{}{}
reaction.

Most of the engulfed \chem{C}{13} is destroyed very efficiently through
\reac{C}{13}{\alpha}{n}{}{} that produces neutrons. The trend of the
remaining mean \chem{C}{13} mass fraction to increase from pulse to pulse is
due to the fact that stronger thermal pulses develop convective tongues
reaching regions always closer to the HBS, i.e. engulfing more material from
the inter-shell zone. As mentioned in Sect. 6.1, the bottom of the HBS
cannot be reached however.

\chem{C}{14} is only produced by the capture of part of the neutrons on the
engulfed \chem{N}{14}, through \reac{N}{14}{n}{p}{}{}. Note that this
reaction contributes to the production of protons. For initial masses above
\mass{4} (\mass{3}) with $Z = 0.02$ (0.005) however, the convective tongue
finally becomes hot enough (base temperature $> 260 \, 10^6$ K) to
significantly destroy the produced \chem{C}{14} by the
\reac{C}{14}{\alpha}{\gamma}{}{} reaction. The relatively low \chem{C}{14}
production inside the \mass{3} with $Z = 0.02$ object is due to the
relatively lower amount of \chem{C}{13} that is ingested (partially due to
the third \chem{C}{13} burning shell; see Sect. 7.1.1), leading to a smaller
quantity of neutrons. However, the last thermal pulses of this star ingest
increasing amounts of \chem{C}{13} and \chem{N}{14} from the inter-shell,
leading to an increasing production of \chem{C}{14}. Due to its short
lifetime ($t_{1/2} = 5730$ yr), \chem{C}{14} partially decays during the
inter-pulse phases of our less massive AGB models. This depletion is however
reduced by the \chem{C}{14} production inside the radiative \chem{C}{13}
burning shell. For the most massive AGB stars that have inter-pulse
durations shorter that $t_{1/2}$, \chem{C}{14} mostly behaves like a stable
nuclide, even if strongly destroyed by $\alpha$ captures. Consequently, as
explained with more details in Forestini et al. (1996), intermediate-mass
AGB stars are globally not important producers of \chem{C}{14}.

Apart from the first two thermal pulses (whatever the initial mass or
metallicity), \chem{N}{14} is almost completely destroyed inside thermal
pulses through \reac{N}{14}{\alpha}{\gamma}{}{}. This reaction, as first
noted by Iben (1976), also significantly contributes to the energetics of
a thermal pulse. Only a small part of \chem{N}{14} is destroyed through
the \reac{N}{14}{n}{p}{C}{14} reaction. Indeed, the inter-shell mean mass
fraction of \chem{N}{14} is ranged between $\sim 10^{-2}$ for the \mass{3}
star with $Z = 0.02$ and $\sim 10^{-3}$ for the \mass{5} star with $Z =
0.005$, while the production level of \chem{C}{14} is $\sim 10^{-5}$
($\sim 10^{-6}$), in mass fraction, for the $Z = 0.02$ (0.005) models
(except for the \mass{3} with $Z = 0.02$ star; see above).

As are \chem{C}{13} and \chem{N}{14}, \chem{N}{15} is engulfed in a thermal
pulse from the inter-shell region. The only significant reaction that
produces this element inside the convective tongue itself is
\reac{O}{18}{p}{\alpha}{}{} (see below for the \chem{O}{18} production).  On
the other hand, \chem{N}{15} is mainly destroyed by the
\reac{N}{15}{\alpha}{\gamma}{}{} reaction. Its rate is lower than that of
\reac{N}{14}{\alpha}{\gamma}{}{} so that \chem{N}{15} is efficiently
destroyed in all the stars only in the asymptotic regime, i.e. when thermal
pulses become very intense. Remember that it arrives earlier for lower $Z$
objects. Note that the \mass{3} stars (whatever $Z$) maintain a relatively
higher \chem{N}{15} abundance, especially during the full amplitude thermal
pulses. Indeed, while these thermal pulses efficiently destroy \chem{N}{15}
whatever the initial mass of the star, its initial abundance inside the
convective tongue is higher for \mass{3} objects.  This is due to the fact
that the engulfed inter-shell region of \mass{3} stars, that contains
\chem{N}{15}, is significantly thicker (in mass) compared to that of more
massive ones.

\chem{O}{16} being partially destroyed in the HBS, the mixing of the
inter-shell region reduces its mean mass fraction. On the other hand, at the
maximum extent of the thermal pulses, it is enhanced, like \chem{C}{12}, due
to the downwards penetration of the convective tongue. From pulse to pulse,
it also becomes significantly produced by \reac{C}{12}{\alpha}{\gamma}{}{}.
This is especially visible for $Z = 0.005$ stars, the initial \chem{O}{16}
content of which is lower. This primary \chem{O}{16} production by the
thermal pulses finally stabilizes around $\sim 10^{-2}$ (in mass fraction).
The decrease observed during the last thermal pulses of the \mass{3} AGB
star with $Z = 0.02$ traduces the more important role of mixing of
inter-shell material (already emphasized for \chem{N}{14} and \chem{N}{15}).

The \chem{O}{17} previously contained in the inter-shell region is also
mixed in the convective tongue. It is not produced inside thermal
pulses. On the contrary, it can be destroyed mainly through
\reac{O}{17}{\alpha}{n}{}{} and \reac{O}{17}{\alpha}{\gamma}{}{}, the
first reaction being in general at least ten times faster than the second
one. This weakly contributes to produce neutrons inside the thermal pulses.
This \chem{O}{17} destruction already operates from the first thermal
pulses, whatever the initial $M$ and $Z$. In later thermal pulses, the
\chem{O}{17} mass fraction stabilizes around 4 to $8 \, 10^{-10}$, that
represents a balance between the amount of \chem{O}{17} brought from the
inter-shell region and destroyed by the thermal pulse. Again, our \mass{3}
models with $Z = 0.02$ indicate a still dominant role of the inter-shell
mixing. It is less true for the $Z = 0.005$ models because at lower $Z$,
thermal pulses are somewhat hotter, that favors \chem{O}{17} depletion.

Contrarily to the case of \chem{O}{17}, the inter-shell mass fraction of
\chem{O}{18} is very low ($\la 10^{-9}$ typically). Its presence inside
thermal pulse consequently results from its production by the
\reac{N}{14}{\alpha}{\gamma}{F}{18} reaction. Note that the rapid
\chem{F}{18} $\beta-$decay to \chem{O}{18} also significantly contributes to
the thermal pulse energetics. This element is destroyed by three reactions
that are, by decreasing order of nuclear time scales,
\reac{O}{18}{\alpha}{\gamma}{}{}, \reac{O}{18}{\alpha}{n}{}{} and
\reac{O}{18}{p}{\alpha}{N}{15}. This last reaction can even become faster
than \reac{O}{18}{\alpha}{n}{}{} when significant amounts of protons are
produced. This especially arrives for the late thermal pulses of the
\mass{3} star with $Z = 0.02$. The convective tongues being hotter for more
massive AGB stars and/or later along the TP-AGB phase of a given star, the
\chem{O}{18} destruction, which is very sensitive to temperature, becomes
more and more efficient from pulse to pulse. This point was already noted
by Boothroyd \& Sackmann (1988b). The thermal pulse equilibrium mean mass
fraction between production and destruction is ranged between $10^{-7}$ and
$10^{-6}$ depending on $M$ and $Z$.

As for \chem{O}{18}, \chem{F}{19} is destroyed in the HBS. It is however
produced inside thermal pulses. As first demonstrated by Forestini et al.
(1992), the \chem{F}{19} synthesis involves many nuclear reactions, so that
it constitutes a good tracer of the nucleosynthesis conditions prevailing
inside thermal pulses. \chem{F}{19} is produced by the
\reac{N}{15}{\alpha}{\gamma}{}{} reaction. The rather rapid increase of the
mean \chem{F}{19} mass fraction during the first thermal pulses of each AGB
star is due to the increasing importance of the inter-shell \chem{N}{15}
destruction. When convective tongues become hot enough to completely destroy
the engulfed \chem{N}{15}, the following thermal pulses only ingest
\chem{N}{15} from matter that has not experienced the preceding thermal
pulse (i.e. $\la 50\%$ of the convective tongue extension in mass typically;
see Table \ref{endagb}). However, as we have explained above, \chem{N}{15}
can be produced by the chain
\reac{N}{14}{\alpha}{\gamma}{F}{18}$(\beta^+)$\reac{O}{18}{p}{\alpha}{N}{15}.
Neutrons are mainly coming from the \reac{C}{13}{\alpha}{n}{}{} reaction,
and to a lower extent, from \reac{O}{18}{\alpha}{n}{}{},
\reac{Ne}{22}{\alpha}{n}{}{}, \reac{Ne}{21}{\alpha}{n}{}{} (see below) and
\reac{O}{17}{\alpha}{n}{}{}. Protons are coming from $\rm{(n , p)}$
reactions, mostly on \chem{Al^g}{26} and \chem{N}{14} (see below). However,
{\it (i)} only part of the neutrons are available to produce protons as many
$\rm{(n , \gamma)}$ reactions also occur (see Sect. 7.5) and {\it (ii)} a
small ($\la 10\%$) part of the protons are also captured by other nuclides
than \chem{O}{18} [mainly by \reac{N}{15}{p}{\alpha}{C}{12}].  The success
of this reaction chain mainly depends on the rate of \chem{O}{18}
destruction, mainly by $\alpha$ captures (see above). This destruction
becomes more efficient from pulse to pulse so that only a small fraction of
\chem{O}{18} is available for the $\rm{(p , \alpha)}$ reaction. On the other
hand, when convective tongues become very hot (base temperature $> 280 \,
10^6$ K), \chem{F}{19} begins to be destroyed by \reac{F}{19}{\alpha}{p}{}{}
and \reac{F}{19}{n}{\gamma}{}{} (if significant amounts of neutrons are
available). In conclusion, three facts concur to considerably decrease the
\chem{F}{19} production after the first thermal pulses (whatever $M$ and
$Z$). Indeed, from pulse to pulse and especially for the most massive AGB
stars,

\begin{itemize}

\item
  the amount of \chem{C}{13} engulfed in the convective tongue of a thermal
  pulse decreases (due to the partial destruction of \chem{C}{13} in the
  inter-shell region during the inter-pulse phase; see Sect. 7.1.1);

\item
  the amount of \chem{N}{15} ingested from the inter-shell region also
  decreases;

\item
  the $\alpha$ captures on \chem{O}{18} are more frequent;

\item
  the \chem{F}{19} destruction becomes significant. This can in particular
  lead to a partial destruction of the \chem{F}{19} synthesized in the
  previous thermal pulses for stars with $M >$ \mass{4}.

\end{itemize}

\subsection {NeNa and MgAl nucleosynthesis}

The mean \chem{Ne}{20} abundance is roughly unchanged in the HBS as well as
by the thermal pulse nucleosynthesis. In number, the contribution of
\reac{O}{17}{\alpha}{n}{Ne}{20} is rather negligible.

\chem{Ne}{21} is significantly destroyed in the HBS. Inside the convective
tongues of quite hot thermal pulses (base temperatures above $270 \, 10^6$
K), it is produced by the small proportion of \chem{O}{18} that is destroyed
through the \reac{O}{18}{\alpha}{n}{}{} reaction. There is also a small
contribution by \reac{O}{17}{\alpha}{\gamma}{}{}. Only a very small
\chem{Ne}{21} fraction can be destroyed by the \reac{Ne}{21}{\alpha}{n}{}{}
at the end of very hot thermal pulses (base temperatures above $290 \, 10^6$
K).

\chem{Ne}{22} is somewhat produced in the HBS. Its abundance is
considerably enhanced inside thermal pulses (by typically a factor of
100) due to its production by \reac{O}{18}{\alpha}{\gamma}{}{}. However,
its abundance from pulse to pulse becomes constant (or even slightly
decreases inside full amplitudes thermal pulses of the most massive stars
and/or those with lower $Z$). At their maximum extent, such convective
tongues are effectively hot enough (base temperatures $> 290 \, 10^6$
K) to activate the \reac{Ne}{22}{\alpha}{\gamma}{Mg}{26} and
\reac{Ne}{22}{\alpha}{n}{Mg}{25} reactions. However, as noted below, only a
small part of the \chem{Ne}{22} nuclides are converted into \chem{Mg}{26}
or \chem{Mg}{25}, respectively (typically one hundredth). Nevertheless,
this consists in a non-negligible neutron source for these thermal pulses,
as first noted by K\"appeler et al. (1990) and emphasized by Straniero et al.
(1995).

Concerning \chem{Na}{23}, one has to distinguish between the 3, 4, \mass{5}
with $Z = 0.02$ and \mass{3} with $Z = 0.005$ stars on the one hand, and the
\mass{6} with $Z = 0.02$, 4 and \mass{5} with $Z = 0.005$ stars on the
other. Inside the former ones, \chem{Na}{23} is significantly produced at
the bottom of the HBS and very slightly produced by the HeBS [during the
inter-pulse phase, through the \reac{Ne}{20}{\alpha}{p}{}{} reaction].
Furthermore, whatever $M$ and $Z$, \chem{Na}{23} is almost unchanged by the
thermal pulse nucleosynthesis. As a consequence, its mean mass fraction
inside the convective tongue does not evolve significantly from pulse to
pulse. At the opposite, inside the latter ones, the bottom of the HBS is hot
enough to somewhat destroy \chem{Na}{23} while it is produced in its upper
part. However, the HeBS substantially produces this nuclide during the
inter-pulse phase. Consequently, when the convective tongues of such stars
penetrate the inter-shell region, they engulf material that is impoverished
in \chem{Na}{23}; this explains its mean mass fraction decrease from pulse
to pulse.

The \chem{Mg}{24} abundance is almost unchanged by the thermal pulse
nucleosynthesis too. Furthermore, the HBS of \mass{3} stars is not hot
enough to destroy it significantly. Its mean mass fraction in the convective
tongue region of such stars consequently remains almost constant from pulse
to pulse. The situation is quite different for more massive stars (whatever
$Z$) as they destroy \chem{Mg}{24} at the bottom of their HBS. Dilution of
inter-shell material by the successive convective tongues thus traduces in a
mean mass fraction decrease with time.

Intermediate nuclide of the MgAl chain, \chem{Mg}{25} is modified by the
proton burning inside the HBS. We refer to Sect. 7.1.1 for more details
(dependence on $M$ and $Z$). However, the evolution of its mean mass
fraction inside the convective tongues of thermal pulses is dominated by its
production through the \reac{Ne}{22}{\alpha}{n}{}{} reaction [operating at a
quite comparable rate than \reac{Ne}{22}{\alpha}{\gamma}{}{}, following our
present knowledge of the corresponding nuclear reaction rates]. Indeed, even
if this reaction is very slow inside thermal pulses of relatively low mass
AGB stars (especially for their first thermal pulses), \chem{Ne}{22} is much
more abundant than \chem{Mg}{25} (by typically a factor of one hundred) in
that region, that explains the sensible increase of the \chem{Mg}{25} mean
mass fraction.

\chem{Mg}{26} is slightly (significantly) depleted in the HBS of the $Z =
0.02$ (0.005) models. On the other hand, it is substantially produced by the
thermal pulse nucleosynthesis {\it (i)} by the
\reac{Ne}{22}{\alpha}{\gamma}{}{} reaction (same remark as for \chem{Mg}{25}
above) and, to a lower extent, {\it (ii)} by the \reac{Al^g}{26}{n}{p}{}{}
reaction (see below). The slope with which its mean mass fraction increases
form pulse to pulse is consequently higher for hotter convective tongues,
i.e. steeper for full amplitude thermal pulses of massive AGB stars.

\chem{Al^g}{26} is one of the most important by-product of the HBS (see
above). With a lifetime $t_{1/2} = 7 \, 10^5$ yr, i.e. much longer than the
inter-pulse duration of intermediate-mass AGB stars, it essentially behaves
like a stable nuclide in the inter-shell region. When engulfed inside the
convective tongue of a thermal pulse, it is mainly destroyed through the
\reac{Al^g}{26}{n}{p}{}{} reaction and, at least ten times slower, by
\reac{Al^g}{26}{n}{\alpha}{}{}. As a result, \chem{Al^g}{26} is the most
important proton source inside thermal pulses, whatever $M$ and $Z$. Its
destruction is however partial, due to the relatively low neutron abundance
(see Sect. 7.5 below). Note that \reac{Al^g}{26}{n}{p}{}{} is also the
principal neutron source in the radiative \chem{C}{13} burning shell, as
reported by Wasserburg et al. (1994) too.

\chem{Al}{27} is significantly produced in the HBS, especially in the most
massive AGB stars. It does not significantly participate to the thermal
pulse nucleosynthesis. As a consequence, it is engulfed by the convective
tongue from the inter-shell region and its corresponding mean mass fraction
slightly increases from pulse to pulse. Note however that with time, most of
the H-burning of the \mass{5} AGB star with $Z = 0.005$ occurs inside its
convective envelope. This considerably reduces the \chem{Al}{27} production.

Si, P and S are almost not concerned by charged particle nuclear reactions
inside the convective tongue of thermal pulses. On the other hand, they are
also rather unchanged in the HBS. The only role played by these elements
concerns neutron captures that we now briefly discuss.

\subsection {Neutron production}

Many evolved AGB stars show considerable overabundances in elements heavier
than iron (see e.g. Smith \& Lambert 1986). The original discovery by
Merrill (1952) of Tc in the spectra of some S stars definitively proved that
these heavy elements are synthesized inside AGB stars. Iben (1975) first
demonstrated how such species can indeed be produced inside the thermal
pulses by slow neutron captures (the so-called $s-$process), those neutrons
coming from the \reac{Ne}{22}{\alpha}{n}{}{} reaction. However, Iben's
computations concerned a very massive AGB star (\mass{7}), while ever since,
all the observed stars showing $s$ elements enrichment have been identified
as being of lower mass (i.e. $<$ \mass{6} and for many of them $<$
\mass{3}).  More recent observations (Aaronson \& Mould 1985) and
evolutionary models (Malaney \& Boothroyd 1987) then strongly suggested that
the major neutron source had to be the \reac{C}{13}{\alpha}{n}{}{} reaction.

\chem{C}{13} is engulfed by the convective tongues of thermal pulses from
the inter-shell region where the HBS ashes accumulate. However, the amount
of \chem{C}{13} spread out by the HBS is by far (i.e. at least a factor of
10 or more) insufficient to explain the $s-$process that requires more than
one neutron by seed \chem{Fe}{56}. This so-called ``$s-$process mystery''
(Sackmann \& Boothroyd 1991) is a common failure of all the recent stellar
evolution models of TP-AGB stars. It is moreover reinforced by the fact that
other lighter nuclides also capture part of the available neutrons (mainly,
by increasing order of atomic mass, \chem{C}{12}, \chem{N}{14},
\chem{O}{17}, \chem{F}{19}, \chem{Mg}{24}, \chem{Mg}{25}, \chem{Al^g}{26},
\chem{Si}{28}, \chem{Si}{29}, \chem{Si}{30}, \chem{P}{31}, \chem{S}{32},
\chem{S}{33}, \dots). One has to note that while the neutron abundance has
to be increased to solve the $s-$process problem, the present computations
probably slightly overestimate the abundances of the above mentioned
nuclides.

More specifically, in our thermal pulse models, the ratio of the engulfed
mass of \chem{C}{13} to the mass of \chem{Fe}{56} inside the convective
tongue at its maximum extent ranges between 0.002 to 0.03.  This upper limit
is reached during the asymptotic regime of all our modeled stars and thus
appears to be rather independent of the initial mass or metallicity. Inside
their first thermal pulses however, \mass{3} stars already produce neutrons
(ratio roughly equal to 0.02) while at the opposite, the corresponding
thermal pulses of \mass{6} stars do not engulf substantial amounts of
\chem{C}{13}. Furthermore, we noted (see Sect. 7.4) that the \chem{Ne}{22}
neutron source operates after the maximum of very intense thermal pulses, at
a rate that is growing with the maximum temperature of the convective
tongue, i.e. with initial mass and/or pulse number.

Let us stress that such low quantities of \chem{C}{13} ingested by thermal
pulses are partly explained by the partial \chem{C}{13} radiative burning,
operating at the bottom of the inter-shell region during the inter-pulse
phase (see Sect. 7.1.1). Consequently, it seems clear that the $s-$process
occuring inside TP-AGB stars has two distinct origins (in time and space):

\begin{itemize}

\item
  during the inter-pulse phase, {\it radiative} $s-$process occurs in a
  radiative zone that is the lower part of the inter-shell region; the only
  neutron source is \chem{C}{13};

\item
  during and inside the convective tongue associated with thermal pulses,
  the {\it classical} $s-$process takes place with two possible neutron
  sources: {\it (i)} the remainder amount of \chem{C}{13} that is engulfed
  by the convective tongue and, to a lower extent (maximum a few percents),
  {\it (ii)} the \chem{Ne}{22}, close to the end of hot thermal pulses.
  Let us emphasize that the $\rm{(\alpha , n)}$ and accompanying
  $\rm{(n , \gamma)}$ reactions have non-negligible energetic effects
  on the structural evolution of the convective tongues.

\end{itemize}

We compute again some thermal pulses by treating together nucleosynthesis
and time-dependent convective mixing through Eqs. (4). We observed an
abundance gradient inside the convective tongue for a few nuclides that are
involved in very fast nuclear reactions (neutrons, protons and some unstable
nuclides like \chem{F}{18} or \chem{Na}{22}). Globally however, the final
abundances resulting from the thermal pulse nucleosynthesis were similar to
within $\sim 10\%$ compared to standard computations. More specifically,
neutrons were systematically found to be much more abundant (by typically
five orders of magnitude) at the base of the convective tongues, meaning
that the ingested \chem{C}{13} is first transported down before it is
destroyed by \reac{C}{13}{\alpha}{n}{}{}. The liberated neutrons are very
rapidly captured, as attested e.g. by a greater proton abundance at the base
of the convective tongue too. This leads us to the conclusion that matter
that is actually irradiated by the neutron flux is only the bottom part of
each convective tongue. The number of free neutrons by seed \chem{Fe}{56} is
so enhanced. Such an approach has already been suggested by Malaney et al.
(1988). The conclusion, however, remains that in order to reproduce the
observed distribution of $s$ elements in the primitive solar system, higher
neutron fluxes are still needed.

The most natural way to conciliate observations and theoretical models
concerning the production of the $s$ elements should be to increase the
amount of \chem{C}{13} in the inter-shell region. This can be done if
protons are transported down, e.g. during the 3DUP. \chem{C}{13} could then
be produced by the \reac{C}{12}{p}{\gamma}{}{} reaction. Such a scenario
would enhance the $s-$process in both sites where it can occur.  In our
opinion, the present failure to build this extra-amount of \chem{C}{13}
could be related to our bad treatment of convective boundaries, especially
when they rapidly penetrates very inhomogeneous regions, like it is the case
during the 3DUP events. The possible occurrence of such a slow-particle
transport process able to transport protons down into the inter-shell region
and its exact efficiency will be quantitatively investigated in the context
of low-mass AGB stars. These stars have indeed longer inter-pulse and 3DUP
durations that could allow to transport significant amounts of hydrogen
downwards (see Sect. 9).

Straniero et al. (1995) have recently shown that such enhanced amounts of
\chem{C}{13} in the inter-shell region could indeed allow the radiative
$s-$process to occur. Most interestingly, they found that the resulting
signatures are rather similar to those coming from classical $s-$process
computations and are even reached faster. This work is very important as
it clearly demonstrates that this so-called ``\chem{C}{13} pocket''
allows to explain the observed $s$ element synthesis in AGB stars.

\subsection {Resulting abundances of each nuclide inside the various
  nucleosynthesis sites}

In order to summarize our detailed discussion of nucleosynthesis in the
previous sections, we now identify which specific nuclear region, inside a
TP-AGB star, mainly contributes to the surface abundance change of each
nuclide, when the third dredge-up occurs.

Concerning the light elements, let us just recall (see Sect. 7.2) that
\chem{Li}{7} can be produced by HBB in the convective envelope of $>$
\mass{4.5} (\mass{4}) AGB stars with $Z = 0.02$ (0.005).

When a third dredge-up occurs, the convective envelope first penetrates the
inter-shell region that is depleted in \chem{C}{12} and then reaches a
deeper region previously enriched in \chem{C}{12} by the convective tongue.
As a result, \chem{C}{12} globally increases in the convective envelope. The
situation is reversed for \chem{C}{13}. After a 3DUP, the envelope is
somewhat impoverished in \chem{C}{13} as this nuclide is almost completely
destroyed inside the convective tongue. Consequently, the isotopic ratio
\chem{C}{12}/\chem{C}{13} significantly increases at each 3DUP, all the more
as the convective envelope mass is reduced. However, in our $\ge$ \mass{5}
(\mass{4}) TP-AGB stars with $Z = 0.02$ (0.005), HBB is strong enough to
partially convert \chem{C}{12} in \chem{C}{13} through the CN cycle so that
the surface \chem{C}{12}/\chem{C}{13} drastically decreases. For these
objects, the 3DUP and HBB have opposite effects.  Last but not least,
\chem{C}{14}, somewhat produced by thermal pulses and the third burning
shell, is up-heaved to the surface during 3DUP events, where it however
substantially decays in the less massive AGB stars.

\chem{N}{14} is produced in the inter-shell region but is strongly destroyed
inside thermal pulses. \chem{N}{15} is destroyed in the inter-shell region
while it is somewhat produced by the first few thermal pulses and then
destroyed inside full amplitude thermal pulses, i.e. those that can be
followed by a 3DUP. As a consequence, the surface isotopic ratio
\chem{N}{14}/\chem{N}{15} is expected to slightly increase from 3DUP to
3DUP. Again, as the CN cycle operates inside the convective envelope of the
most massive AGB stars, \chem{N}{14} is produced from \chem{C}{12} and
\chem{N}{15} is destroyed, thus leading to very high
\chem{N}{14}/\chem{N}{15} ratios. If HBB occurs at very high temperatures
(i.e. $\ga 10^8$ K) a rather long time enough (i.e. for the lowest mass loss
rate), the approach towards the CNO bi-cycle equilibrium leads to an
increase of the \chem{N}{15} abundance and consequently, a decreasing
\chem{N}{14}/\chem{N}{15} ratio. This mainly concerns our \mass{6}
(\mass{5}) AGB star with $Z = 0.02$ (0.005).

\chem{O}{16}, like \chem{C}{12}, is depleted in the inter-shell region, but
is produced inside the thermal pulses. During 3DUP events, its surface
abundance very slightly increases. \chem{O}{17} is always considerably
depleted in both the inter-shell and thermal pulse regions, so that a 3DUP
decreases its surface abundance. Finally, at the end of full amplitude
thermal pulses, \chem{O}{18} has a lower abundance as inside the convective
envelope. As it is also completely destroyed in the inter-shell region, its
abundance drastically decreases from 3DUP to 3DUP. Consequently, both
\chem{O}{16}/\chem{O}{17} and \chem{O}{16}/\chem{O}{18} surface isotopic
ratios significantly increase during each dredge-up. If HBB is strong enough
however, the operation of the ON cycle increases the amount of \chem{O}{17}
to the expense of \chem{O}{16} and substantially destroys \chem{O}{18}. So,
in massive AGB stars, \chem{O}{16}/\chem{O}{17} can decrease while
\chem{O}{16}/\chem{O}{18} further increases much more than in stars that do
not experience HBB.

\chem{F}{19} is destroyed in the HBS but it is significantly produced by
thermal pulses. As a consequence, 3DUP events must lead to correlated
\chem{F}{19} and \chem{C}{12} surface enhancements, since both elements are
produced in the same region.

As already explained, \chem{Ne}{20} is almost unchanged in the HBS and
thermal pulse nucleosynthesis. \chem{Ne}{21} is destroyed inside the
inter-shell region but produced inside thermal pulses, this production
slightly increasing from pulse to pulse, whatever $M$ or $Z$. Third
dredges-up consequently slightly decreases the \chem{Ne}{20}/\chem{Ne}{21}
isotopic ratio. In our \mass{6} (\mass{5}) AGB models with $Z = 0.02$
(0.005) however, \chem{Ne}{21} is partially destroyed by HBB. On the other
hand, the \chem{Ne}{20}/\chem{Ne}{22} ratio substantially decreases from
3DUP to 3DUP due to large \chem{Ne}{22} production in the HBS and especially
inside the thermal pulses.

During 3DUP events, the surface abundance of \chem{Na}{23} slightly
increases for AGB stars with initial masses up to \mass{4} (\mass{3}) and
$Z = 0.02$ (0.005), due to its production in the HBS. The hotter HBS of
more massive AGB stars do not produce \chem{Na}{23}, globally. Our
\mass{6} (\mass{5}) models for the same respective $Z$ even slightly
destroy \chem{Na}{23} by HBB.

\chem{Mg}{24} is somewhat destroyed in the HBS of our most massive AGB
stars, especially those with $Z = 0.005$, while its abundance does not
change significantly after the thermal pulse nucleosynthesis. \chem{Mg}{25}
is depleted in the inter-shell region of our \mass{3} models with $Z =
0.02$ as well as in the $Z = 0.005$ models of all masses. It is however
produced in the HBS of $>$ \mass{4} AGB stars with $Z = 0.02$.
Nevertheless, its surface abundance evolution following 3DUP events is
mainly conditioned by its large abundance (compared to the envelope one)
in the region where thermal pulses occur, especially inside the most
massive AGB stars. Therefore, the surface \chem{Mg}{24}/\chem{Mg}{25}
isotopic ratio necessarily decreases in all AGB stars. The same is true
for the \chem{Mg}{24}/\chem{Mg}{26} ratio that roughly behaves in a
similar way. Reasons are also rather similar, even if \chem{Mg}{26} is
less modified than \chem{Mg}{25} in the HBS. At the surface of our
\mass{6} (\mass{5}) with $Z = 0.02$ (0.005) AGB stars, the abundances of
both \chem{Mg}{25} and \chem{Mg}{26} can be enhanced to the expense of
\chem{Mg}{24}, if HBB is strong enough, leading to very low
\chem{Mg}{24}/\chem{Mg}{25} and \chem{Mg}{24}/\chem{Mg}{26} ratios. As we
will see in Sect. 8.1 however, this only concerns the end of the TP-AGB
phase in extreme situations allowing HBB to occur a long time enough (e.g.
in case of relatively low mass loss rates).

Only a very small amount of \chem{Al^g}{26} is present in the convective
envelope of E-AGB stars as a result of the first (and eventually the second)
dredge(s)-up. Along the TP-AGB phase, \chem{Al^g}{26} is largely produced in
the HBS and most of it accumulates in the inter-shell region. Inside full
amplitude thermal pulses however, it is somewhat destroyed. For its part,
\chem{Al}{27} is slightly produced in the HBS, especially inside the most
massive AGB stars. Consequently, \chem{Al^g}{26}/\chem{Al}{27} significantly
increases with time during the TP-AGB phase, due to the repetitive 3DUP
episodes. Of course, dilution by convective mixing being more important for
increasing $M$, the highest ratios are expected at the surface of the less
massive AGB stars. However, the same extreme cases mentioned in the case of
the Mg isotopic ratios can lead to very high \chem{Al^g}{26}/\chem{Al}{27}
ratios for the most massive AGB stars.

>From Si, the 3DUP does not change significantly the surface isotopic ratios.
However, as stressed in Sect. 7.5, when models will be able to reproduce the
$s$ element enhancements at the surface of evolved AGB stars, Si and S
isotopic ratios could be slightly modified by the high neutron flux.

\section {Wind composition along TP-AGB phase}

In this last section, we present our predictions concerning {\it (i)} the
evolution of all the surface isotopic ratios along the TP-AGB phase (by
including our evolutionary models and extrapolations up to the convective
envelope removal) and {\it (ii)} the total amount of material ejected at
different ages for all the species up to \chem{Si}{28}. We refer to the
previous sections for detailed explanations. Thus, in the following, we will
just comment these predictions.

\subsection {Surface isotopic ratios along TP-AGB}

\begin{figure*}
  \psfig{file=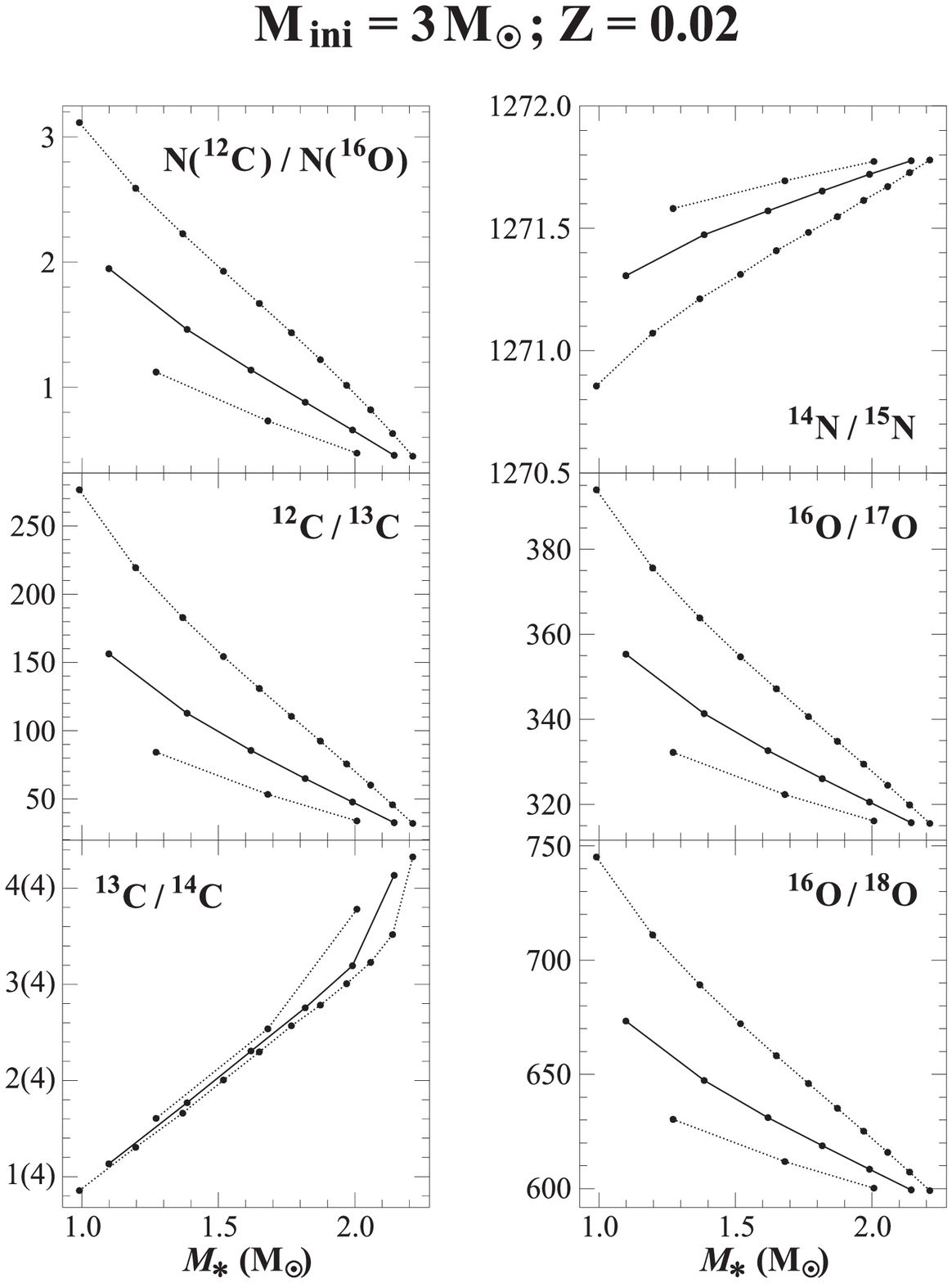,width=18.0cm}
  \caption[ ]{\label{xsurf302a}
    Surface isotopic ratios as a function of the remaining total stellar
    mass, that decreases due to the mass loss along the TP-AGB.  The solid
    line refers to the standard mass loss rate (see Sect. 6.3.2 for the
    corresponding $\eta$ parameter values in the Reimers relation).  Dotted
    lines correspond to mass loss rates increased or decreased by a factor
    of two. The extrapolated thermal pulses (and 3DUP episodes) are
    indicated by the filled circles, and are of course more numerous for
    lower mass loss rates. Here are indicated the \chem{C}{12}/\chem{O}{16},
    \chem{C}{12}/\chem{C}{13}, \chem{C}{13}/\chem{C}{14},
    \chem{N}{14}/\chem{N}{15}, \chem{O}{16}/\chem{O}{17} and
    \chem{O}{16}/\chem{O}{18} ratios, in the case of the \mass{3} ($Z =
    0.02$) star. All ratios are in mass fractions, except for the
    \chem{C}{12}/\chem{O}{16} ratio (given in number abundances) }
\end{figure*}

\begin{figure*}
  \psfig{file=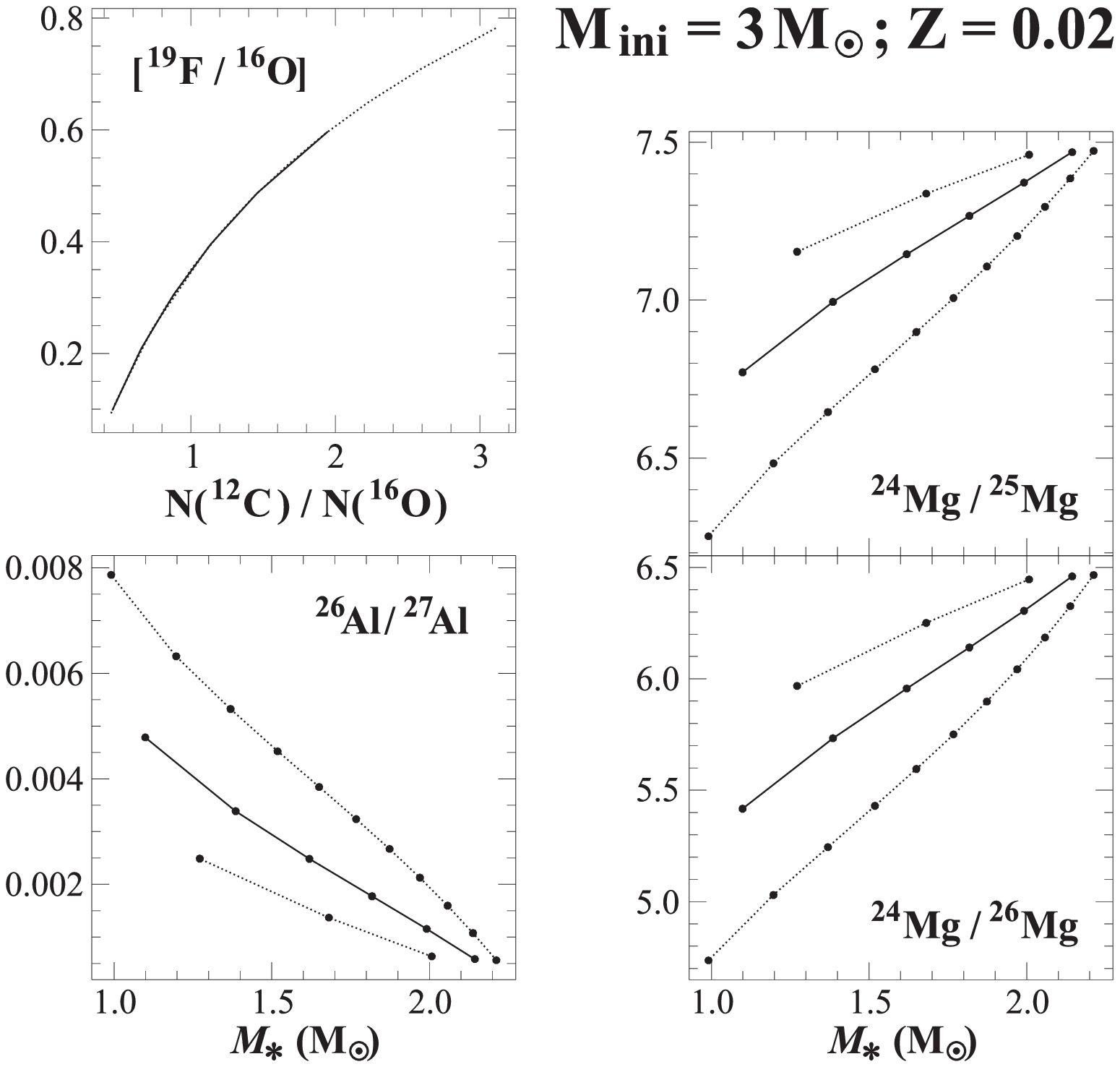,width=18.0cm}
  \caption[ ]{\label{xsurf302b}
    \chem{Mg}{24}/\chem{Mg}{25}, \chem{Mg}{24}/\chem{Mg}{26},
    \chem{Al^g}{26}/\chem{Al}{27} ratios as well as
    [\chem{F}{19}/\chem{O}{16}] as a function of \chem{C}{12}/\chem{O}{16},
    in the case of the \mass{3} ($Z = 0.02$) star. All ratios are in mass
    fractions and [A/B] means $\log {\rm (A/B)}_{*} - \log {\rm (A/B)}_{\sun}$
    }
\end{figure*}

Figs. \ref{xsurf302a} and \ref{xsurf302b} display, for the \mass{3} AGB star
with $Z = 0.02$, the evolution of the isotopic ratios as a function of the
total remaining mass that decreases due to mass loss. Similar figures for
the other six AGB stars are only displayed in the electronic version.

Let us stress that we stop our calculations at the last thermal pulse that
occurs before the convective envelope mass is reduced below \mass{0.2}.
This stage indeed roughly corresponds to the PN ejection that we do not
maintain to model here. This explains the quite large remaining total stellar
masses in the figures, especially for those computed with the highest
mass loss rates.

The first comment concerns the formation of carbon stars. \mass{3} AGB
stars with $Z = 0.02$ and 3 and \mass{4} AGB stars with $Z = 0.005$
become carbon stars during their TP-AGB phase. The surface
\chem{C}{12}/\chem{O}{16} becomes greater than unity more rapidly for
{\it (i)} decreasing total mass and/or {\it (ii)} decreasing $Z$. This
metallicity dependence is confirmed by the increase of C stars compared
to M stars from the galactic bulge to LMC and SMC (Blanco et al. 1978).
Our \mass{6} (\mass{5}) AGB stars with $Z = 0.02$ (0.005) can also become
C stars with our lowest mass loss rates. In such cases indeed, in spite
of very efficient HBB transforming \chem{C}{12} in \chem{C}{13} and
\chem{N}{14}, many 3DUP events can still occur with very reduced envelope
masses, so that the relatively short inter-pulse phases are not long enough
to substantially destroy \chem{C}{12}. We will see however, together with
other isotopic ratios, that these very low mass loss rates seem actually
extreme. Let us emphasize that the only super-lithium-rich star (see
Sect. 7.2) we found to be also a carbon star is the \mass{5} with $Z =
0.005$ one (with our medium mass loss rate). However, as already
mentioned in Sect. 6.2, observations indicate that $Z = 0.02$ carbon
stars are formed easier (i.e. at lower luminosities) than predicted by
all the models. Let us also stress the very high
\chem{C}{12}/\chem{C}{13} isotopic ratios that can be expected at the
surface of $\le$ \mass{4} AGB stars, whatever $Z$. On the other hand,
more massive AGB stars experience HBB that progressively leads the
\chem{C}{12}/\chem{C}{13} isotopic ratio close to its CN cycle
equilibrium value (formation of J stars). The case of \mass{4} stars
with $Z = 0.005$ very clearly shows the competition between the 3DUP episodes
(increasing \chem{C}{12}/\chem{C}{13}) and HBB (decreasing it). For lower
mass loss rates, the 3DUP events are more numerous for the same total mass
decrease, leading to a more rapid \chem{C}{12}/\chem{C}{13} increase.
However, the corresponding longer TP-AGB phase allows the base of the
convective envelope to reach higher temperatures that finally destroy
\chem{C}{12} through the CN cycle. This efficient transformation of
\chem{C}{12} to \chem{C}{13} also reduces the \chem{C}{12}/\chem{O}{16}
ratio, i.e. the possibility to form carbon stars. It happens if the
temperatures at the base of the convective envelopes become higher than
$\sim 75 \, 10^6$ K. Such stars have corresponding surface luminosities
$>$ \lum{25000} (\lum{30000}) for $Z = 0.02$ (0.005), i.e. $M_{bol} <
-6.25$ (-6.45). We consequently predict that C stars cannot be formed at
higher luminosities than this threshold, due to strong HBB. This is in
agreement with Boothroyd et al. (1993) predictions. This also corresponds
to the observed threshold for AGB stars of the LMC (Cohen et al. 1981).

Finally, note the surface pollution in \chem{C}{14} that is more important,
compared to that of \chem{C}{13}, for decreasing total masses. For \mass{3}
AGB stars, the surface \chem{C}{13}/\chem{C}{14} ratio is quite independent
of the mass loss rate. This is due to the inter-pulse duration that is
significantly greater than the \chem{C}{14} lifetime, so that the
\chem{C}{14} decay is the dominant factor for the \chem{C}{13}/\chem{C}{14}
ratio evolution at the surface of such stars. The quite complicated behavior
of the \chem{C}{13}/\chem{C}{14} isotopic ratio at the surface of our most
massive models is again due to the competition between the 3DUP that adds
\chem{C}{14} into the convective envelope and HBB that can partially destroy
it. By extrapolation, it is reasonable to expect even lower
\chem{C}{13}/\chem{C}{14} ratios in low-mass AGB stars. As mentioned in
Forestini et al. (1996), \chem{C}{14} could become detectable for $<$
\mass{3} evolved AGB stars.

Due to HBB, those stars showing very low surface \chem{C}{12}/\chem{C}{13}
ratios are also expected to have very high \chem{N}{14}/\chem{N}{15} ones.
It has to be stressed indeed that in our most massive AGB stars, the CN
cycle operating inside the convective envelope produces large amounts of
\chem{N}{14}.

The \chem{O}{16}/\chem{O}{17} and \chem{O}{16}/\chem{O}{18} surface isotopic
ratios also evolve as expected from our discussions in Sect. 7.  Note that
quite low \chem{O}{16}/\chem{O}{17} and very high \chem{O}{16}/\chem{O}{18}
ratios, which are clear signatures of the operation of the ON cycle at the
base of the convective envelope (HBB), appear rather simultaneously with the
CN cycle signatures in our most massive objects. Also remark the progressive
\chem{O}{16}/\chem{O}{18} decrease in our \mass{6} (\mass{5}) AGB star with
$Z = 0.02$ (0.005).  For these objects, the temperature at the base of the
convective envelope finally becomes high enough for \chem{O}{16} to
partially burn.

The \chem{F}{19} surface enhancements we predict are displayed in a special
form, corresponding to the observation presentations adopted by Jorissen et
al. (1992). Compared to their Fig. 8, two conclusions can be stated.

\begin{itemize}

\item
  There is a clear observed correlation between the \chem{F}{19} excess
  and the degree of \chem{C}{12} surface pollution, that is perfectly
  understandable from the identified nucleosynthetic processes and the
  mixing episodes (see Sect. 7). Furthermore, our models clearly indicate a
  trend: the less massive the AGB star, the more efficient the
  \chem{F}{19} surface enhancement. In fact, our most massive models even
  show a surface \chem{F}{19} depletion due to its partial destruction by
  HBB.

\item
  Our $Z = 0.02$ models well reproduce only the less \chem{F}{19} enriched
  stars in the [\chem{F}{19}/\chem{O}{16}] versus
  \chem{C}{12}/\chem{O}{16} diagram. Based on the trend just mentioned, we
  predict that the AGB stars showing very large \chem{F}{19} surface
  enhancements have initial total masses $<$ \mass{3}. We plan to verify
  this statement, i.e. that low-mass AGB stars are the most efficient
  \chem{F}{19} galactic producers (see Sect. 9).

\end{itemize}

As expected, the \chem{Mg}{24}/\chem{Mg}{25} and \chem{Mg}{24}/\chem{Mg}{26}
surface isotopic ratios are much less sensitive to the 3DUP episodes and
HBB. More precisely, only our most massive models with the lowest mass loss
rates show large \chem{Mg}{24} depletions due to a very strong HBB. Again,
these are however quite extreme situations.

Last but not least, we predict substantial \chem{Al^g}{26} surface
enhancements due to the repetitive dredges-up of the HBS and inter-shell
regions. The \chem{Al^g}{26}/\chem{Al}{27} isotopic ratio tends to decrease
with increasing total mass due to {\it (i)} the greater dilution as mixing
occurs in a more massive convective envelope and {\it (ii)} the thinner HBS
and inter-shell regions (in mass). Nevertheless,
\chem{Al^g}{26}/\chem{Al}{27} could well be increasing again if
\chem{Al^g}{26} was produced inside the convective envelope itself by strong
HBB. Note however that such a very high \chem{Al^g}{26} production level is
actually expected to be accompanied by very low Mg isotopic ratios, as
\chem{Al^g}{26} is produced by the MgAl chain in the extreme situations
mentioned above. Let us finally mention that Wasserburg et al. (1994)
presented computations in which the evolution of the \chem{Al^g}{26} surface
abundance has been followed self-consistently, by including the effect of
neutron irradiation due to an enhanced amount of \chem{C}{13} in the
inter-shell region (to engender a $s-$process).

Our results compare very well with those obtained by Boothroyd et al.
(1994, 1995) concerning the O isotopic ratios in intermediate-mass stars.
They also mention the problem to reproduce the \chem{O}{16}/\chem{O}{17}
ratio, especially in low-mass carbon AGB stars. We cannot go further in this
comparison without having yet computed the TP-AGB phase of low-mass stars.
Boothroyd et al. (1995) already suggested that to conciliate all the data
with observations of low-mass stars, one probably has to take into account
of slow-particle transports inside radiative zones, what they called
``cool-bottom precessing''. That is what we plan to do (see Sect. 9).

\subsection {Other comparisons with observations}

Evolutionary models of TP-AGB stars can be strongly constrained by the
confrontation between predicted and observed surface isotopic ratios,
especially if different ratios are observed for the same stars. For most of
the observed AGB stars, only the C and O isotopic ratios have been
determined (see e.g. Harris \& Lambert 1984; Harris et al. 1985; Lambert et
al. 1986; Harris et al. 1987; Harris \& Lambert 1987; Kahane et al. 1992).
Globally, we found agreements. General features come out.

\begin{itemize}

\item
  As the temperatures required at the base of the convective envelope to
  allow \chem{Li}{7} production (see Sect. 7.2) are significantly lower
  than those allowing efficient \chem{C}{12} destruction by the CN cycle,
  we predict that luminous C stars enriched in \chem{Li}{7} can exist,
  especially for lower $Z$ objects (that have lower \chem{O}{16} content).
  There are indeed a few \% of the observed C stars that are also
  super-lithium-rich and there are more numerous in the LMC compared to our
  galaxy (see e.g. Smith \& Lambert 1990b for the LMC and Feast 1974 for the
  Galaxy).

\item
  There is no observed C star with \chem{C}{12}/\chem{C}{13} $< 15$. Most of
  the J stars show \chem{C}{12}/\chem{C}{13} ratios between 3 et 4, i.e.
  very close to the CN cycle equilibrium value.

\item
  J stars also show rather low \chem{O}{16}/\chem{O}{17} isotopic ratios
  while only upper limits have been found for the \chem{O}{16}/\chem{O}{18}
  ratio. Most of the other AGB stars present \chem{O}{16}/\chem{O}{17}
  ratios ranged between 300 and 4000 and \chem{O}{16}/\chem{O}{18} ratios
  ranged between 500 and 5000. Only stars with rather high
  \chem{O}{16}/\chem{O}{17} ratios are difficult to explain with our
  intermediate-mass AGB models (see also Boothroyd et al. 1995). Also note
  that the corresponding O isotopic ratios in the ISM are
  \chem{O}{16}/\chem{O}{17} $= 2460^{+750}_{-750}$ and
  \chem{O}{16}/\chem{O}{18} $= 675^{+200}_{-200}$. Again, there is a problem
  with \chem{O}{17}. Let us however stress that important uncertainties
  remain concerning the nuclear cross sections of proton captures on
  \chem{O}{17}. On the other hand, low-mass AGB stars are present among
  those observations.

\end{itemize}

Presently, there is only one star for which all the isotopic ratios have
been determined. It is Cw Leo. This C star is one of the nearest evolved
AGB star ever observed. Its distance is estimated to $d = 150 \pm 50$ pc
(see e.g. Claussen et al. 1987; Keady et al. 1988; Griffin 1990). This
corresponds to a surface luminosity $L \simeq$ \lum{20000 \pm 5000} and a
present mass loss rate of roughly $(2.5 \pm 1.0) \, 10^{-5}$ \mass{}$\,
\rm{yr^{-1}}$. Such a high wind makes it quite optically thick. A very
extended circumstellar envelope has been detected to surround Cw Leo,
IRC+10216. IRC+10216 shows evidences for a bi-polar structure (see e.g.
Gu\'elin et al. 1993), attesting that Cw Leo is very evolved and close to
the end of its TP-AGB phase. Consequently, its present convective envelope
mass is probably very reduced ($\la$ \mass{0.5} typically).

The circumstellar envelope of Cw Leo allowed very accurate determinations of
the isotopic ratios (see Kahane et al. 1988 and 1992; Gu\'elin et al.  1995;
Forestini et al. 1996): \chem{C}{12}/\chem{C}{13} $= 44^{+3}_{-3}$,
\chem{C}{13}/\chem{C}{14} $> 1400$, \chem{N}{14}/\chem{N}{15} $> 5300$,
\chem{O}{16}/\chem{O}{17} $= 840^{+230}_{-170}$, \chem{O}{16}/\chem{O}{18}
$= 1260^{+315}_{-240}$, \chem{Mg}{24}/\chem{Mg}{25} $= 7.6^{+0.8}_{-1.1}$,
\chem{Mg}{24}/\chem{Mg}{26} $= 6.5^{+0.7}_{-0.7}$ and
\chem{Al^g}{26}/\chem{Al}{27} $\ga 0.01$ (to be confirmed). Furthermore, the
Si isotopic ratios appear to have interstellar medium values, i.e.
unaltered by the star. As derived in Gu\'elin et al. (1995) and Forestini et
al. (1996), almost all these isotopic ratios seem to be consistent (within
the error bars), following our predictions, with an AGB star that had a main
sequence total mass between 4 and \mass{5}. However,
\chem{O}{16}/\chem{O}{17} is again not well explained. As various surface
isotopic ratios (especially \chem{N}{14}/\chem{N}{15} and
\chem{O}{16}/\chem{O}{18}) are significantly changing between 4 and \mass{5}
AGB models, we are presently computing the TP-AGB phase of a \mass{4.5} star
with $Z = 0.02$ in order to improve this last prediction.

Finally, another very interesting way to constraint stellar evolution models
of AGB stars is to compare the predicted surface isotopic ratios with those
determined in primitive meteorites (mainly through grains included inside
carbon rich chondrites). Most of the \chem{C}{12}/\chem{C}{13},
\chem{N}{14}/\chem{N}{15} and \chem{Al^g}{26}/\chem{Al}{27} isotopic
anomalies measured in grains (see e.g. Ott 1991; Zinner et al. 1991a) are
consistent with those coming from the surface of evolved TP-AGB stars of
intermediate-mass. This is of course in favor of the presence, when the
solar system has been formed, of one or a few AGB stars in its surroundings.
Such grains are indeed well known to be formed in the cool atmosphere of
evolved AGB stars (i.e. when mass loss rates become rather high). Concerning
the measured O isotopic ratios, the situation is not so clear as many grains
have ratios significantly out of the corresponding predicted ranges. Some of
these grains at least can have been formed in low-mass AGB stars. Again, as
mentioned in Sect. 8.1, Boothroyd et al. (1995) recently suggested that in
such stars, cool-bottom processing can help to reduce these discrepancies.
On the contrary, some other grains perfectly agree with our predictions.
This is for example the case of a $\sim 3 \,\rm{\mu m}$ $\rm{Al_2 O_3}$
grain of the Bishunpur LL3.1 chondrite by Huss et al.  (1994). The derived
ratios, \chem{O}{16}/\chem{O}{17} $= 385 \pm 5$, \chem{O}{16}/\chem{O}{18}
$= 853 \pm 30$ and \chem{Al^g}{26}/\chem{Al}{27} $= (1.7 \pm 0.2) \,
10^{-3}$, are in perfect agreement with ratios corresponding to a $3 -$
\mass{4} AGB star of nearly solar metallicity.  Finally, other SiC
\chem{Al^g}{26}-rich grains raise problems, like that discovered by Zinner
et al. (1991b), exhibiting a very high \chem{Al^g}{26}/\chem{Al}{27} ratio
of $\simeq 0.23$.  It seems hard to explain such a value from our models.
However, this grain also exhibits an anomalously large
\chem{Ca}{44}/\chem{Ca}{40} isotopic ratio that seems to exclude its AGB
star origin.

\subsection {Yields}

Tables \ref{yield1} to \ref{yield3} give the total mass ejected by the wind
(in \mass{}) for various elements {\it (i)} at the top of the first RGB,
{\it (ii)} at the beginning of the TP-AGB phase and {\it (iii)} at the end
of the TP-AGB phase (thus including our extrapolated nucleosynthesis
calculations). As such data mainly concerns the chemical evolution of
galaxies, we also present the net yield (positive or negative) for each
element. The data of Tables \ref{yield1} to \ref{yield3} correspond to the
medium mass loss rates (see Sect. 8.1), i.e. the solid lines of Figs.
\ref{xsurf302a} and \ref{xsurf302b}. Brief comments are made about these
predictions in the electronic version.

\begin{table*}
  \caption[ ]{\label{yield1}
    Total mass ejected (in \mass{}) in the wind for some nuclides at the RGB
    top, at the end of the E-AGB phase and at the AGB tip, just before the
    ejection of the planetary nebula. Also given is the net contribution for
    these elements, i.e. the total final mass ejected reduced by the amount
    of matter that would have been ejected with the initial composition.
    Here are presented elements from He to N }
  \begin{flushleft}
    \begin{tabular}{llccccccc}
      \noalign{\smallskip}
      \hline
      \noalign{\smallskip}
      & &\multicolumn{4}{c}{$Z = 0.02$}&\multicolumn{3}{c}{$Z = 0.005$} \\
      \noalign{\vspace{-2.0mm}}
      & &\multicolumn{4}{c}{\makebox[73mm]{\downbracefill}}&\multicolumn{3}{c}{\makebox[53mm]{\downbracefill}}\\
      & &\mass{3}&\mass{4}&\mass{5}&\mass{6}&\mass{3}&\mass{4}&\mass{5} \\
      \noalign{\smallskip}
      \hline \hline
      \noalign{\smallskip}
      $M \, (\rm{M_{\odot}})$&{\it at RG top}&2.9978&3.9985&4.9970&5.9980&2.9996&3.9999&4.9997 \\
      &{\it at $t_{0}$}&2.9151&3.7732&4.7009&5.7167&2.8053&3.5067&4.3149 \\
      &{\it at AGB tip}&1.0989&1.0143&1.0750&1.3735& 0.9471&1.2519&1.3818 \\
      \noalign{\smallskip}
      \noalign{\smallskip}
      \chem{He}{3}&{\it at RG top}&2.59$\, 10^{ -7}$&1.51$\, 10^{ -7}$&2.93$\, 10^{ -7}$&1.59$\, 10^{ -7}$&7.73$\, 10^{ -8}$&9.70$\, 10^{ -9}$&3.54$\, 10^{ -8}$ \\
      &{\it at $t_{0}$}&1.30$\, 10^{ -5}$&2.35$\, 10^{ -5}$&2.41$\, 10^{ -5}$&1.96$\, 10^{ -5}$&3.26$\, 10^{ -5}$&5.70$\, 10^{ -5}$&6.52$\, 10^{ -5}$ \\
      &{\it at AGB tip}&2.87$\, 10^{ -4}$&3.03$\, 10^{ -4}$&1.57$\, 10^{ -4}$&5.00$\, 10^{ -5}$&3.30$\, 10^{ -4}$&3.04$\, 10^{ -4}$&9.19$\, 10^{ -5}$ \\
      &{\it net yield}& 2.25$\, 10^{ -4}$& 2.07$\, 10^{ -4}$& 3.46$\, 10^{ -5}$&-9.44$\, 10^{ -5}$& 2.74$\, 10^{ -4}$& 2.29$\, 10^{ -4}$&-4.27$\, 10^{ -6}$ \\
      \noalign{\smallskip}
      \chem{He}{4}&{\it at RG top}&6.53$\, 10^{ -4}$&4.28$\, 10^{ -4}$&8.69$\, 10^{ -4}$&5.84$\, 10^{ -4}$&1.15$\, 10^{ -4}$&1.89$\, 10^{ -5}$&8.10$\, 10^{ -5}$ \\
      &{\it at $t_{0}$}&2.59$\, 10^{ -2}$&6.91$\, 10^{ -2}$&9.59$\, 10^{ -2}$&9.13$\, 10^{ -2}$&5.11$\, 10^{ -2}$&0.13&0.18 \\
      &{\it at AGB tip}&0.59&0.91&1.32&1.59&0.55&0.72&1.28 \\
      &{\it net yield}& 3.76$\, 10^{ -2}$& 3.20$\, 10^{ -2}$& 0.17& 0.23& 3.56$\, 10^{ -2}$& 3.70$\, 10^{ -2}$& 0.37 \\
      \noalign{\smallskip}
      \chem{Li}{7}&{\it at RG top}&1.59$\, 10^{-11}$&1.02$\, 10^{-11}$&2.49$\, 10^{-11}$&1.45$\, 10^{-11}$&1.42$\, 10^{-14}$&2.67$\, 10^{-15}$&1.26$\, 10^{-14}$ \\
      &{\it at $t_{0}$}&2.63$\, 10^{-11}$&3.35$\, 10^{-11}$&5.18$\, 10^{-11}$&3.87$\, 10^{-11}$&4.47$\, 10^{-12}$&1.15$\, 10^{-11}$&1.96$\, 10^{-11}$ \\
      &{\it at AGB tip}&2.92$\, 10^{-10}$&6.35$\, 10^{-11}$&7.33$\, 10^{ -8}$&1.08$\, 10^{ -7}$&8.23$\, 10^{-11}$&4.60$\, 10^{-10}$&3.57$\, 10^{-11}$ \\
      &{\it net yield}&-1.85$\, 10^{ -8}$&-2.95$\, 10^{ -8}$& 3.44$\, 10^{ -8}$& 6.23$\, 10^{ -8}$&-5.00$\, 10^{ -9}$&-6.34$\, 10^{ -9}$&-8.92$\, 10^{ -9}$ \\
      \noalign{\smallskip}
      \chem{C}{12}&{\it at RG top}&6.44$\, 10^{ -6}$&4.25$\, 10^{ -6}$&9.11$\, 10^{ -6}$&5.84$\, 10^{ -6}$&2.14$\, 10^{ -7}$&4.01$\, 10^{ -8}$&1.90$\, 10^{ -7}$ \\
      &{\it at $t_{0}$}&1.72$\, 10^{ -4}$&4.53$\, 10^{ -4}$&5.90$\, 10^{ -4}$&5.56$\, 10^{ -4}$&9.01$\, 10^{ -5}$&2.34$\, 10^{ -4}$&3.43$\, 10^{ -4}$ \\
      &{\it at AGB tip}&1.26$\, 10^{ -2}$&9.71$\, 10^{ -3}$&5.97$\, 10^{ -3}$&2.74$\, 10^{ -3}$&1.06$\, 10^{ -2}$&4.40$\, 10^{ -3}$&9.68$\, 10^{ -4}$ \\
      &{\it net yield}& 6.53$\, 10^{ -3}$& 1.33$\, 10^{ -4}$&-6.62$\, 10^{ -3}$&-1.21$\, 10^{ -2}$& 9.00$\, 10^{ -3}$& 2.19$\, 10^{ -3}$&-1.93$\, 10^{ -3}$ \\
      \noalign{\smallskip}
      \chem{C}{13}&{\it at RG top}&1.26$\, 10^{ -7}$&8.52$\, 10^{ -8}$&1.48$\, 10^{ -7}$&1.14$\, 10^{ -7}$&1.14$\, 10^{ -8}$&2.17$\, 10^{ -9}$&1.04$\, 10^{ -8}$ \\
      &{\it at $t_{0}$}&8.75$\, 10^{ -6}$&2.41$\, 10^{ -5}$&3.23$\, 10^{ -5}$&3.09$\, 10^{ -5}$&4.90$\, 10^{ -6}$&1.31$\, 10^{ -5}$&1.91$\, 10^{ -5}$ \\
      &{\it at AGB tip}&1.95$\, 10^{ -4}$&3.12$\, 10^{ -4}$&1.00$\, 10^{ -3}$&6.67$\, 10^{ -4}$&5.05$\, 10^{ -5}$&9.09$\, 10^{ -5}$&1.93$\, 10^{ -4}$ \\
      &{\it net yield}& 1.22$\, 10^{ -4}$& 1.97$\, 10^{ -4}$& 8.50$\, 10^{ -4}$& 4.89$\, 10^{ -4}$& 3.06$\, 10^{ -5}$& 6.43$\, 10^{ -5}$& 1.58$\, 10^{ -4}$ \\
      \noalign{\smallskip}
      \chem{C}{14}&{\it at RG top}&4.89$\, 10^{-30}$&4.77$\, 10^{-29}$&3.22$\, 10^{-21}$&5.69$\, 10^{-19}$&2.99$\, 10^{-37}$&9.59$\, 10^{-27}$&3.23$\, 10^{-21}$ \\
      &{\it at $t_{0}$}&6.92$\, 10^{-26}$&4.77$\, 10^{-29}$&3.62$\, 10^{-21}$&5.56$\, 10^{-16}$&1.68$\, 10^{-32}$&4.92$\, 10^{-24}$&3.38$\, 10^{-18}$ \\
      &{\it at AGB tip}&6.20$\, 10^{ -9}$&8.93$\, 10^{ -9}$&2.90$\, 10^{ -9}$&1.16$\, 10^{ -9}$&8.54$\, 10^{-10}$&1.43$\, 10^{ -9}$&4.35$\, 10^{-12}$ \\
      &{\it net yield}& 6.20$\, 10^{ -9}$& 8.93$\, 10^{ -9}$& 2.90$\, 10^{ -9}$& 1.16$\, 10^{ -9}$& 8.54$\, 10^{-10}$& 1.43$\, 10^{ -9}$& 4.35$\, 10^{-12}$ \\
      \noalign{\smallskip}
      \chem{N}{14}&{\it at RG top}&3.41$\, 10^{ -6}$&2.22$\, 10^{ -6}$&3.92$\, 10^{ -6}$&2.99$\, 10^{ -6}$&3.25$\, 10^{ -7}$&4.47$\, 10^{ -8}$&1.69$\, 10^{ -7}$ \\
      &{\it at $t_{0}$}&2.35$\, 10^{ -4}$&6.53$\, 10^{ -4}$&9.26$\, 10^{ -4}$&9.01$\, 10^{ -4}$&1.53$\, 10^{ -4}$&3.60$\, 10^{ -4}$&4.66$\, 10^{ -4}$ \\
      &{\it at AGB tip}&5.24$\, 10^{ -3}$&8.52$\, 10^{ -3}$&1.84$\, 10^{ -2}$&2.89$\, 10^{ -2}$&1.58$\, 10^{ -3}$&2.07$\, 10^{ -3}$&1.04$\, 10^{ -2}$ \\
      &{\it net yield}& 3.01$\, 10^{ -3}$& 5.03$\, 10^{ -3}$& 1.38$\, 10^{ -2}$& 2.35$\, 10^{ -2}$& 9.75$\, 10^{ -4}$& 1.26$\, 10^{ -3}$& 9.33$\, 10^{ -3}$ \\
      \noalign{\smallskip}
      \chem{N}{15}&{\it at RG top}&8.89$\, 10^{ -9}$&5.83$\, 10^{ -9}$&1.28$\, 10^{ -8}$&8.02$\, 10^{ -9}$&2.40$\, 10^{-10}$&4.36$\, 10^{-11}$&2.02$\, 10^{-10}$ \\
      &{\it at $t_{0}$}&1.92$\, 10^{ -7}$&4.93$\, 10^{ -7}$&6.33$\, 10^{ -7}$&5.89$\, 10^{ -7}$&1.01$\, 10^{ -7}$&2.56$\, 10^{ -7}$&3.70$\, 10^{ -7}$ \\
      &{\it at AGB tip}&4.13$\, 10^{ -6}$&6.32$\, 10^{ -6}$&1.97$\, 10^{ -6}$&2.66$\, 10^{ -6}$&1.02$\, 10^{ -6}$&9.13$\, 10^{ -7}$&5.13$\, 10^{ -6}$ \\
      &{\it net yield}&-4.65$\, 10^{ -6}$&-7.47$\, 10^{ -6}$&-1.62$\, 10^{ -5}$&-1.87$\, 10^{ -5}$&-1.35$\, 10^{ -6}$&-2.26$\, 10^{ -6}$& 9.53$\, 10^{ -7}$ \\
      \noalign{\smallskip}
      \hline
      \noalign{\smallskip}
    \end{tabular}
  \end{flushleft}
\end{table*}

\begin{table*}
  \caption[ ]{\label{yield2}
    Same as Table 8 for elements from O to Na
    }
  \begin{flushleft}
    \begin{tabular}{llccccccc}
      \noalign{\smallskip}
      \hline
      \noalign{\smallskip}
      & &\multicolumn{4}{c}{$Z = 0.02$}&\multicolumn{3}{c}{$Z = 0.005$} \\
      \noalign{\vspace{-2.0mm}}
      & &\multicolumn{4}{c}{\makebox[73mm]{\downbracefill}}&\multicolumn{3}{c}{\makebox[53mm]{\downbracefill}}\\
      & &\mass{3}&\mass{4}&\mass{5}&\mass{6}&\mass{3}&\mass{4}&\mass{5} \\
      \noalign{\smallskip}
      \hline \hline
      \noalign{\smallskip}
      \chem{O}{16}&{\it at RG top}&2.23$\, 10^{ -5}$&1.47$\, 10^{ -5}$&3.00$\, 10^{ -5}$&2.00$\, 10^{ -5}$&1.10$\, 10^{ -6}$&1.92$\, 10^{ -7}$&8.22$\, 10^{ -7}$ \\
      &{\it at $t_{0}$}&8.33$\, 10^{ -4}$&2.20$\, 10^{ -3}$&2.84$\, 10^{ -3}$&2.67$\, 10^{ -3}$&4.68$\, 10^{ -4}$&1.21$\, 10^{ -3}$&1.69$\, 10^{ -3}$ \\
      &{\it at AGB tip}&1.92$\, 10^{ -2}$&2.86$\, 10^{ -2}$&3.61$\, 10^{ -2}$&3.96$\, 10^{ -2}$&5.34$\, 10^{ -3}$&6.74$\, 10^{ -3}$&4.17$\, 10^{ -3}$ \\
      &{\it net yield}&-1.36$\, 10^{ -4}$&-1.66$\, 10^{ -3}$&-3.72$\, 10^{ -3}$&-7.33$\, 10^{ -3}$& 1.27$\, 10^{ -4}$&-2.29$\, 10^{ -4}$&-5.01$\, 10^{ -3}$ \\
      \noalign{\smallskip}
      \chem{O}{17}&{\it at RG top}&2.18$\, 10^{ -8}$&1.15$\, 10^{ -8}$&1.51$\, 10^{ -8}$&1.26$\, 10^{ -8}$&4.19$\, 10^{ -9}$&2.09$\, 10^{-10}$&4.38$\, 10^{-10}$ \\
      &{\it at $t_{0}$}&2.62$\, 10^{ -6}$&5.56$\, 10^{ -6}$&5.45$\, 10^{ -6}$&4.51$\, 10^{ -6}$&1.87$\, 10^{ -6}$&3.63$\, 10^{ -6}$&3.17$\, 10^{ -6}$ \\
      &{\it at AGB tip}&5.87$\, 10^{ -5}$&7.26$\, 10^{ -5}$&7.14$\, 10^{ -5}$&1.17$\, 10^{ -4}$&1.94$\, 10^{ -5}$&2.03$\, 10^{ -5}$&1.05$\, 10^{ -5}$ \\
      &{\it net yield}& 5.09$\, 10^{ -5}$& 6.03$\, 10^{ -5}$& 5.52$\, 10^{ -5}$& 9.80$\, 10^{ -5}$& 1.73$\, 10^{ -5}$& 1.75$\, 10^{ -5}$& 6.78$\, 10^{ -6}$ \\
      \noalign{\smallskip}
      \chem{O}{18}&{\it at RG top}&4.74$\, 10^{ -8}$&3.13$\, 10^{ -8}$&6.61$\, 10^{ -8}$&4.29$\, 10^{ -8}$&1.79$\, 10^{ -9}$&3.36$\, 10^{-10}$&1.59$\, 10^{ -9}$ \\
      &{\it at $t_{0}$}&1.42$\, 10^{ -6}$&3.76$\, 10^{ -6}$&4.87$\, 10^{ -6}$&4.57$\, 10^{ -6}$&7.55$\, 10^{ -7}$&1.97$\, 10^{ -6}$&2.86$\, 10^{ -6}$ \\
      &{\it at AGB tip}&3.10$\, 10^{ -5}$&4.84$\, 10^{ -5}$&2.09$\, 10^{ -5}$&5.32$\, 10^{ -6}$&7.66$\, 10^{ -6}$&1.06$\, 10^{ -5}$&3.81$\, 10^{ -6}$ \\
      &{\it net yield}&-1.26$\, 10^{ -5}$&-2.01$\, 10^{ -5}$&-6.91$\, 10^{ -5}$&-1.01$\, 10^{ -4}$&-4.11$\, 10^{ -6}$&-5.16$\, 10^{ -6}$&-1.69$\, 10^{ -5}$ \\
      \noalign{\smallskip}
      \chem{F}{19}&{\it at RG top}&9.54$\, 10^{-10}$&6.23$\, 10^{-10}$&1.27$\, 10^{ -9}$&8.46$\, 10^{-10}$&4.88$\, 10^{-11}$&8.38$\, 10^{-12}$&3.55$\, 10^{-11}$ \\
      &{\it at $t_{0}$}&3.70$\, 10^{ -8}$&9.54$\, 10^{ -8}$&1.22$\, 10^{ -7}$&1.13$\, 10^{ -7}$&2.07$\, 10^{ -8}$&5.27$\, 10^{ -8}$&7.34$\, 10^{ -8}$ \\
      &{\it at AGB tip}&1.87$\, 10^{ -6}$&1.68$\, 10^{ -6}$&2.07$\, 10^{ -6}$&1.75$\, 10^{ -6}$&4.31$\, 10^{ -7}$&3.41$\, 10^{ -7}$&1.62$\, 10^{ -7}$ \\
      &{\it net yield}& 1.06$\, 10^{ -6}$& 3.96$\, 10^{ -7}$& 3.84$\, 10^{ -7}$&-2.33$\, 10^{ -7}$& 2.11$\, 10^{ -7}$& 4.65$\, 10^{ -8}$&-2.26$\, 10^{ -7}$ \\
      \noalign{\smallskip}
      \chem{Ne}{20}&{\it at RG top}&3.79$\, 10^{ -6}$&2.49$\, 10^{ -6}$&5.07$\, 10^{ -6}$&3.39$\, 10^{ -6}$&1.92$\, 10^{ -7}$&3.24$\, 10^{ -8}$&1.39$\, 10^{ -7}$ \\
      &{\it at $t_{0}$}&1.46$\, 10^{ -4}$&3.89$\, 10^{ -4}$&5.12$\, 10^{ -4}$&4.85$\, 10^{ -4}$&8.34$\, 10^{ -5}$&2.11$\, 10^{ -4}$&2.93$\, 10^{ -4}$ \\
      &{\it at AGB tip}&3.25$\, 10^{ -3}$&5.05$\, 10^{ -3}$&6.69$\, 10^{ -3}$&7.90$\, 10^{ -3}$&8.64$\, 10^{ -4}$&1.17$\, 10^{ -3}$&1.58$\, 10^{ -3}$ \\
      &{\it net yield}&-1.87$\, 10^{ -6}$&-6.51$\, 10^{ -5}$&-3.03$\, 10^{ -5}$&-2.98$\, 10^{ -5}$&-1.51$\, 10^{ -5}$&-4.53$\, 10^{ -6}$& 2.59$\, 10^{ -5}$ \\
      \noalign{\smallskip}
      \chem{Ne}{21}&{\it at RG top}&9.81$\, 10^{ -9}$&6.45$\, 10^{ -9}$&1.31$\, 10^{ -8}$&8.76$\, 10^{ -9}$&5.31$\, 10^{-10}$&8.75$\, 10^{-11}$&3.71$\, 10^{-10}$ \\
      &{\it at $t_{0}$}&3.83$\, 10^{ -7}$&1.03$\, 10^{ -6}$&1.35$\, 10^{ -6}$&1.27$\, 10^{ -6}$&2.37$\, 10^{ -7}$&5.86$\, 10^{ -7}$&7.85$\, 10^{ -7}$ \\
      &{\it at AGB tip}&1.01$\, 10^{ -5}$&1.42$\, 10^{ -5}$&9.29$\, 10^{ -6}$&2.71$\, 10^{ -6}$&2.83$\, 10^{ -6}$&3.36$\, 10^{ -6}$&1.15$\, 10^{ -6}$ \\
      &{\it net yield}& 1.79$\, 10^{ -6}$& 1.21$\, 10^{ -6}$&-7.85$\, 10^{ -6}$&-1.75$\, 10^{ -5}$& 5.92$\, 10^{ -7}$& 3.60$\, 10^{ -7}$&-2.80$\, 10^{ -6}$ \\
      \noalign{\smallskip}
      \chem{Ne}{22}&{\it at RG top}&2.99$\, 10^{ -7}$&1.97$\, 10^{ -7}$&4.06$\, 10^{ -7}$&2.69$\, 10^{ -7}$&1.38$\, 10^{ -8}$&2.54$\, 10^{ -9}$&1.11$\, 10^{ -8}$ \\
      &{\it at $t_{0}$}&1.06$\, 10^{ -5}$&2.81$\, 10^{ -5}$&3.66$\, 10^{ -5}$&3.45$\, 10^{ -5}$&5.85$\, 10^{ -6}$&1.54$\, 10^{ -5}$&2.21$\, 10^{ -5}$ \\
      &{\it at AGB tip}&9.45$\, 10^{ -4}$&6.81$\, 10^{ -4}$&7.75$\, 10^{ -4}$&8.94$\, 10^{ -4}$&2.40$\, 10^{ -4}$&1.39$\, 10^{ -4}$&1.44$\, 10^{ -4}$ \\
      &{\it net yield}& 6.83$\, 10^{ -4}$& 2.69$\, 10^{ -4}$& 2.35$\, 10^{ -4}$& 2.56$\, 10^{ -4}$& 1.69$\, 10^{ -4}$& 4.44$\, 10^{ -5}$& 1.98$\, 10^{ -5}$ \\
      \noalign{\smallskip}
      \chem{Na}{23}&{\it at RG top}&8.40$\, 10^{ -8}$&5.48$\, 10^{ -8}$&1.07$\, 10^{ -7}$&7.41$\, 10^{ -8}$&5.66$\, 10^{ -9}$&7.33$\, 10^{-10}$&2.92$\, 10^{ -9}$ \\
      &{\it at $t_{0}$}&4.11$\, 10^{ -6}$&1.13$\, 10^{ -5}$&1.55$\, 10^{ -5}$&1.49$\, 10^{ -5}$&2.60$\, 10^{ -6}$&6.08$\, 10^{ -6}$&7.76$\, 10^{ -6}$ \\
      &{\it at AGB tip}&9.47$\, 10^{ -5}$&1.49$\, 10^{ -4}$&2.16$\, 10^{ -4}$&2.51$\, 10^{ -4}$&2.87$\, 10^{ -5}$&3.50$\, 10^{ -5}$&2.80$\, 10^{ -5}$ \\
      &{\it net yield}& 2.76$\, 10^{ -5}$& 4.30$\, 10^{ -5}$& 7.72$\, 10^{ -5}$& 8.80$\, 10^{ -5}$& 1.06$\, 10^{ -5}$& 1.07$\, 10^{ -5}$&-3.98$\, 10^{ -6}$ \\
      \noalign{\smallskip}
      \hline
      \noalign{\smallskip}
    \end{tabular}
  \end{flushleft}
\end{table*}

\begin{table*}
  \caption[ ]{\label{yield3}
    Same as Table 8 for elements from Mg to Si
    }
  \begin{flushleft}
    \begin{tabular}{llccccccc}
      \noalign{\smallskip}
      \hline
      \noalign{\smallskip}
      & &\multicolumn{4}{c}{$Z = 0.02$}&\multicolumn{3}{c}{$Z = 0.005$} \\
      \noalign{\vspace{-2.0mm}}
      & &\multicolumn{4}{c}{\makebox[73mm]{\downbracefill}}&\multicolumn{3}{c}{\makebox[53mm]{\downbracefill}}\\
      & &\mass{3}&\mass{4}&\mass{5}&\mass{6}&\mass{3}&\mass{4}&\mass{5} \\
      \noalign{\smallskip}
      \hline \hline
      \noalign{\smallskip}
      \chem{Mg}{24}&{\it at RG top}&1.20$\, 10^{ -6}$&7.90$\, 10^{ -7}$&1.61$\, 10^{ -6}$&1.08$\, 10^{ -6}$&6.09$\, 10^{ -8}$&1.03$\, 10^{ -8}$&4.40$\, 10^{ -8}$ \\
      &{\it at $t_{0}$}&4.62$\, 10^{ -5}$&1.23$\, 10^{ -4}$&1.63$\, 10^{ -4}$&1.54$\, 10^{ -4}$&2.64$\, 10^{ -5}$&6.68$\, 10^{ -5}$&9.29$\, 10^{ -5}$ \\
      &{\it at AGB tip}&1.03$\, 10^{ -3}$&1.60$\, 10^{ -3}$&2.12$\, 10^{ -3}$&2.31$\, 10^{ -3}$&2.74$\, 10^{ -4}$&3.70$\, 10^{ -4}$&2.28$\, 10^{ -4}$ \\
      &{\it net yield}&-1.35$\, 10^{ -6}$&-2.55$\, 10^{ -5}$&-2.09$\, 10^{ -5}$&-2.09$\, 10^{ -4}$&-6.05$\, 10^{ -6}$&-4.73$\, 10^{ -6}$&-2.65$\, 10^{ -4}$ \\
      \noalign{\smallskip}
      \chem{Mg}{25}&{\it at RG top}&1.60$\, 10^{ -7}$&1.05$\, 10^{ -7}$&2.14$\, 10^{ -7}$&1.42$\, 10^{ -7}$&8.26$\, 10^{ -9}$&1.40$\, 10^{ -9}$&5.96$\, 10^{ -9}$ \\
      &{\it at $t_{0}$}&6.12$\, 10^{ -6}$&1.63$\, 10^{ -5}$&2.11$\, 10^{ -5}$&1.99$\, 10^{ -5}$&3.57$\, 10^{ -6}$&9.04$\, 10^{ -6}$&1.24$\, 10^{ -5}$ \\
      &{\it at AGB tip}&1.42$\, 10^{ -4}$&2.22$\, 10^{ -4}$&2.84$\, 10^{ -4}$&5.17$\, 10^{ -4}$&3.88$\, 10^{ -5}$&5.13$\, 10^{ -5}$&1.73$\, 10^{ -4}$ \\
      &{\it net yield}& 6.17$\, 10^{ -6}$& 7.72$\, 10^{ -6}$& 3.19$\, 10^{ -6}$& 1.85$\, 10^{ -4}$& 2.00$\, 10^{ -6}$& 2.06$\, 10^{ -6}$& 1.08$\, 10^{ -4}$ \\
      \noalign{\smallskip}
      \chem{Mg}{26}&{\it at RG top}&1.82$\, 10^{ -7}$&1.20$\, 10^{ -7}$&2.44$\, 10^{ -7}$&1.63$\, 10^{ -7}$&9.31$\, 10^{ -9}$&1.56$\, 10^{ -9}$&6.69$\, 10^{ -9}$ \\
      &{\it at $t_{0}$}&7.00$\, 10^{ -6}$&1.87$\, 10^{ -5}$&2.51$\, 10^{ -5}$&2.39$\, 10^{ -5}$&4.05$\, 10^{ -6}$&1.03$\, 10^{ -5}$&1.44$\, 10^{ -5}$ \\
      &{\it at AGB tip}&1.68$\, 10^{ -4}$&2.57$\, 10^{ -4}$&3.48$\, 10^{ -4}$&4.06$\, 10^{ -4}$&4.50$\, 10^{ -5}$&5.98$\, 10^{ -5}$&5.43$\, 10^{ -5}$ \\
      &{\it net yield}& 1.23$\, 10^{ -5}$& 1.22$\, 10^{ -5}$& 2.52$\, 10^{ -5}$& 2.57$\, 10^{ -5}$& 2.84$\, 10^{ -6}$& 3.39$\, 10^{ -6}$&-1.99$\, 10^{ -5}$ \\
      \noalign{\smallskip}
      \chem{Al^{g}}{26}&{\it at RG top}&1.20$\, 10^{-16}$&2.83$\, 10^{-15}$&1.01$\, 10^{-15}$&1.93$\, 10^{-14}$&1.29$\, 10^{-17}$&4.29$\, 10^{-21}$&1.11$\, 10^{-21}$ \\
      &{\it at $t_{0}$}&6.15$\, 10^{-16}$&1.39$\, 10^{-10}$&5.63$\, 10^{ -9}$&2.44$\, 10^{ -8}$&3.10$\, 10^{-14}$&6.12$\, 10^{-10}$&2.73$\, 10^{ -9}$ \\
      &{\it at AGB tip}&2.03$\, 10^{ -7}$&8.87$\, 10^{ -8}$&4.03$\, 10^{ -7}$&3.67$\, 10^{ -6}$&4.34$\, 10^{ -8}$&9.02$\, 10^{ -9}$&7.69$\, 10^{ -6}$ \\
      &{\it net yield}& 2.03$\, 10^{ -7}$& 8.87$\, 10^{ -8}$& 4.03$\, 10^{ -7}$& 3.67$\, 10^{ -6}$& 4.34$\, 10^{ -8}$& 9.02$\, 10^{ -9}$& 7.69$\, 10^{ -6}$ \\
      \noalign{\smallskip}
      \chem{Al}{27}&{\it at RG top}&1.36$\, 10^{ -7}$&8.91$\, 10^{ -8}$&1.82$\, 10^{ -7}$&1.22$\, 10^{ -7}$&6.89$\, 10^{ -9}$&1.16$\, 10^{ -9}$&4.97$\, 10^{ -9}$ \\
      &{\it at $t_{0}$}&5.21$\, 10^{ -6}$&1.39$\, 10^{ -5}$&1.84$\, 10^{ -5}$&1.74$\, 10^{ -5}$&2.99$\, 10^{ -6}$&7.57$\, 10^{ -6}$&1.05$\, 10^{ -5}$ \\
      &{\it at AGB tip}&1.18$\, 10^{ -4}$&1.83$\, 10^{ -4}$&2.45$\, 10^{ -4}$&2.94$\, 10^{ -4}$&3.27$\, 10^{ -5}$&4.38$\, 10^{ -5}$&7.42$\, 10^{ -5}$ \\
      &{\it net yield}& 1.49$\, 10^{ -6}$&-5.26$\, 10^{ -8}$& 3.78$\, 10^{ -6}$& 1.00$\, 10^{ -5}$& 1.24$\, 10^{ -6}$& 1.60$\, 10^{ -6}$& 1.87$\, 10^{ -5}$ \\
      \noalign{\smallskip}
      \chem{Si}{28}&{\it at RG top}&1.53$\, 10^{ -6}$&1.00$\, 10^{ -6}$&2.05$\, 10^{ -6}$&1.37$\, 10^{ -6}$&7.76$\, 10^{ -8}$&1.31$\, 10^{ -8}$&5.60$\, 10^{ -8}$ \\
      &{\it at $t_{0}$}&5.87$\, 10^{ -5}$&1.57$\, 10^{ -4}$&2.07$\, 10^{ -4}$&1.96$\, 10^{ -4}$&3.36$\, 10^{ -5}$&8.52$\, 10^{ -5}$&1.18$\, 10^{ -4}$ \\
      &{\it at AGB tip}&1.31$\, 10^{ -3}$&2.04$\, 10^{ -3}$&2.70$\, 10^{ -3}$&3.18$\, 10^{ -3}$&3.49$\, 10^{ -4}$&4.73$\, 10^{ -4}$&6.07$\, 10^{ -4}$ \\
      &{\it net yield}&-1.47$\, 10^{ -6}$&-2.76$\, 10^{ -5}$&-1.35$\, 10^{ -5}$&-1.84$\, 10^{ -5}$&-5.95$\, 10^{ -6}$&-2.20$\, 10^{ -6}$&-1.85$\, 10^{ -5}$ \\
      \noalign{\smallskip}
      \hline
      \noalign{\smallskip}
    \end{tabular}
  \end{flushleft}
\end{table*}

\section {Conclusions and prospects}

We tried to confirm and improve our present knowledge of the structural and
nucleosynthetic evolution of intermediate-mass AGB stars. To do that, we
have first presented in detail {\it (i)} the physics and numerical aspects
of our stellar evolution code, {\it (ii)} results concerning evolution
phases prior to the AGB one and {\it (iii)} comparisons with other works.
These informations are essential to better understand and appraise the large
set of predictions we discuss about the thermally pulsing AGB stage. We also
compare our predictions with various observations (at different evolutionary
stages) in order to clearly identify the processes that should be included
in future models.

Let us just mention global key features stemming from our intermediate-mass
evolution models.

\begin{itemize}

\item
  Concerning the structure (intensities, temperatures) of the thermal
  pulses as well as the thermal properties at the base of the convective
  envelope, the most decisive quantities are the total mass and core mass.
  Both change with time in a way that is governed by the increasing mass
  loss rate. This translates into a great sensitivity of the various
  nucleosynthesis processes to the adopted mass loss rate. Consequently,
  the evolution of the thermal pulse characteristics crucially depend on
  this last quantity, unfortunately still rather badly known. In addition,
  the lower the mass loss rates, the longer the TP-AGB phase, the more
  third dredge-up events have time to occur and the greater are the surface
  abundance changes due to HBB. This is clearly demonstrated by the high
  sensitiveness of the predicted surface evolution of almost all the
  isotopic ratios to the mass loss rate dependence on time.

\item
  On the other hand, at a given total stellar mass, core masses are greater
  for lower metallicities, as a result of the central burning phases.
  As a consequence, for the reason mentioned at the former point, TP-AGB
  stars with $Z = 0.005$ mostly behave like $Z = 0.02$ ones with a total
  mass greater by $\sim$ \mass{1}. However, due to important differences in
  initial compositions that also influence the nuclear reactions, this is
  not completely true as far as isotopic ratios are concerned. At the end
  of the TP-AGB phase indeed, some yields are quite different for both
  $Z$ we have studied.

\item
  The neutron captures on intermediate-mass and heavy nuclides occur in
  two distinct sites inside TP-AGB stars, namely {\it (i)} at the base of
  the inter-shell zone during the inter-pulse phase and, as already known,
  {\it (ii)} inside the convective tongues associated with thermal pulses.
  Along the asymptotic TP-AGB, the amount of produced neutrons is rather
  independent of the initial total mass and metallicity (at least for
  intermediate-mass objects).

\item
  Concerning chemical evolution of galaxies, intermediate-mass stars, at
  the end of their existence, appear to be rather significant producers of
  \chem{He}{3}, \chem{He}{4}, \chem{Li}{7} (most massive and/or higher $Z$
  stars), \chem{C}{13}, \chem{N}{14}, \chem{O}{17}, \chem{F}{19},
  \chem{Ne}{22}, \chem{Na}{23}, \chem{Mg}{26} and \chem{Al}{27}. A found
  trend, compared to observations, however clearly supports the idea that
  most of the \chem{F}{19} present in the interstellar medium has to come
  from low-mass AGB stars (that point has to be confirmed by forthcoming
  evolutionary models).

  Intermediate-mass AGB stars also substantially produce radio-nuclides,
  namely \chem{C}{14} and \chem{Al}{26}. The former one mostly comes from
  lower mass stars while it is the contrary for the last one.

  These stars partially deplete the interstellar medium content in
  \chem{Li}{7} (less massive stars), \chem{N}{15}, \chem{O}{16},
  \chem{O}{18}, \chem{Ne}{20}, \chem{Mg}{24} and \chem{Si}{28}.

\end{itemize}

We finish by recalling the major problems that, among all the present
intermediate-mass AGB models, remain to be solved.

\begin{itemize}

\item
  We have not yet identified the \chem{C}{13} additional source needed to
  quantitatively explain the $s-$process. This process is however required
  in order to account for many observations unquestionably indicating that
  AGB stars are responsible for the main component of the solar system heavy
  element distribution. Some suggestions have been made. Radiative diffusion
  operating during the TP-AGB phase of low-mass stars could substantially help.
  Although already invoked, such a process has never been yet included in AGB
  star evolution models.

\item
  The rather high level of \chem{Li}{7} production detected at the surface
  of some evolved but relatively faint AGB stars of our galaxy still
  remains unaccountable. Thus, the Cameron-Fowler mechanism does not
  explain all the super-lithium-rich AGB stars. As these galactic objects
  have relatively low initial masses, the explanation could also come from
  the radiative slow-transport processes above mentioned.

\item
  Last but not least, some C stars are observed, especially in the galactic
  bulge, with low luminosities at which the models fail at predicting the
  occurrence of the third dredge-up. This is probably due to our bad knowledge
  (and treatment) of finely tuned convection motions in boundaries of
  stellar regions having strong chemical composition gradients. Clearly,
  some kind of extra-mixing (deeper inside nuclearly processed regions)
  has to be invoked. However, no self-consistent mechanism has been settled
  and tested through complete evolution models up to now.

\end{itemize}

Other confrontations with observations require the modeling of low-mass AGB
stars. A clear distinction between low- and intermediate-mass stars is
justified. Indeed, numerous observations indicate that ``non-standard''
particle transport processes are acting inside low-mass stars at different
phases of their evolution, that substantially modify the chemical structure
compared to what is obtained in classical models. During the various
dredge-up episodes, matter up-heaved to the surface has consequently a
different composition. This is not the case in intermediate-mass stars.  In
conclusion, low-mass AGB stellar models, maintaining to make detailed
predictions concerning the evolution of surface isotopic ratios, have to
include such slow-particle transport processes. They are currently being
calculated and will be presented in a next-coming paper.

\begin{acknowledgements}
  We are first indebted to Maurizio Busso, the referee, for its very careful
  reading of our manuscript that helped us to significantly improve and
  clarify it. We also thank him and Roberto Gallino for the very instructive
  discussions we already had together about these very complex AGB stars.
  Let us finally thank Lionel Siess for his always valuable contributions to
  improve the stellar evolution code and related utilities.  Part of the
  computations presented in this paper (roughly representing 7 months of CPU
  time) were performed at the ``Center de Calcul Intensif de l'Observatoire
  de Grenoble''. Most of them have been realized at ``IMAG'' on a IBM SP1
  computer financed by the MESR, CNRS and R\'egion Rh\^one-Alpes.  This work
  was supported by grants from the GDR ``Structure Interne des Etoiles et
  des Plan\`etes G\'eantes'' (CNRS).
\end{acknowledgements}

\end{document}